\documentclass[journal,draftclsnofoot,onecolumn,12pt,twoside]{IEEEtran}
\IEEEoverridecommandlockouts

\usepackage[tbtags]{amsmath} 
\usepackage{amssymb}  
\usepackage{verbatim} 
\usepackage{amsxtra}  
\usepackage{comment}

\usepackage{mathabx}

\usepackage{enumerate}

\usepackage{amsfonts}
\usepackage{color}

\usepackage{subcaption}
\usepackage{cite}

\usepackage{amsthm}

\newtheorem{thm}{Theorem}
\newtheorem{cor}{Corollary}
\newtheorem{prop}{Proposition}

\theoremstyle{definition}
\newtheorem*{defn*}{Definition}
\newtheorem{scheme}{Scheme}
\newtheorem*{scheme*}{Scheme}

\theoremstyle{remark}
\newtheorem{remark}{Remark}

\numberwithin{thm}{section}
\numberwithin{lem}{section}
\numberwithin{claim}{section}
\numberwithin{assert}{section}
\numberwithin{cor}{section}
\numberwithin{prop}{section}
\numberwithin{defn}{section}
\numberwithin{remark}{section}
\numberwithin{algo}{section}
\numberwithin{scheme}{section}
\numberwithin{example}{section}

\usepackage{fdsymbol}

\usepackage{tikz}
\usetikzlibrary{shapes,arrows,decorations.markings,positioning,calc}

\tikzstyle{plant} = [draw, fill=red!5, rectangle, 
    minimum height=3em, minimum width=6em]
\tikzstyle{block} = [draw, fill=blue!5, rectangle, 
    minimum height=3em, minimum width=18mm]
\tikzstyle{sum} = [draw, fill=yellow!10, circle, node distance=1cm]
\tikzstyle{coord} = [coordinate]
\tikzstyle{gain} = [draw, fill=red!5, regular polygon, regular polygon sides=3, shape border rotate=-90]
\tikzstyle{pinstyle} = [pin edge={to-,thick,black}]

\tikzstyle{BitPipe} = [thick, decoration={markings,mark=at position
   1 with {\arrow[semithick]{open triangle 60}}},
   double distance=1.4pt, shorten >= 5.5pt,
   preaction = {decorate},
   postaction = {draw,line width=1.4pt, white,shorten >= 4.5pt}]

\usetikzlibrary{fit,backgrounds}

\usepackage{hyperref}

\usepackage{epsfig, graphicx, psfrag}

\usepackage[cal=boondox]{mathalfa}

\newcommand{\khina}{A.~Khina}

\newcommand{\MyStudent}{}

\providecommand{\thmref}[1]{Theorem~\ref{#1}}

\providecommand{\secref}[1]{Section~\ref{#1}}

\providecommand{\propref}[1]{Proposition~\ref{#1}}
\providecommand{\remref}[1]{Remark~\ref{#1}}
\providecommand{\figref}[1]{Figure~\ref{#1}}
\providecommand{\corref}[1]{Corollary~\ref{#1}}
\providecommand{\appref}[1]{Appendix~\ref{#1}}

\providecommand{\schemeref}[1]{Scheme~\ref{#1}}

\usepackage{refcount}

\newcommand{\reals}{\mathbb{R}}

\newcommand{\nats}{\mathbb{N}}

\newcommand{\floor}[1]{\lfloor{#1}\rfloor}

\let\limsup\relax
\DeclareMathOperator*{\limsup}{\overline{lim}}

\DeclareMathOperator*{\argmax}{arg\, max}

\providecommand{\Exp}[1]{{\operatorname{exp} \left\{ #1 \right\}}}
\providecommand{\E}[1]{\ensuremath{\mathbb{E} \left[ #1 \right]}}
\providecommand{\CE}[2]{\ensuremath{\mathbb{E} \left[ #1 \middle| #2 \right]}}

\providecommand{\EE}{E}

\newcommand{\bm}[1]{\mbox{\boldmath{$#1$}}}

\newcommand{\SNR}{S}
\newcommand{\ENR}{\gamma}

\newcommand{\e}{\mathrm{e}}

\newcommand{\Comment}[1]{}
\newcommand{\old}[1]{}
\newcommand{\rem}[1]{}

\newcommand{\eps}{{\epsilon}}

\providecommand{\hu}{{\hat{u}}}

\providecommand{\hU}{{\hat{U}}}

\providecommand{\hW}{{\hat{W}}}

\newcommand{\tb}{\tilde{b}}

\providecommand{\bU}{{\bm U}}

\newcommand{\bX}{{\bf X}}

\providecommand{\bu}{{\bm u}}

\newcommand{\bY}{{\bf Y}}
\newcommand{\bZ}{{\bf Z}}

\newcommand{\hbU}{\hat{\bU}}

\newcommand{\calB}{{\cal B}}

\newcommand{\calE}{{\cal E}}

\newcommand{\calI}{{\cal I}}

\newcommand{\calU}{{\cal U}}

\providecommand{\bA}{\mathbf{A}}

\providecommand{\bU}{\mathbf{U}}

\providecommand{\bX}{\mathbf{X}}
\providecommand{\bY}{\mathbf{Y}}
\providecommand{\bZ}{\mathbf{Z}}

\newcommand{\abs}[1]{\left| #1 \right|}
\newcommand{\Norm}[1]{\left\| #1 \right\|}

\providecommand{\e}{{\rm e}}
\providecommand{\comment}[1]{}

\providecommand{\norm}[1]{\Norm{#1}}

\providecommand{\BW}{\mathrm{B}}

\newcommand{\beqn}[1]{\begin{eqnarray}\label{#1}}
\newcommand{\eeqn}{\end{eqnarray}}
\newcommand{\beq}[1]{\begin{equation}\label{#1}}
\newcommand{\eeq}{\end{equation}}

\newcommand{\tbZ}{{\tilde{ \bf Z}}}

\providecommand{\half}{\frac{1}{2}}

\providecommand{\half}{\frac{1}{2}}

\makeatletter
\newcommand{\vast}{\bBigg@{4}}
\newcommand{\Vast}{\bBigg@{5}}
\makeatother

\providecommand{\E}[1]{\bbE \left[ #1 \right]}
\providecommand{\CE}[2]{\bbE \left[ #1 \middle| #2 \right]}

\newcommand{\PR}[1]{\Pr\left( #1 \right)}

\newcommand{\Eh}{\EE_\infty}
\newcommand{\Rh}{R_\mathrm{h}}

\newcommand {\Lh} {L_\mathrm{h}}
\newcommand {\Mh} {M_\mathrm{h}}
\newcommand {\wh} {w_\mathrm{m}}
\newcommand {\wm} {w_\mathrm{m}}
\newcommand {\wl} {w_\ell}

\newcommand {\hWh} {\hat{W}_\mathrm{m}}
\newcommand{\Ch}{C_\mathrm{h}}
\newcommand {\Mm} {M_\mathrm{m}}
\newcommand {\Ml} {M_\ell}
\newcommand {\Lm} {L_\mathrm{m}}
\newcommand {\Ll} {L_\ell}

\providecommand{\Esp}{\EE_\mathrm{sp}}
\providecommand{\Ewsp}{\EE_\mathrm{wsp}}

\providecommand{\Exp}[1]{\ensuremath{{\operatorname{exp} \left\{ #1
\right\}}}}

\providecommand{\Q}[1]{\ensuremath{Q \left( #1 \right)}}

\providecommand{\lrp}[1]{\left( #1 \right)}
\providecommand{\lrs}[1]{\left[ #1 \right]}
\providecommand{\lrc}[1]{\left\{ #1 \right\}}

\providecommand{\lrcm}[2]{\left\{ #1 \middle| #2 \right\}}

\providecommand{\dist}{MP$\alpha$E}

\usetikzlibrary{shapes,arrows,decorations.markings,positioning}

\tikzstyle{plant} = [draw, fill=red!5, rectangle, 
    minimum height=3em]
\tikzstyle{block} = [draw, fill=blue!5, rectangle, 
    minimum height=3em, minimum width = 20mm]
\tikzstyle{sum} = [draw, fill=yellow!10, circle, node distance=1cm]
\tikzstyle{pinstyle} = [pin edge={to-,thick,black}]

\tikzstyle{BitPipe} = [thick, decoration={markings,mark=at position
   1 with {\arrow[semithick]{open triangle 60}}},
   double distance=1.4pt, shorten >= 5.5pt,
   preaction = {decorate},
   postaction = {draw,line width=1.4pt, white,shorten >= 4.5pt}]

\providecommand{\Pc}{\mathrm{P}}
\providecommand{\Cc}{\mathrm{C}}
\providecommand{\Rc}{\mathrm{R}}
\providecommand{\Rhc}{\mathrm{R_h}}
\newcommand {\Rmc}{\mathrm{R_m}}
\providecommand{\Nzero}{\mathrm{N}_0}
\providecommand{\sinc}{\mathrm{sinc}}
\providecommand{\rect}{\mathrm{rect}}
\providecommand{\Helper}{V}
\providecommand{\Time}{\mathrm{T}}
\providecommand{\Energy}{E}
\providecommand{\EEc}{\mathrm{\EE}}

\usepackage{ifthen}

\usepackage{mathtools}

\mathtoolsset{showonlyrefs=true}

\usepackage{comment}
\newcommand{\ignore}[1]{}

\providecommand{\ColumnNum}{1}
\newcommand{\col}{\ifthenelse{\equal{\ColumnNum}{1}}}

\newcommand{\figwidth}{3.2in}

\begin{document}

\title{Modulation and Estimation with a Helper}

\author{Anatoly Khina\textit{, Member, IEEE}, and Neri Merhav\textit{, Life Fellow, IEEE}
    \thanks{A.~Khina is with the Department of Electrical Engineering--Systems, Tel Aviv University, Tel Aviv~6997801, Israel (e-mail: \texttt{anatolyk@tau.ac.il}).}
    \thanks{N.~Merhav is with the Viterby Faculty of Electrical and Computer Engineering, Technion--Israel Institute of Technology, Haifa 3200004, Israel (e-mail: \texttt{merhav@ee.technion.ac.il}).}
    \thanks{The work of A.~Khina was supported by 
    the \textsc{Israel Science Foundation} (grant 2077/20) and by a grant from the Tel Aviv University Center for AI and Data Science (TAD).}
}

\maketitle


\begin{abstract}
    The problem of transmitting a parameter value over an additive white Gaussian noise (AWGN) channel is considered, 
    where, in addition to the transmitter and the receiver, there is a helper that observes the noise non-causally and provides a description of limited rate $\Rh$ to the transmitter and/or the receiver.
    We derive upper and lower bounds on the optimal achievable $\alpha$-th moment of the estimation error and show that they coincide for small values of $\alpha$ and for high values of $\Rh$.
    The upper bound relies on a recently proposed channel-coding scheme that 
    effectively conveys $\Rh$ bits essentially error-free and the rest of the rate---over the same AWGN channel without help, with the error-free bits being allocated to the most significant bits of the quantized parameter.
    We then concentrate on the setting with a total transmit energy constraint, for which we derive achievability results for both channel coding and parameter modulation for several scenarios: when the helper assists only the transmitter or only the receiver and knows the noise, and when the helper assists the transmitter and/or the receiver and knows both the noise and the message. In particular, for the message-informed helper that assists both the receiver and the transmitter, it is shown that the error probability in the channel-coding task decays doubly exponentially.
    Finally, we translate these results to those for continuous-time power-limited AWGN channels with unconstrained bandwidth. As a byproduct, we show that the capacity with a message-informed helper that is available only at the transmitter can exceed the sum of the capacity without help and the help rate $\Rh$, when the helper knows only the noise but not the message.
\end{abstract}

\begin{IEEEkeywords}
	AWGN channel, helper, parameter estimation, channel coding, PPM.
\end{IEEEkeywords}

\allowdisplaybreaks

\section{Introduction}
\label{s:intro}

Conveying a parameter over an additive white Gaussian noise (AWGN) channel is a fundamental problem in both information theory and estimation theory, 
which has received much attention over the years \cite{Shannon49,KotelnikovJSCC,Elias57:JSCC:BW-expansion,Goblick65,GallagerBook1968,WozencraftJacobsBook,Seidman:ParamterEstimationBound:IT1970,ZivZakai:ParamterEstimationBound:IT1969,ChazanZivZakai:ParamterEstimationBound:IT1975,Zehavi:TOAbound,BurnashevInfiniteBandwidthExponent_precursor,BurnashevInfiniteBandwidthExponent1,BurnashevInfiniteBandwidthExponent,ViterbiOmuraBook,WeissWeinstein_OrigPaper,Merhav:OptimumJSCC_InfBW,WeinbergerMerhav_FiniteBWJSCC,ChungPhD,ShannonKotelnikovMaps_Ramstad,ShannonKotelnikovMaps_MMSE_Dec,Merhav_WeakNoiseAnamolyTradeoff_Scalar,Merhav_WeakNoiseAnamolyTradeoff_Vector,Merhav:JSCC:scalar-parameter:IT2019,TridenskiZamirIngber,KokenInfiniteBandwidth_Broadcast,MinimumEnergyBound_Tuncel,TuncelInfinitedBW_SeparationCompanding:Journal,EnergyLimitedJSCC:PPM:Lev_Khina:TCOM2022,EnergyLimitedJSCC:Universal:Lev_Khina:ISIT2022,baniasadi-tuncel:ZeroBW:ISIT2020,baniasadi-tuncel-SI:ISIT2021,Baniasadi-Koken-Tuncel:energy-limited-JSCC:universal:IT2022}.
In particular, consider the problem of transmitting a scalar parameter over a continuous-time AWGN channel 
with two-sided power spectral density $N_0/2$ subject to a total energy constraint $E$. 
For this problem, Burnashev~\cite{BurnashevInfiniteBandwidthExponent} proved that the minimum achievable mean square error, $D$, decays exponentially with $E$ \cite{BurnashevInfiniteBandwidthExponent,BurnashevInfiniteBandwidthExponent_precursor,BurnashevInfiniteBandwidthExponent1}:
\begin{align}
\label{eq:Burnashev}
    D \doteq \Exp{-\frac{1}{6} \cdot \frac{E}{N_0}} ,
\end{align}
where `$\doteq$' denotes equality up to sub-exponential multiplicative terms in $\frac{E}{N_0}$.
The decay exponent of $\frac{E}{6N_0}$ in \eqref{eq:Burnashev} is significantly smaller than the exponent dictated by the data-processing theorem (DPT) \cite[Chapter~9.3, Problem~10.8, and Theorem~10.4.1]{CoverBook2Edition}, \cite[Chapter~3]{ElGamalKimBook} by a factor of 6, to wit 
\begin{align}
\label{eq:DPT:asymptotics}
    D \doteq \Exp{-\frac{E}{N_0}}.
\end{align} 
This gap in performance stems from the constraint of processing and transmitting a single sample in \eqref{eq:Burnashev} as opposed to the large source block lengths used in the achievability proof of the DPT.
The results of~\eqref{eq:Burnashev} and \eqref{eq:DPT:asymptotics} hold also for the transmission of a scalar parameter over a discrete-time AWGN channel with a total energy constraint $E$ and noise variance $N_0$ (see, e.g., \cite{MinimumEnergyBound_Tuncel,TuncelInfinitedBW_SeparationCompanding:Journal,EnergyLimitedJSCC:PPM:Lev_Khina:TCOM2022}).
For the finite-bandwidth continuous-time setting and for the 
power-constraint discete-time setting, bounds on the performance that are akin to Burnashev's bound~\eqref{eq:Burnashev} can be obtained from the work of Weinberger and Merhav~\cite{WeinbergerMerhav_FiniteBWJSCC}.

To facilitate the communication between the transmitter and the receiver,  a third party that has a correlated description of the source and/or the channel noise can act as a helper by providing its description over a rate-limited channel to the transmitter and/or the receiver.
Wyner \cite{Wyner:WAK:TIT1975}, and Ahlswede and K\"orner \cite{Ahlswede-Korner:WAK:TIT1975} studied the source coding problem in which the helper has a correlated description of the source and can provide it over a noiseless rate-limited channel to the receiver (see also \cite[Chapter~10.4]{ElGamalKimBook}).
An extension of this scenario to the case of a transmitter and a helper that transmit to the receiver over a multiple-access channel was studied by Ahlswede and Han~\cite{Ahlswede-Han:WAK-SW-overMAC:TIT1983}.

Channel coding over a state-dependent channel where the state is known to the helper, and the helper can provide a rate-limited description of the state to the transmitter was studied in \cite{HeegardElGamal83,Rosenzweig-Steinberg-Shamai:rate-limited-SI:TIT2005,LapidothNarayan} (see also \cite[Chapter~3.5]{Keshet-Steinberg-Merhav:side-info:NOW2008}, \cite[Chapter~7.8]{ElGamalKimBook}).
An interesting special case of the latter scenario, which has been studied more recently \cite{Marti:MSc,Lapidoth-Marti:Tx-assisted:TIT2020,Bross-Lapidoth-Marti:Rx-assisted:TCOM2020,Merhav:Tx-assisted-EE:TIT2021,Wang-Lapidoth:Tx-assisted:ArXiv2023}, 
is that of an additive noise channel with a helper that knows the noise sequence and can share a rate-limited description of
the channel noise with the transmitter and/or the receiver.
This scenario is motivated by network scenarios, such as the following.
Consider a scenario in which both the transmitter and the helper 
act as transmitters, 
where the transmit signal of the helper interferes with the communication between the transmitter and the receiver. To mitigate this interference, the helper provides information about its transmit signal via a rate-limited channel to the transmitter and/or the receiver. 
Lapidoth and Marti \cite{Lapidoth-Marti:Tx-assisted:TIT2020} (see also \cite{Bross-Lapidoth-Marti:Rx-assisted:TCOM2020}, \cite[Chapter~5.2]{Marti:MSc}) proved that the capacity of a discrete-time AWGN channel with a helper of rate (\textit{help rate}) $\Rh$ that is subject to a power constraint $P$
equals the capacity without assistance $C_0 \triangleq \half \log \lrp{1 + \SNR}$ plus the help rate $\Rh$: 
\begin{align}
\label{eq:Tx-assisted-capacity:Lapidoth-Marti}
    C = C_0 + \Rh ,
\end{align}
both when the noise is known causally and non-causally to the helper,\footnote{All logarithms and exponents are to the base 2 in this work unless otherwise specified.}
where $S \triangleq P/N$ denotes the signal-to-noise ratio (SNR).
Merhav \cite{Merhav:Tx-assisted-EE:TIT2021} derived upper and lower bounds on the error exponent at all rates $R < C$ for the non-causal helper setting; in particular, he showed that $\Rh$ nats may be conveyed essentially error-free, 
whereas the remaining $R - \Rh$ nats are conveyed over an AWGN channel without help.
When the helper has access to the message of the transmitter (``cribbing''), but has only a limited channel to the transmitter and/or the receiver, it can tailor its help with accordance to the message that the transmitter wishes to send, and potentially attain better results.

There exists a rich body of literature on source and channel coding with a helper. However, parameter transmission over a noisy channel with a helper, that knows the noise realization and provides a rate-limited description to the transmitter and/or the receiver, has not been hitherto addressed.

In this work, we study the problem of conveying a parameter over an AWGN channel with a helper that knows the noise non-causally and can provide a description over a rate-limited channel to the transmitter and/or receiver. The problem is formalized in \secref{s:model}.
We provide the necessary background about channel coding over AWGN channels with a helper that knows the noise sequence in \secref{ss:background:comm-w-helper}.
In \secref{s:main}, 
we derive upper and lower bounds on the achievable $\alpha$-th moment of the absolute error---which we term mean power-$\alpha$ error (\dist) following \cite{WeinbergerMerhav_FiniteBWJSCC}---and show that these bounds coincide for low $\alpha$ values as well as for high $\Rh$ values.

To derive a lower (impossibility) bound on the \dist, we use the extension of the technique of Ziv and Zakai \cite{ZivZakai:ParamterEstimationBound:IT1969}, and Chazan, Zakai and Ziv \cite{ChazanZivZakai:ParamterEstimationBound:IT1975} with
exponentially many hypotheses (rather than~$2$) \cite{Brown-Liu:Ziv-Zakai-bound:Multiple-Hypohtheses:TIT1993,Bell:PhDl:Ziv-Zakai-Extension}.
For the upper (achievability) bound, we judiciously employ the  
aforementioned achievability result of \cite{Merhav:Tx-assisted-EE:TIT2021} for transmitter-assisted channel coding over an AWGN channel: 
We apply uniform quantization to the parameter value where the quantization outputs are naturally labeled and allocate the most significant bits (MSBs) of rate $\Rh$ nats of the quantized description essentially error-free.

In \secref{s:fixed-E}, we concentrate on the setting where the input is subject to a total energy constraint. 
To derive upper bounds on the \dist, 
we refine, in \secref{ss:fixed-E:Rh}, the analysis of the channel-coding scheme with a helper with a fixed help rate of Merhav~\cite{Merhav:Tx-assisted-EE:TIT2021} for this setting, as the original scheme and analysis do not carry over straightforwardly.

In the remainder of 
\secref{s:fixed-E},
we consider the energy-constrained setting in which the helper is cognizant of both the noise and the message; we refer to such a helper as being \textit{cribbed}.
When the cribbed helper assists only the transmitter, we show in \secref{ss:crib:Tx-only} that a channel-coding scheme that is based on pulse position modulation (PPM), and is reminiscent of the dirty-paper coding scheme of Liu and Viswanath \cite{Liu-Viswanath:Dirty-Paper:Opportunistic-PPM:TIT2006}, 
outperforms the scheme with a helper that knows only the noise but not the message for certain help rates. 
In \secref{ss:crib:two-sided}, we consider a helper that assists both the transmitter and the receiver. In this scenario, the helper can ``simulate'' several PPM schemes (without help) and then send the index of the best one to the transmitter and the receiver, to indicate which one of them to use. We show that the error probability of this scheme decays doubly exponentially with the total help budget. 

In \secref{s:CT}, we translate the results of Sections~\ref{s:main} and~\ref{s:fixed-E} to the continuous-time transmission-assisted AWGN channel that is subject to a power constraint but has unconstrained bandwidth and compare the resulting bounds.
In particular, we prove that a helper that is cognizant of both the noise and the message allows outperforming \eqref{eq:Tx-assisted-capacity:Lapidoth-Marti} for certain help rates,\footnote{Here the rates are measured in nats per second.}
where \eqref{eq:Tx-assisted-capacity:Lapidoth-Marti} is achievable when the helper knows only the noise but not the message~\eqref{eq:Tx-assisted-capacity:Lapidoth-Marti}.

We conclude the paper with a summary and discussion of future directions in \secref{s:discussion}.

\section{Problem Setup}
\label{s:model}


The sets of natural and real numbers are denoted by $\nats$ and $\reals$, respectively.
We denote indexed sequences between $i \in \nats$ and $j \in \nats$ by boldface letters: $\bA_{i:j} = \lrp{A_i, A_{i+1}, \ldots, A_j}$, and $\bA = \bA_{1:n}$ for conciseness. 
We denote random variables by uppercase letters and realizations thereof by lowercase letters.
$\calB_n(r) \triangleq \lrcm{x \in \reals^n}{\, \norm{x}_2 \leq r}$ is the $n$-dimensional ball of radius $r$ and $\norm{\cdot}_2$ denotes the Euclidean norm. $\mathrm{Vol}\{g\}$ denotes the volume of $g \subset \reals^n$. 
The limit and the limit superior are denoted by $\lim$ and $\limsup$, respectively.
We further denote by $\lim\limits_{x \downarrow a} f(x)$ the limit of a function $f(\cdot)$ as $x$ decreases in value approaching $a$ (limit from the right).

We make use of small $o$ notation, viz., $f(x) = o\lrp{g(x)}$ means that $\lim_{x \to \infty} {f(x)}/{g(x)} = 0$; if the argument is not clear [e.g., in the case of $o(1)$], we specify the argument in the subscript, namely, $o_x(\cdot)$.
$\mod$ and $*$ denote the modulo and convolution operations, respectively.



\begin{figure}[t]
    \centering
    {\center
    \resizebox{\figwidth}{!}{\pgfmathsetseed{4}

\begin{tikzpicture}[auto, arrow/.style={very thick, ->, >=stealth'},node distance=15mm,>=latex']
    
    \node [coord] (input) {};
    \node[block, right = of input] (Tx) {Transmitter};
    
    \draw[arrow] (input) -- node[above] (x) {$u$} (Tx) {};


    \node[sum, right = 13mm of Tx] (channel) {\bf \Large +};
    \node[coord, above = 20mm of channel] (noise) {};

    \draw[arrow] (noise) -- node[above, pos=0] {$\bZ$} (channel);


    \node[block, above = 10mm of Tx] (helper) {Helper};
    \draw[BitPipe] (helper) -- node [left] {$\Helper$} (Tx);
    \draw[arrow] (channel) |- (helper);


    \node[block, right = 13mm of channel] (Rx) {Receiver};
    
    \draw[arrow] (Tx) -- node {$\bX$} (channel);
    \draw[arrow] (channel) -- node {$\bY$} (Rx);

    \node[coord, right = of Rx] (output) {};

    \draw[arrow] (Rx) -- node[above] (hx) {$\hu$} (output) {};
\end{tikzpicture}
    }}
    \caption{Parameter estimation--modulation over an AWGN channel with an assisted transmitter.}
    \label{fig:model}
\end{figure}
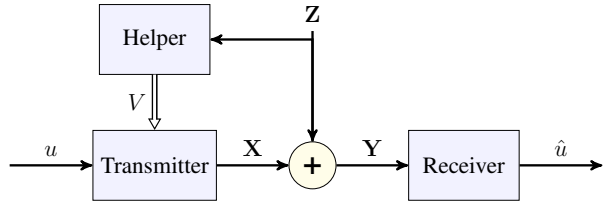   

We now formalize the modulation--estimation setting of this work; see also \figref{fig:model}. 
We concentrate here on the transmitter-assisted setting, namely, the setting in which the helper assists only the transmitter. The setting of a helper that assists the receiver or both the transmitter and the receiver is defined similarly.

\textit{Parameter.} 
The parameter to be conveyed is $u \in \left[ -1/2, 1/2 \right)$.

\textit{Transmitter.}
Maps the parameter $u$ and the helper's description $\Helper \in \lrc{1, \ldots, M}$ to a channel input sequence $\bX = \lrp{X_1, X_2, \ldots, X_n}$ that is subject to either a fixed (in $n$) power constraint~$P$:
\begin{align}
\label{eq:constraint:power}
    \sum_{t=1}^n \E{X_t^2} &\le nP &\forall u \in \left[ -1/2, 1/2 \right), 
\end{align}
or to a fixed (in $n$) energy constraint $\Energy$:
\begin{align}
\label{eq:constraint:energy}
    \sum_{t=1}^n \E{X_t^2} &\le \Energy &\forall u \in \left[ -1/2, 1/2 \right). 
\end{align}

\textit{Channel.}
The input sequence $\bX$ is transmitted over an AWGN channel:
\begin{align}
    Y_t &= X_t + Z_t, & t \in \lrc{1,2,\ldots,n},
\end{align}
where $\bY \triangleq \lrp{Y_1, Y_2, \ldots, Y_n}$ is the output sequence, and $\bZ=(Z_1, Z_2, \ldots,Z_n)$ is the noise sequence whose entries 
are independent identically distributed (i.i.d.) Gaussian of zero mean and variance~$\sigma^2$. 

For the power-constrained setting, we define the SNR to be 
    $\SNR \triangleq P/\sigma^2$, 
and in the energy-constrained setting, we define 
$\ENR \triangleq \Energy / \sigma^2$.

\textit{Helper.} 
Knows (non-causally) the noise sequence $\bZ$ and maps it to a finite-rate description $\Helper \in \lrc{1, 2, \ldots, \Mh}$.
$\Helper$ is revealed to the transmitter and/or receiver prior to the beginning of transmission.
In the fixed power setting, we will assume that $\Mh = \Exp{n \Rh}$ where the help rate $\Rh$ is fixed, namely, $\Mh$ grows exponentially with $n$.
In the energy-limited setting, we will consider both the fixed help rate case and the total nat budget $\Lh = \log \Mh$ which remains fixed, independently of $n$.

\textit{Receiver.}
Constructs an estimate $\hU$ of the parameter $u$ given the output $\bY$.

\textit{Objective.}
Achieving the minimal \dist, 
\begin{align} 
\label{eq:def:distortion}
    \epsilon(\alpha) \triangleq \sup_{u \in [-1/2, 1/2)} \E{|\hat{U}-u|^\alpha} ,
\end{align}
where the expectation is with respect to the channel noise.

In this work, we will derive upper and lower bounds on the minimum achievable \dist~\eqref{eq:def:distortion}.
In particular, for the power-limited scenario \eqref{eq:constraint:power}, we will concentrate on bounding the optimal achievable \dist\ exponent\footnote{Both the energy in \eqref{eq:constraint:energy} and the error exponent are denoted by $E$. However, since the latter is a function, the distinction between them is clear.}
\begin{align}
\label{eq:def:distortion:EE}
    \EE(\alpha) \triangleq \limsup_{n \to \infty} - \frac{1}{n} \log \inf \eps(\alpha) ,
\end{align}
where the infimum is taken over all transmitter--receiver--helper triplets.


\section{Background: Channel Coding over the AWGN Channel with a Helper}

\label{ss:background:comm-w-helper}

Consider the problem of reliable communication over an AWGN channel. The setting is as follows.

\textit{Message.} A message $W$ is uniformly distributed over $\lrc{1, 2, \ldots, M}$, with $M = \Exp{n R}$.

\textit{Transmitter.}
Maps the message $W$ and the helper's description $\Helper \in \lrc{1, \ldots, \Mh}$ to a channel input sequence $\bX$ of length $n$ 
that is subject to the power constraint \eqref{eq:constraint:power}.

The channel and helper descriptions are the same as in \secref{s:model}.

\textit{Receiver.} Constructs an estimate $\hW$ of $W$ given the output $\bY$.

\textit{Objectives.}
The rate and the error probability are defined as 
\begin{align} 
    R &\triangleq \frac{1}{n} \log M & \text{and} &&
    P_e &\triangleq \PR{\hW \neq W} ,
\end{align} 
respectively. 
For a given rate $R$, we define the optimal channel-coding error exponent as
\begin{align}
\label{eq:def:EE:DT}
    \EE_e(R) \triangleq 
    \limsup_{n \to \infty}
    - \frac{1}{n} \log \inf P_e ,
\end{align}
where the infimum is taken over all transmitter--receiver--helper triplets.

Denote by $C_0 = \half \log \lrp{1 + \SNR}$ the capacity of an AWGN with SNR $\SNR$ without help.
Then, the following results for the transmitter-assisted setting are known.

\begin{thm}[\!\cite{Lapidoth-Marti:Tx-assisted:TIT2020}]
\label{thm:FEC:capacity}
    The capacity of an AWGN channel with SNR $\SNR$ and help rate $\Rh$~is 
    \begin{align}
        \Ch = C_0 + \Rh .
    \end{align}
\end{thm}

\begin{thm}[\!\cite{Merhav:Tx-assisted-EE:TIT2021}]
\label{thm:FEC:EE:LB}
    The optimal achievable error exponent at rate $R$ over an AWGN channel with SNR $\SNR$ and help rate $\Rh$ is bounded from below as 
    \begin{align}
        \EE_e(R) \geq 
        \begin{cases}
            \infty, & R < \Rh
         \\ \EE_a \lrp{R - \Rh}, & \Rh \leq R < C_0 + \Rh
         \\ 0, & R \geq C_0 + \Rh,
        \end{cases}
    \end{align}
    where 
    $\EE_e(R) = \infty$ for $R < \Rh$ means that an arbitrarily large exponent is achievable, 
    and where $\EE_a (\cdot)$ is any achievable error-exponent function over the AWGN channel without help.
    
    Moreover, for $R \in [\Rh, C_0 + \Rh)$, $R' < \Rh$ nats per time step can be conveyed 
    with an arbitrarily large error exponent and where $\Rh - R'$ is arbitrarily small for a sufficiently large $n$, whereas the remaining $R - R'$ nats per time step---with an error exponent that is arbitrarily close to $\EE_a \lrp{R - \Rh}$.
\end{thm}

\begin{thm}[\!\cite{Merhav:Tx-assisted-EE:TIT2021}]
\label{thm:FEC:EE:UB}
    The optimal achievable error exponent at rate $R$ over an AWGN channel with SNR $\SNR$ and help rate $\Rh$ is bounded from above by the weak sphere-packing bound  
    \begin{align}
        \Ewsp (R) &\triangleq 
        \begin{cases} 
            \infty, & R < \Rh
         \\ \frac{1}{2}\left[\zeta(R) - \log \zeta(R) - 1\right], & \Rh < R < C_0 + \Rh
         \\ 0, & R\ge C_0 + \Rh
        \end{cases}
    \end{align}
    with 
    \begin{align} 
    \label{eq:zeta}
        \zeta(R) \triangleq \frac{\Exp{2C_0}-1}{\Exp{2(R - \Rh)}-1} = \frac{\SNR}{\Exp{2(R - \Rh)}-1} \,.
    \end{align} 
\end{thm}
As the name suggests, the weak sphere-packing bound is weaker (larger upper bound) than the ordinary sphere-packing bound~\cite[Eq.~(7.4.33)]{GallagerBook1968} (see also \cite[Example~1]{Nakibouglu:sphere-packing-bound;PPI2020}, \cite{Tridenski--Somekh-Baruch:AWGN-method-of-types:Arxiv2023}), which is given by
\begin{align}
\label{eq:sphere-packing:ordinary}
    \Esp(R) &\triangleq 
    \begin{cases}
        \infty, & R \leq \Rh
     \\ \frac{\SNR}{4\beta}\left[\beta+1-(\beta-1)\sqrt{1+\frac{4\beta}{\SNR(\beta-1)}}\right] 
     \\ + \frac{1}{2}\log\left(\beta-\frac{\SNR(\beta-1)}{2}\left[\sqrt{1+\frac{4\beta}{\SNR(\beta-1)}}-1\right]\right), 
        & 0 \leq R - \Rh \leq C_0
     \\ 0, & R - \Rh \geq C_0
    \end{cases}
\end{align}
where $\beta=\Exp{2(R-\Rh)}$.
In particular, $\lim\limits_{R \downarrow \Rh} \Ewsp(R) = \infty$, whereas \mbox{$\lim\limits_{R \downarrow \Rh} \Esp(R) = \SNR / 2 < \infty$}.

Indeed, whereas the ordinary sphere-packing bound relies on the independence between the input $\bX$ and the noise $\bZ$, 
in our setting of interest, the two can be correlated via the helper description $\Helper$, which is a function of the noise $\bZ$ and is provided to the transmitter to generate the input $\bX$. This, in turn, suggests a larger upper bound; see \cite{Merhav:Tx-assisted-EE:TIT2021} for more details and a discussion.

We further note that the weak sphere-packing bound of \thmref{thm:FEC:EE:UB} shares important properties with the achievability bound of \thmref{thm:FEC:EE:LB}: 
both are infinite for $R < \Rh$, strictly positive for $R < C_0 + \Rh$, and zero above the capacity, i.e., for $R > C_0 + \Rh$.

\begin{remark}
    While channel coding under an average error probability was considered in \cite{Merhav:Tx-assisted-EE:TIT2021}, 
    discarding half of all the codewords that have the highest error probabilities achieves the same error exponent with respect to the maximal error probability;
    see \cite[Chapter~7.7]{CoverBook2Edition}.
\end{remark}


\section{Power-Limited Input}
\label{s:main}

In this section, we derive upper and lower bounds on the achievable \dist\ exponent~\eqref{eq:def:distortion:EE} in \secref{ss:main:LB} and \secref{ss:main:UB}, respectively, for the power-limited setting \eqref{eq:constraint:power}, and extend them for vector parameters in \secref{ss:main:compare}.
We further compare these bounds 
in \secref{ss:main:UB} and show that they coincide 
when $\alpha$ is small or $\Rh \gg C_0$.
While we focus in this section on the setting of a helper that assists the transmitter only, the results readily apply to the setting of a helper that assists the receiver only, \textit{mutatis mutandis}.

\subsection{\dist~Exponent Upper Bounds}
\label{ss:main:LB}

We first introduce a bound 
for the transmitter-assisted setting, and then a tighter bound for the receiver-assisted setting.
Both bounds are based on the bounding technique 
of  Ziv and Zakai \cite{ZivZakai:ParamterEstimationBound:IT1969}, and Chazan, Zakai, and Ziv \cite{ChazanZivZakai:ParamterEstimationBound:IT1975}, 
but with a number of hypotheses that grows exponentially with $n$ (rather than~$2$) \cite{Brown-Liu:Ziv-Zakai-bound:Multiple-Hypohtheses:TIT1993,Bell:PhDl:Ziv-Zakai-Extension}.

\begin{thm}
\label{thm:Ziv-Zakai-based-LB}
    The \dist\ exponent \eqref{eq:def:distortion:EE}, in the transmitter-assisted scenario,
    is bounded from above as
    $\EE(\alpha) \leq \alpha \left( \Rh + C_0^* \right)$,
    where 
    \begin{subequations}
    \begin{align}
        C_0^* &\triangleq 
        \frac{1}{2}\left(\log\left(1+\frac{\SNR}{s^*}\right)+\frac{s^*-\log
s^*-1}{\alpha}\right),
    \label{eq:def:C0*}
     \\ s^* &\triangleq \frac{2(\alpha+1)\SNR}{\sqrt{(\SNR+1)^2+4\alpha \SNR}+\SNR-1}.
    \label{eq:def:s*}
    \end{align}
    \end{subequations}
\end{thm} 

\begin{IEEEproof}
    Define $U$ to be uniformly distributed over the finite set 
    \begin{align}
    \label{eq:Ziv-Zakai:proof:discrete-set}
        \calU_M \triangleq
         \left\{\frac{1}{M} \lrp{i - \frac{M-1}{2}} \middle|\, i \in \left\{0, \ldots, M-1\right\} \right\} ,
    \end{align}
    such that all the points in this set belong to $[-1/2, 1/2)$. Then 
    \begin{subequations}
    \label{eq:Ziv-Zakai-based-LB:proof}
    \noeqref{eq:Ziv-Zakai-based-LB:proof:Bayesian}
    \begin{align}
        \epsilon(\alpha) &\triangleq \sup_{u \in [-1/2, 1/2)} \E{\abs{\hat{U}-u}^\alpha} 
    \label{eq:Ziv-Zakai-based-LB:proof:def}
     \\ &\geq \E{\abs{\hat{U}-U}^\alpha} 
    \label{eq:Ziv-Zakai-based-LB:proof:Bayesian}
     \\ &\geq \frac{1}{\lrp{2M}^\alpha} \PR{ \abs{\hU - U}^\alpha > \frac{1}{\lrp{2M}^\alpha}}
    \label{eq:Ziv-Zakai-based-LB:proof:Markov}
     \\ &\geq \frac{1}{\lrp{2M}^\alpha} \inf_{\hU} \PR{\hU \neq U}
    \label{eq:Ziv-Zakai-based-LB:proof:dist2error}
     \\ &\geq \Exp{-\alpha R n}\, \Exp{-n \Ewsp(R) + o(n)}
    \label{eq:Ziv-Zakai-based-LB:proof:sphere-packing}
    \end{align}
    \end{subequations}
    where 
    \eqref{eq:Ziv-Zakai-based-LB:proof:def} holds by definition \eqref{eq:def:distortion};
    \eqref{eq:Ziv-Zakai-based-LB:proof:Markov} follows from the Markov inequality;
    the infimum in \eqref{eq:Ziv-Zakai-based-LB:proof:dist2error} is over all estimators of $U$ given $\bY$, and holds since a point $\hU \in \reals$ can be within a distance $\frac{1}{2M}$ of at most a single point in $\calU_M$;
    and \eqref{eq:Ziv-Zakai-based-LB:proof:sphere-packing} follows from 
    \thmref{thm:FEC:EE:UB}.
    
    To tighten the bound in \eqref{eq:Ziv-Zakai-based-LB:proof}, 
    let us optimize the bound in \eqref{eq:Ziv-Zakai-based-LB:proof:sphere-packing} over $R$, namely, 
    \begin{subequations}
    \label{eq:aR+Ewsp}
    \begin{align}
        \inf_{R \in \reals} \left\{  \alpha R + \Ewsp(R) \right\}
        &= \min_{R \in [\Rh}, C_0 + \Rh] \left\{ \alpha R + \frac{1}{2}\left[\zeta(R) - \log \zeta(R) - 1\right] \right\}
    \label{eq:aR+Ewsp:substitute_Ewsp}
     \\ &\:\: = \alpha \Rh + \half \min_{s \geq 1} \left\{\alpha \log \lrp{1 + \frac{\SNR}{s}} + s - \log s - 1 \right\} \quad\ 
    \label{eq:aR+Ewsp:substitute_zeta}
     \\ &\:\: = \alpha \Rh + \half \left\{\alpha \log \lrp{1 + \frac{\SNR}{s^*}} + s^* - \log s^* - 1 \right\}
    \label{eq:aR+Ewsp:s_opt}
    \end{align}
    \end{subequations}
    where 
    \eqref{eq:aR+Ewsp:substitute_Ewsp} follows by the definition of $\Ewsp$ in \thmref{thm:FEC:EE:UB},
    \eqref{eq:aR+Ewsp:substitute_zeta} follows by the definition of $\zeta(R)$ in \thmref{thm:FEC:EE:UB} by substituting $s = \zeta(R)$, 
    and 
    \eqref{eq:aR+Ewsp:s_opt} holds by solving the minimization in \eqref{eq:aR+Ewsp:substitute_zeta}, resulting in 
    \begin{align} 
        s^* &\triangleq \frac{\sqrt{(1+\SNR)^2 + 4\alpha \SNR}}{2} - \frac{\SNR - 1}{2} 
     \\ &= \frac{2(\alpha+1)\SNR}{\sqrt{(\SNR+1)^2+4\alpha \SNR}+\SNR-1} 
    \end{align} 
    and by noting that $s^* \geq 1$ for all $\SNR \geq 0$.
    
    Substituting \eqref{eq:aR+Ewsp} into \eqref{eq:Ziv-Zakai-based-LB:proof} completes the proof. 
\end{IEEEproof}

\begin{remark}[DPT-based upper bound on the \dist\ exponent]
\label{rem:LBs:compare}
    The bound in \thmref{thm:Ziv-Zakai-based-LB} is valid for any choice of $s \geq 1$ in lieu of $s^*$.
    In particular, the choice $s = 1$ yields
    \begin{align}
    \label{eq:DPT:exp}
        \EE(\alpha) \leq \alpha \lrp{C_0 + \Rh},
    \end{align}
    which coincides with an upper bound on the \dist\ exponent resulting from the DPT. This implies, in turn, that the bound of \thmref{thm:Ziv-Zakai-based-LB} cannot be worse than this DPT-based bound.
\end{remark}

    To prove \thmref{thm:Ziv-Zakai-based-LB}, we used the (impossibility) \textit{weak sphere-packing bound} of \thmref{thm:FEC:EE:UB}, which was derived for transmitter-assisted communication. 
    As explained \secref{ss:background:comm-w-helper}, 
    this bound is weaker than the ordinary sphere-packing bound \eqref{eq:sphere-packing:ordinary} but when the receiver obtains help in lieu of the transmitter, the channel input and the noise are independent, and hence the weak sphere-packing bound can be replaced by the ordinary sphere-packing bound \eqref{eq:sphere-packing:ordinary}.
    This, in turn, yields the following upper bound for the receiver-assisted scenario.

\begin{thm}
\label{thm:Ziv-Zakai-based-LB:Rx-helper}
    The \dist\ exponent \eqref{eq:def:distortion:EE}, in the receiver-assisted scenario,
    is bounded from above as
    $\EE(\alpha) \leq \alpha \left( \Rh + C_0^* \right)$,
    where $C_0^*$ is the solution $R_0 \in \lrp{0, C_0}$ of  
    \begin{align}
    \label{eq:def:C0*:Rx}
        \alpha R_0 = \Esp(R_0 + \Rh) ,
    \end{align}
    with $\Esp(\cdot)$ defined in \eqref{eq:sphere-packing:ordinary}.
\end{thm} 

\begin{IEEEproof}[Proof sktech] 
    Define $U$ as in the proof of \thmref{thm:Ziv-Zakai-based-LB}. Then, following the steps of \eqref{eq:Ziv-Zakai-based-LB:proof}, yields the bound 
    \begin{align}
        \eps(\alpha) 
        \geq \Exp{-\alpha R n}\, \Exp{-n \Esp(R) + o(n)} ,
    \label{eq:Ziv-Zakai-based-LB:Rx-helper:proof:sphere-packing}
    \end{align}
    with the weak sphere-packing error exponent $\Ewsp(\cdot)$ of \thmref{thm:FEC:EE:UB} replaced by the ordinary sphere-packing error exponent $\Esp(\cdot)$ \eqref{eq:sphere-packing:ordinary}.
    This is possible since in the case of an assisted-receiver (only) scenario, the input $\bX$ is independent of the noise $\bZ$.

    Paralleling \eqref{eq:aR+Ewsp}, to minimize $\alpha R + \Esp(R) = \alpha \Rh + (R - \Rh) + \Esp(R)$ for $S > 0$, one may easily verify that $\alpha (R - \Rh)$ and $\Esp(R)$ are strictly monotonically increasing and decreasing with $R \in (\Rh, \Rh + C_0)$, respectively, and 
    \begin{align}
        \lim_{R \downarrow \Rh} \Esp (R) &= S/2 > 0, 
     \\ \lim_{R \uparrow C_0 + \Rh} \Esp(R) &= 0,
     \\ \lim_{R \uparrow \Rh} \alpha (R - \Rh) &= 0,
     \\ \lim_{R \uparrow C_0 + \Rh} \alpha (R - \Rh) &= \alpha C_0 > 0.
    \end{align}
    Thus, the minimum of $\alpha R + \Esp(R)$ is achieved for 
    the unique solution $R^*$ of the equation $\alpha (R - \Rh) = \Esp(R)$ with respect to $R \in (\Rh, \Rh + C_0)$.
    Denoting $C_0^* = R^* - \Rh$ completes the proof.
\end{IEEEproof}

\subsection{\dist~Exponent Lower Bounds}
\label{ss:main:UB}

We derive a lower bound on the \dist\ exponent by constructing a separation-based scheme that quantizes the parameter and communicates the quantization index using a transmitter-assisted code. 
Using natural labeling of the quantization index and conveying the $n\Rh / \log 2$ MSBs of the index essentially error-free by utilizing the result of \thmref{thm:FEC:EE:LB}.

More precisely, let
$M', M \in \nats$ such that $M'$ is a divisor of $M$, namely, $\frac{M}{M'} \in \nats$. Denote the corresponding rates by $R = \frac{1}{n}\log M$ and $R' = \frac{1}{n}\log M'$ respectively, where we use $R' = \Rh - \eps$ with $\eps > 0$ that can be chosen to be arbitrarily small.
We make use of a uniform scalar quantizer of the unit interval that takes one of $M$ possible values, 
with the quantized value of $u \in \left[ -1/2, 1/2 \right)$ being 
\begin{align}
\label{eq:quantizer}
    Q_M(u) \triangleq \frac{1}{M} \lrp{ \left\lfloor M u \right\rfloor + \frac{1}{2}} .
\end{align}

\begin{scheme}
\label{scheme:unequal-error-protection}
\ 

    \textit{Transmitter.}
    \begin{itemize}
    \item 
        Quantize $u$ using a uniform quantizer \eqref{eq:quantizer}, resulting in a quantized value $\hu_w = Q_M(u)$, 
        and a corresponding index $w \in \left\{ 1, \ldots, M \right\}$ that is assigned using natural labeling, 
        namely, $\hu_1 < \hu_2 < \cdots < \hu_M$.
        Decompose $w$ as 
        \begin{align}
            w &= \frac{M}{M'} \cdot (\wh-1) + w_\ell, 
        \end{align}
        for $\wh \in \left\{ 1, \ldots, M' \right\}$ and $w_\ell \in \left\{ 1, \ldots, \frac{M}{M'} \right\}$.
    \item 
        Encode $w$ into a transmit signal $\bX$ using the channel-coding scheme of \cite{Merhav:Tx-assisted-EE:TIT2021} with a helper $\Helper$ of rate $\Rh$,
        such that the sub-message $w_h$ is conveyed error-free (more precisely, with arbitrarily large error exponent) and 
        the sub-message $w_\ell$ is conveyed with an error exponent of (close to) $\EE_a(R - \Rh)$. 
    \end{itemize}

    \textit{Receiver.}
    \begin{itemize}
    \item
        Receive $\bY$.
    \item 
        Using $\bY$, decode $\hWh$ and $\hW_\ell$, which are the reconstructions of $\wh$ and $w_\ell$, respectively.
    \item 
        Construct $\hW = \frac{M}{M'} \cdot (\hWh-1) + \hW_\ell$.
    \item
        Construct the estimate $\hu = \hu_{\hW}$ of $u$.
    \end{itemize}
\end{scheme}

This scheme attains the following \dist.

\begin{thm}
\label{thm:nat-label-scheme} 
    The \dist\ exponent \eqref{eq:def:distortion:EE} is bounded from below as
    \begin{align}
    \label{eq:thm:nat-label-scheme:distortion} 
    \begin{aligned}
        \EE(\alpha) &\geq 
        \alpha \Rh + \max_{R_0 \in \lrs{0, C_0}} \min \lrc{\alpha R_0, \EE_a \lrp{R_0}},
    \end{aligned}
    \end{align}
    where $\EE_a (\cdot)$ denotes any achievable channel-coding error exponent for coding over 
    the same transmitter-assisted AWGN channel, 
    and the maximization is attained for the solution $R_0^*$ of
    \begin{align}
    \label{eq:thm:nat-label-scheme:opt-R} 
        \alpha R_0 = \EE_a \lrp{R_0},
    \end{align}
    with respect to $R_0 \in [0, C_0]$,
    assuming $\EE_a(R)$ is monotonically decreasing with $R$.\footnote{\label{foot:EE:monotonic}Otherwise, $\EE_a(R)$ may be easily improved via $\tilde{\EE}_a(R) \triangleq \sup\limits_{x \geq R} \EE_a(x)$.}
\end{thm}

\begin{IEEEproof}
    Since according to the communication scheme in \cite{Merhav:Tx-assisted-EE:TIT2021},
    $\wh$ is conveyed essentially without error, the \dist\ in conveying $u$ reduces to that of transmitting an effective parameter whose support is an interval of size $1/M' = \Exp{-n R'} = \Exp{- n \lrp{\Rh - \eps}}$ over an AWGN channel at rate $R - \Rh \in [0, C_0)$ without help. 
    Therefore, the \dist\ of the scheme is bounded for any $u \in [-1/2, 1/2)$ as follows.
    \begin{subequations}
    \label{eq:proof:nat-label-scheme}
    \noeqref{eq:proof:nat-label-scheme:min-exp}
    \begin{align}
        \E{\abs{\hat{U}-u}^\alpha}
        &= \CE{\abs{\hat{U}-u}^\alpha}{\hW = w} \PR{\hW = w}
        + \CE{\abs{\hat{U}-u}^\alpha}{\hWh \neq \wh} \PR{\hWh \neq \wh}
    \nonumber
    \\* &\col{\qquad}{} + \CE{\abs{\hat{U}-u}^\alpha}{\hW_\ell \neq w_\ell, \hWh = \wh} \PR{\hW_\ell \neq w_\ell, \hWh = \wh \!}
    \label{eq:proof:nat-label-scheme:smooth}
     \\ &\leq \frac{1}{M^\alpha} + \PR{\hWh \neq \wh} + \frac{1}{\lrp{M'}^\alpha} \PR{\hW_\ell \neq w_\ell}
    \label{eq:proof:nat-label-scheme:bound_by_1}
     \\ &\leq \Exp{-n R \alpha} + \Exp{-n \Eh}
     \\ &\col{\qquad}{\quad} + \Exp{-n \lrp{\Rh - \eps} \alpha} \cdot \Exp{-n \lrp{\EE_a \lrp{R - \Rh} - \eps} + o(n)} 
    \label{eq:proof:nat-label-scheme:Neri-bound}
     \\ &= \Exp{-n \lrs{\min \lrc{ \alpha R, \EE_a \lrp{R - \Rh} + \alpha \Rh} - \eps} + o(n)} ,
    \label{eq:proof:nat-label-scheme:min-exp}
    \end{align} 
    \end{subequations}
    where 
    \eqref{eq:proof:nat-label-scheme:smooth} follows from the law of total expectation
    by noting that the events 
    \begin{align}
        &\lrc{\hW = w} = \lrc{\hWh = \wh, \hW_\ell = w_\ell}, 
     \\ &\lrc{\hWh \neq \wh} = \lrc{\hWh \neq \wh, \hW_\ell = w_\ell} \cup \lrc{\hWh \neq \wh, \hW_\ell \neq w_\ell}, 
     \\ &\text{and } \lrc{\hWh = \wh, \hW_\ell \neq w_\ell} 
    \end{align}
    are disjoint and their union is the entire sample space;
    \eqref{eq:proof:nat-label-scheme:bound_by_1} holds since the estimation error and the probability of correct decoding of $w$ are bounded by 1 and since the estimation error given (correct) $\wh$ is bounded by $1/M'$;
    and \eqref{eq:proof:nat-label-scheme:Neri-bound} holds for any $\eps > 0$, however small, for a sufficiently large $n$ \cite{Merhav:Tx-assisted-EE:TIT2021} for $R < C_0+\Rh$, with $\Eh$ arbitrarily large for a sufficiently large $n$, 
    since the estimation error given correct decoding of $w$ is bounded by $\Exp{-nR}$, and since the estimation error given correct decoding of $\wh$ is bounded by $\Exp{-n R'} = \Exp{-n \lrp{\Rh - \eps}}$; 
    \eqref{eq:proof:nat-label-scheme:min-exp} holds since $\Eh$ is arbitrarily large for a sufficiently large $n$.

    Optimizing \eqref{eq:proof:nat-label-scheme} over $R$ and denoting $R_0 = R - \Rh$ concludes the proof.
\end{IEEEproof}

\begin{remark}
    \schemeref{scheme:unequal-error-protection} assumes assistance at the transmitter and relies on the 
    transmitter-assisted channel coding scheme of Merhav~\cite{Merhav:Tx-assisted-EE:TIT2021} and its achievability bound which is stated in \thmref{thm:FEC:EE:LB}. 
    This channel-coding scheme allows communicating $n \lrp{\Rh-\eps}$ bits with arbitrarily large error exponent for any $\eps > 0$. 
    For the receiver-assisted scenario, in lieu of the transmitter-assisted channel-coding scheme of Merhav,
    one may use the scheme of Lapidoth and Yan~\cite{Lapidoth-Yan:Entropy:Rx-helper:listsize-capacity:Entropy2021} to convey \textit{exactly} $n\Rh$ bits.
\end{remark}

We next derive explicit bounds using specific choices of $\EE_a$.
To that end, consider first the case of a very noisy channel, namely, $\SNR \ll 1$.
For this case, the capacity without assistance equals $C_0 = \SNR/2 + o_{1/\SNR}(\SNR)$, and the error exponent to \cite{GallagerBook1968,ViterbiOmuraBook}
\begin{align}
\label{eq:EE:very-noisy}
    \EE_e(R) = o_{1/\SNR}(\SNR) +
    \begin{cases}
        \frac{C_0}{2}-R, & R < \frac{C_0}{4}
     \\ (\sqrt{C_0}-\sqrt{R})^2, & \frac{C_0}{4}\le R\le C_0
     \\ 0, & R\ge C_0 ,
    \end{cases} 
\end{align}
where $g(S) = o_{1/S}(S)$ means $\lim\limits_{S \to 0} g(S)/S = 0$.

The solution of \eqref{eq:thm:nat-label-scheme:opt-R} of 
\thmref{thm:nat-label-scheme} 
for a very noisy channel results in the following bound.
\begin{cor}
    The \dist\ exponent is bounded from below as 
    \begin{align}
    \label{eq:very-noisy:D}
        \EE(\alpha) \geq \alpha (R_0^* + \Rh) + o_{1/S}(S),
    \end{align}
    where
    \begin{align} 
    \label{eq:nat-label-scheme:R*}
        R_0^* = 
        \begin{cases}
            \frac{C_0}{ 2 (1 + \alpha)}, & \alpha \ge 1
         \\ \frac{C_0}{\left( 1 + \sqrt{\alpha} \right)^2}, & \alpha < 1 .
        \end{cases}
    \end{align}
\end{cor}

Relaxing the assumption of the very noisy channel and substituting the (achievable) random-coding and expurgated error exponents \cite[Chapter~7.4]{GallagerBook1968} in \thmref{thm:nat-label-scheme} yields the following achievability result.
\begin{cor}
\label{cor:expurgated}
    The \dist\ exponent is bounded from below as 
    \begin{align}
    \label{eq:achievability-improved}
        \EE(\alpha) \geq \alpha \lrp{ \Rh+\max \lrc{R_0^*,R_0^{**}} },
    \end{align}
    where 
    $R_0^*$ and $R_0^{**}$ are the solutions of the equations $\alpha R_0 = \EE_r(R_0)$ and $\alpha R_0 = \EE_\mathrm{ex} (R_0)$, respectively, 
    where
    \begin{align}
        \EE_r(R_0) &= 
        \begin{cases}
            1 - \xi + \frac{\SNR}{2} + \half \log \lrp{\xi \lrs{\xi - \frac{\SNR}{2}}} - R_0, & R_0 < \half \log \xi 
         \\ \frac{\SNR}{4\beta}\left[\beta+1-(\beta-1)\sqrt{1+\frac{4\beta}{\SNR(\beta-1)}}\right] 
         \\ + \frac{1}{2}\log\left(\beta-\frac{\SNR(\beta-1)}{2}\left[\sqrt{1+\frac{4\beta}{\SNR(\beta-1)}}-1\right]\right), 
            & \half \log \xi \leq R_0 < C_0
         \\ 0, & R_0 \geq C_0
        \end{cases}
    \end{align}
    with $\beta = \Exp{2R_0}$ and $\xi = \half \lrp{1 + \frac{\SNR}{2} + \sqrt{1 + \frac{\SNR^2}{4}}}$,
    and 
    \begin{align}
        \EE_\mathrm{ex} \lrp{R_0} = 
        \begin{cases}
            \frac{\SNR}{4} \lrp{1-\sqrt{1 - \Exp{-2 R_0}}}, & R_0 < \half \log \lrp{\half + \half \sqrt{1 + \frac{\SNR^2}{4}}}
         \\ 0, & R_0 \geq \half \log \lrp{\half + \half \sqrt{1 + \frac{\SNR^2}{4}}}
        \end{cases}
    \label{eq:expurgated}
    \end{align}
     are the random-coding and expurgated error exponents of the Gaussian channel \cite[Chapter~7.4]{GallagerBook1968}, respectively.
\end{cor}

\begin{remark}
    By \cite[Appendix~B]{Merhav:JSCC:DMC:TIT2013}, $R_0^*$ can also be found via 
        $\displaystyle R_0^* = \sup_{0\le\rho\le 1}\frac{\EE_0(\rho)}{\rho+\alpha}$,
    where 
    \begin{align}
        \EE_0(\rho) &= \sup_{0\le s < \frac{1}{2\SNR}}\bigg[s(1+\rho)\SNR + \frac{1}{2}\log(1-2s\SNR) 
     \col{}{\\ &\qquad\qquad\qquad} + \frac{\rho}{2}\log\left(1-2s\SNR 
        + \frac{\SNR}{1+\rho}\right)\bigg],
    \end{align}
     is the Gallager function~\cite{GallagerBook1968}.
\end{remark}

Consider now the limit of a very small $\alpha > 0$ (and arbitrary $\SNR$). The next result states that the upper bound of \thmref{thm:Ziv-Zakai-based-LB} [and in fact even the DPT-based bound~\eqref{eq:DPT:exp}] and the lower bound of \thmref{thm:nat-label-scheme} coincide in the limit of very small $\alpha$.
\begin{cor}
\label{cor:alpha-->0}
    The \dist\ exponent is bounded from below as 
    \begin{align}
        \EE(\alpha) \geq \alpha \lrp{C_0 + \Rh} + o(\alpha) .
    \end{align}
    Consequently, the \dist\ exponent, in the limit of $\alpha \downarrow 0$, satisfies
    \begin{align}
    \label{eq:alpha-->0}
        \lim_{\alpha \downarrow 0} \frac{\EE(\alpha)}{\alpha} = C_0 + \Rh .
    \end{align}
\end{cor}
\begin{IEEEproof}
    Consider the achievability result of \thmref{thm:nat-label-scheme} in the limit of small $\alpha$
    with any $\EE_a(R_0)$ that satisfies $\EE_a(R_0) > 0$ for $R_0 \in [0, C_0)$.
    Then, in the limit of $\alpha \downarrow 0$, the solution $R_0^*$ of \eqref{eq:thm:nat-label-scheme:opt-R}, approaches $C_0$; consequently, the lower bound in \eqref{eq:thm:nat-label-scheme:distortion} approaches $\alpha \lrp{\Rh + C_0}$.
    
    Combinning this result with the upper bound of \eqref{eq:DPT:exp} yields \eqref{eq:alpha-->0}.
\end{IEEEproof}

Similarly, the ratio between the upper and lower bounds on the \dist\ exponent tends to 1 in the limit of large $\Rh$ for a fixed $\SNR$.
\begin{cor}
\label{cor:Rh-->oo}
    The \dist\ exponent, in the limit of $\Rh \to \infty$, satisfies
    \begin{align}
    \label{eq:Rh-->oo}
        \lim_{\Rh \to \infty} \frac{\EE(\alpha)}{\alpha \lrp{C_0 + \Rh}} = 1.
    \end{align}
\end{cor}

\begin{IEEEproof}
    By \remref{rem:LBs:compare} and 
    \begin{align}
        \alpha \Rh + \alpha \max\lrp{R_0^*, R_0^{**}} \leq \EE(\alpha) \leq \alpha \Rh + \alpha C_0,
    \end{align}
    where, for a fixed $S$, $\max\lrp{R_0^*, R_0^{**}}$ and $C_0$
    are constant in $\Rh$. 
    Hence, the squeeze theorem yields \eqref{eq:Rh-->oo} by noting that
    \begin{align}
        \lim_{\Rh \to \infty} \frac{\alpha \Rh + \alpha \max\lrp{R_0^*, R_0^{**}}}{\alpha \Rh + \alpha C_0} = 1 .
    \end{align}
\end{IEEEproof}

\subsection{Vector Parameter}
\label{ss:main:vec-param}

Consider now the case where the parameter to be modulated is a vector
$\bu = (u_1,\ldots,u_d)\in[-1/2,1/2)^d$,
and define the $d$-dimensional \dist\ and the \dist\ exponent, respectively, by 
    \begin{align} 
    \label{eq:def:vec-param:distortion}
    \begin{aligned} 
        \epsilon_d(\alpha) &\triangleq \sup_{\bu \in [-1/2, 1/2)^d} \E{\norm{\hbU - \bu}_\alpha^\alpha}
     \\ &= \sum_{i=1}^d \E{|\hU_i-u_i|^\alpha}
    \end{aligned}
    \end{align} 
and 
\begin{align}
    \EE_d(\alpha) \triangleq \limsup_{n \to \infty} - \frac{1}{n} \log \inf \eps_d(\alpha) ,
\end{align}
where $\hbU$ is the estimate of vector parameter $\bu$ given the output $\bY$.
Then, the results of the scalar-parameter case carry over to the vector-parameter setting with $\alpha$ replaced by $\alpha/d$, as follows.

\begin{cor}
\label{cor:vec-param}
    The $d$-dimensional \dist\ exponent is bounded from above as 
    \begin{align}
        \EE_d(\alpha)
        \leq \frac{\alpha}{d} \left( \Rh + C_0^* \right)  ,
    \end{align}
    where $C_0^*$ is defined in \eqref{eq:def:C0*}, 
    and from below as
    \begin{align}
        \EE_d(\alpha) &\geq 
        \frac{\alpha}{d} \Rh + \max_{R_0 \in \lrs{0, C_0}} \min \lrc{\frac{\alpha}{d} R_0, \EE_a \lrp{R_0}}
    \label{eq:thm:vec-param:distortion} 
    \end{align}
    with the minimum achieved by the solution of 
    \begin{align}
    \label{eq:thm:vec-param:opt-R} 
        \frac{\alpha}{d} R_0 = \EE_a \lrp{R_0}, 
    \end{align}
    where $\EE_a(\cdot)$ is as in \thmref{thm:nat-label-scheme}.
\end{cor}

\begin{IEEEproof}
    \textit{Impossibility.}
    Define $U$ to be uniformly distributed over the discrete finite set 
    $\calU_M^d$ where $\calU_M$ was defined in \eqref{eq:Ziv-Zakai:proof:discrete-set}.
    Then, the \dist\ is bounded from below as follows.
    \begin{subequations}
    \label{eq:vec-param:LB}
    \noeqref{eq:vec-param:LB:max>mean,eq:vec-param:LB:max>mean:explicit}
    \begin{align}
        \eps_d(\alpha) &\triangleq \sup_{\bu \in [-1/2, 1/2)^d} \E{\norm{\hbU - \bu}_\alpha^\alpha}
    \label{eq:vec-param:LB:def}
     \\ &\geq \E{\norm{\hbU - \bU}_\alpha^\alpha}
    \label{eq:vec-param:LB:max>mean}
     \\ &\geq \sum_{i=1}^d \E{|\hU_i-U_i|^\alpha}
    \label{eq:vec-param:LB:max>mean:explicit}
     \\ &\ge \sum_{i=1}^d \frac{1}{M^\alpha} \PR{|\hU_i-U_i|\ge \frac{1}{M}}
    \label{eq:vec-param:LB:Markov}
     \\ &\ge \Exp{-n\alpha r}\PR{\bigcup_{i=1}^d\left\{|\hU_i-U_i|\ge
    \Exp{-nr}\right\}} \quad\ \ 
    \label{eq:vec-param:LB:UniBound}
     \\ &\geq \Exp{-n\alpha r}\exp\{-n\Ewsp (r\cdot d) + o(n)\}
    \label{eq:vec-param:LB:Merhav}
     \\ &= \Exp{-n\frac{\alpha}{d} R}\exp\{-n\Ewsp (R) + o(n)\}
    \label{eq:vec-param:LB:substit}
    \end{align}
    \end{subequations}
    where
    \eqref{eq:vec-param:LB:def} holds by definition~\eqref{eq:def:vec-param:distortion},
    \eqref{eq:vec-param:LB:Markov} follows from the Markov inequality;
    \eqref{eq:vec-param:LB:UniBound} follows from the union bound by taking $M = \Exp{n r}$;
    \eqref{eq:vec-param:LB:Merhav} 
    follows from 
    \thmref{thm:FEC:EE:LB};
    and \eqref{eq:vec-param:LB:substit} holds for $R \triangleq r \cdot d$.
    Noting that \eqref{eq:vec-param:LB:substit} is identical to 
    \eqref{eq:Ziv-Zakai-based-LB:proof:sphere-packing} with $\alpha/ d$ in lieu of $\alpha$, completes the proof of the lower bound.
    
    \textit{Achievability.}
    The achievability part is the same as before except that the quantization stage is carried out by applying uniform quantization of each parameter dimension at rate $R/d$. \hfill 
\end{IEEEproof}

\corref{cor:vec-param} means that all the results of Sections~\ref{ss:main:UB} and \ref{ss:main:LB} readily apply to the $d$-dimnesional case with $\alpha$ replaced by $\alpha/d$. In particular, we attain the following refinement of \corref{cor:alpha-->0}, which states that the upper and lower bounds of \corref{cor:vec-param} coincide in the limit of large $d$ and/or small $\alpha$.
\begin{cor}
\label{cor:d-->oo}
    The $d$-dimensional \dist\ exponent is bounded from below as 
    \begin{align}
        \EE_d(\alpha) \geq \frac{\alpha}{d} \lrp{C_0 + \Rh} + o\lrp{\frac{\alpha}{d}} .
    \end{align}
    Consequently, the \dist\ exponent, in the limit of $\frac{\alpha}{d} \downarrow 0$, satisfies
    \begin{align}
    \label{eq:d-->oo}
        \lim_{\frac{\alpha}{d} \downarrow 0} \frac{d}{\alpha} \EE_d(\alpha) = C_0 + \Rh .
    \end{align}
\end{cor}

\subsection{Numerical Comparison of Upper and Lower Bounds}
\label{ss:main:compare}

\begin{figure}[t]
\centering
    \begin{subfigure}{\figwidth}
        \psfrag{A}[t]{$\SNR$}
        \psfrag{error exponents}[bl]{Error exponent}
        \includegraphics[width = \figwidth]{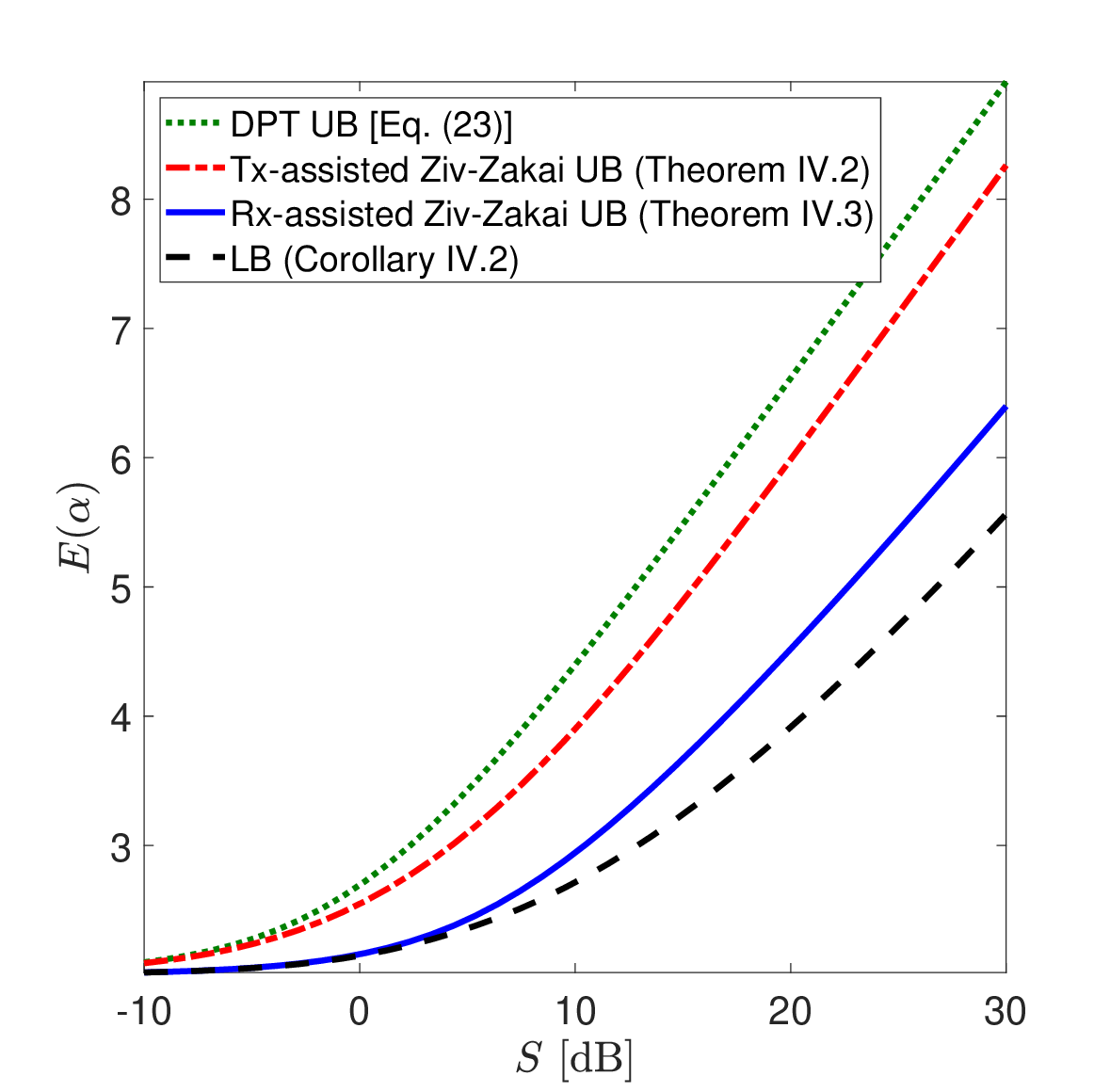}
        \caption{$\alpha = 2$}
        \label{graph1:a=2}
    \end{subfigure}
    \begin{subfigure}{\figwidth}
        \psfrag{A}[t]{$\SNR$}
        \psfrag{error exponents}[bl]{Error exponent}
        \includegraphics[width = \figwidth]{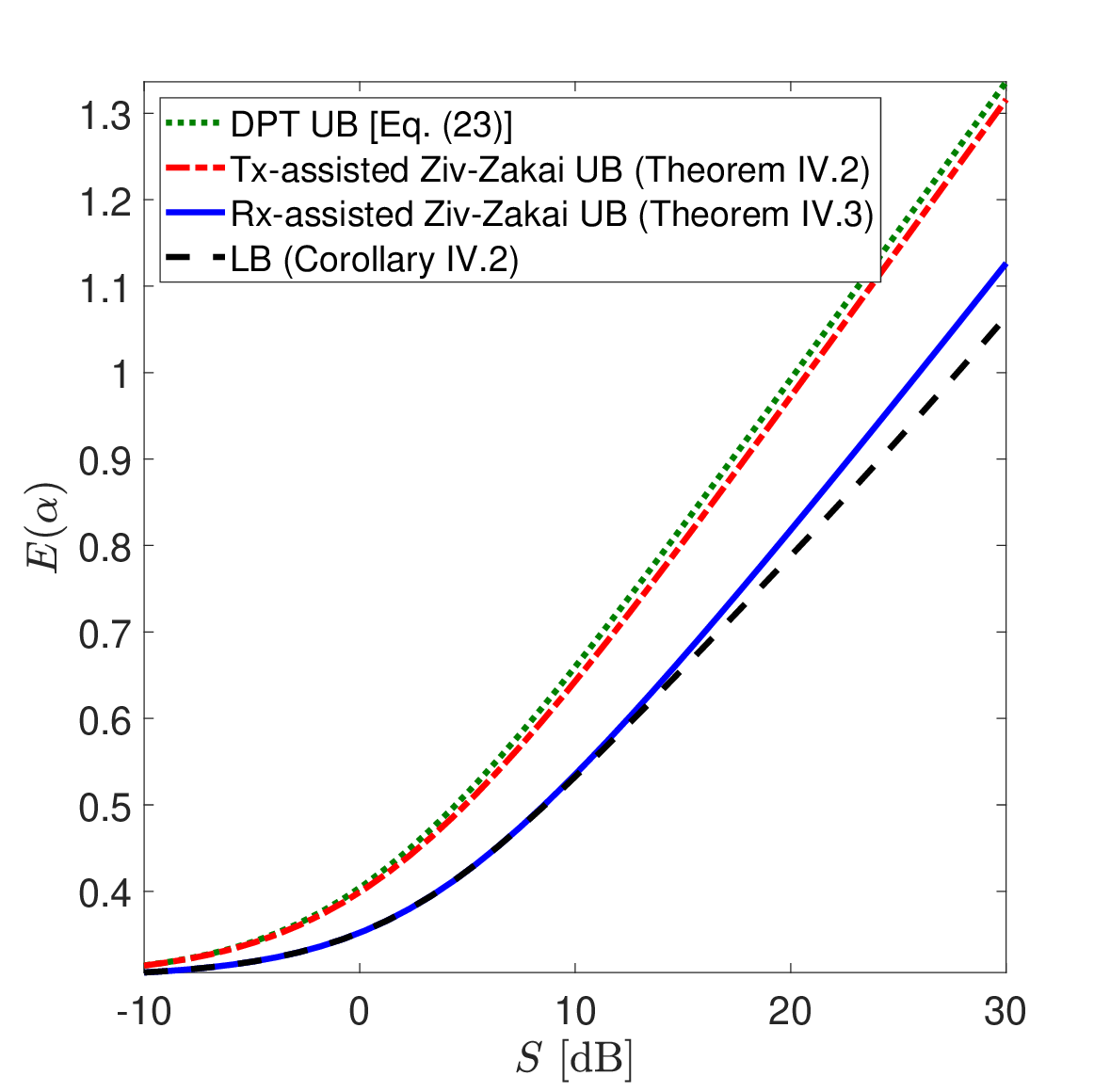}
        \caption{$\alpha = 0.3$}
        \label{graph1:a=0.3}
    \end{subfigure}
    \caption{Graphs of the converse and achievability bounds as functions of $\SNR$,
        for $\Rh=1$ and
        $\alpha=0.3, 2$: The DPT impossibility bound \eqref{eq:DPT:exp}, the transmitter- and receiver-assisted Ziv--Zakai based impossibility bounds  of Theorems~\ref{thm:Ziv-Zakai-based-LB} and~\ref{thm:Ziv-Zakai-based-LB:Rx-helper}, respectively, and the
        achievability bound \corref{cor:expurgated}.}
\label{graph1}
\end{figure}

\begin{figure}[t]
\centering
    \begin{subfigure}{\figwidth}
        \psfrag{A}[t]{$\SNR$}
        \psfrag{error exponents}[bl]{Error exponent}
        \includegraphics[width = \figwidth]{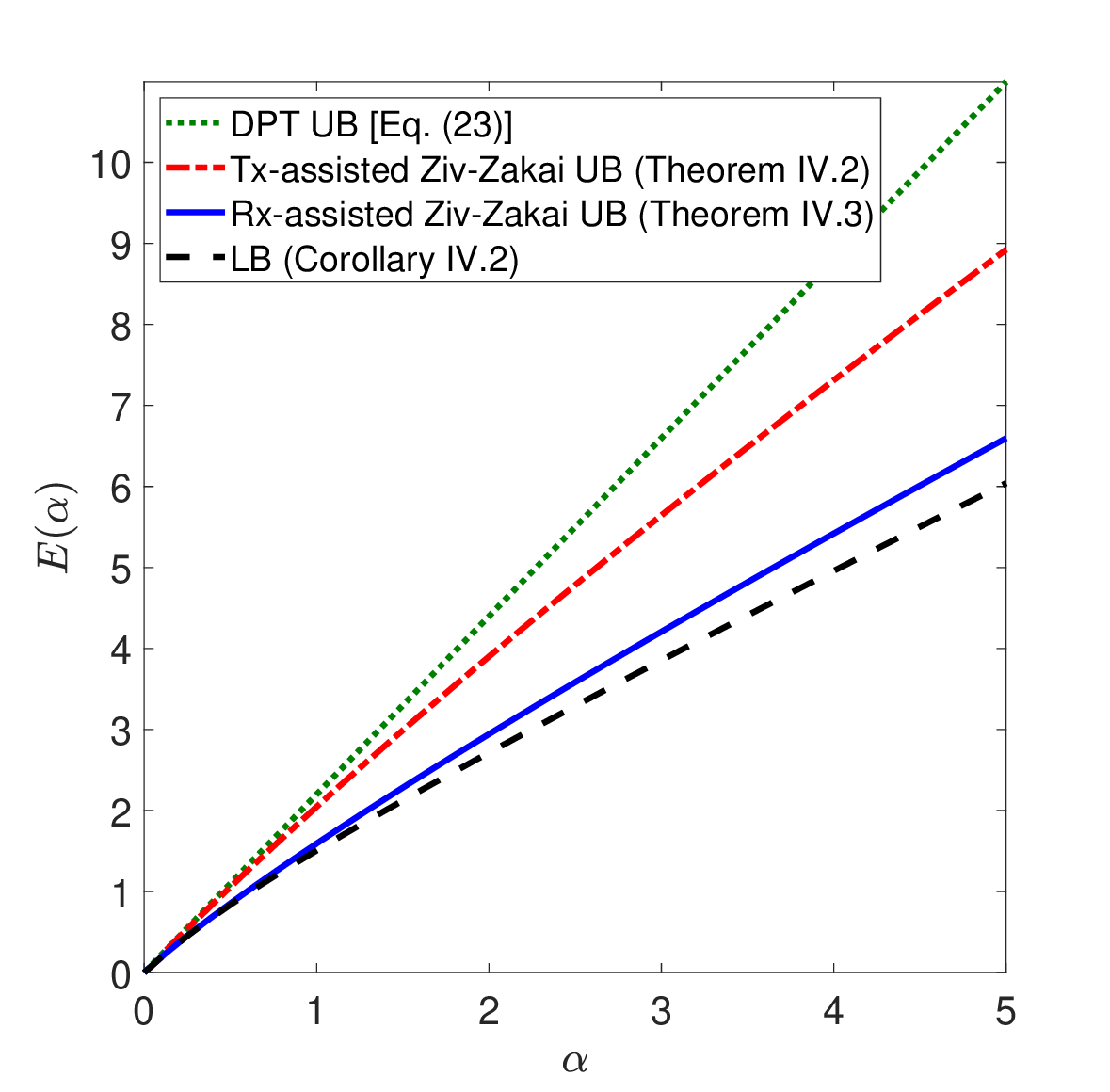}
        \caption{$\SNR = 10$}
        \label{graph2:SNR=10}
    \end{subfigure}
    \begin{subfigure}{\figwidth}
        \psfrag{A}[t]{$\SNR$}
        \psfrag{error exponents}[bl]{Error exponent}
        \includegraphics[width = \figwidth]{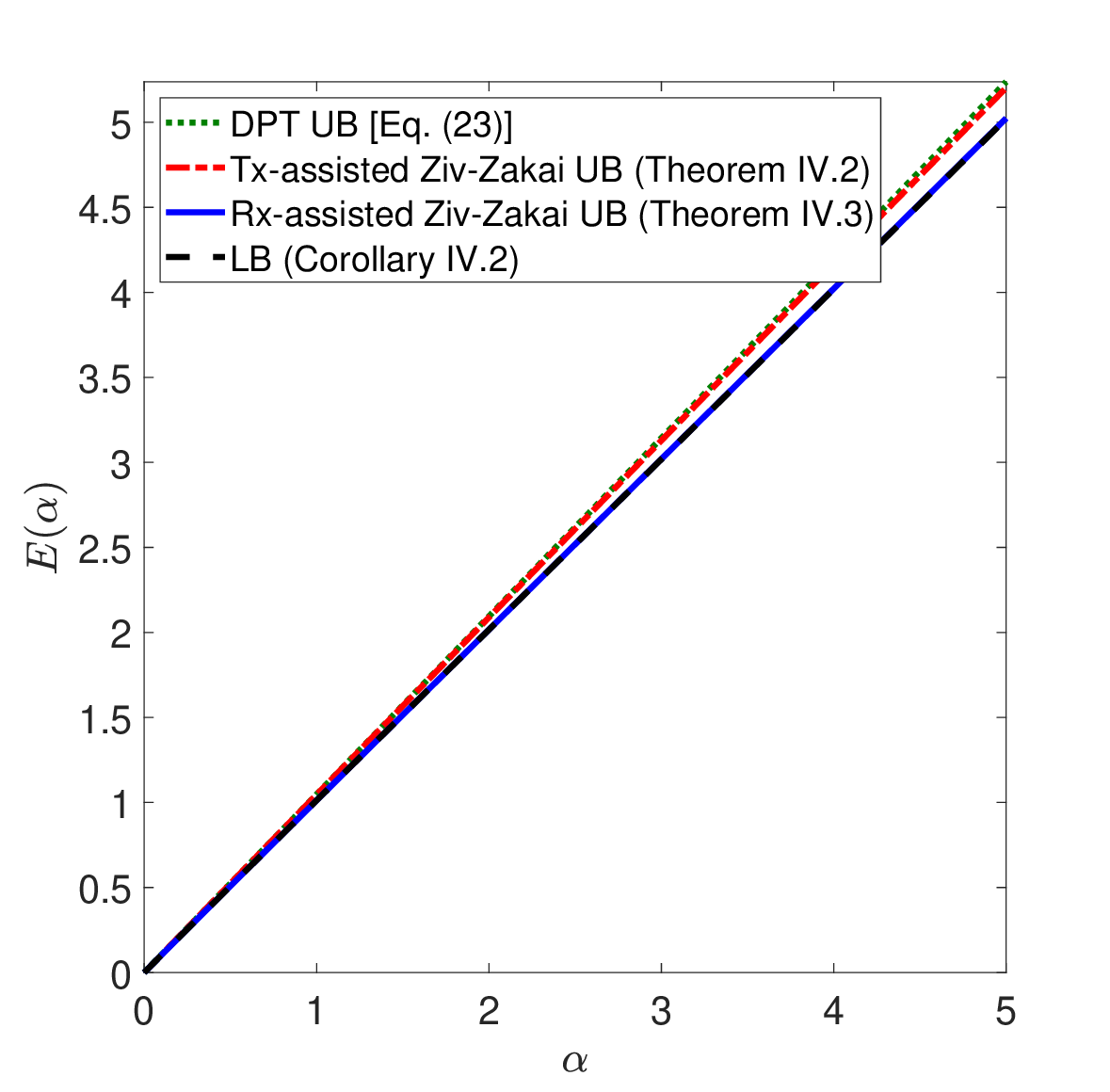}
        \caption{$\SNR = 0.1$}
        \label{graph2:SNR=.1}
    \end{subfigure}
    \caption{Graphs of the converse and achievability bounds as functions of $\alpha$,
        for $\Rh=1$ and
        $\SNR=0.1, 10$: The DPT impossibility bound \eqref{eq:DPT:exp}, the transmitter- and receiver-assisted Ziv--Zakai based impossibility bounds  of Theorems~\ref{thm:Ziv-Zakai-based-LB} and~\ref{thm:Ziv-Zakai-based-LB:Rx-helper}, respectively, and the
        achievability bound \corref{cor:expurgated}.}
\label{graph2}
\end{figure}

We depict the upper and lower bounds on the \dist\ exponent of Sections~\ref{ss:main:LB} and~\ref{ss:main:UB}, respectively, as a function of $\SNR$ for $\alpha = 0.3, 2$ in \figref{graph1}, 
and as a function of $\alpha$ for $\SNR = 1, 10$ in \figref{graph2}.

We observe that the DPT-based bound of \eqref{eq:DPT:exp} is indeed weaker than the transmitter-assisted Ziv--Zakai-based bound of \thmref{thm:Ziv-Zakai-based-LB}, in agreement with \remref{rem:LBs:compare}. The latter is weaker, in turn, than the receiver-assisted Ziv--Zakai-based bound of \thmref{thm:Ziv-Zakai-based-LB:Rx-helper}.

These figures further
demonstrate that 
while there is a gap between the lower and upper bounds, 
it diminishes in the limit of small $\alpha$ and for $C_0 \ll \Rh$, 
in agreement with Corollaries~\ref{cor:alpha-->0} and~\ref{cor:Rh-->oo}; in particular, note that in \figref{graph2:SNR=.1}, 
\begin{align}
    C_0 = \half \log(1 + 0.1) \approx 0.0477 \ll 1 = \Rh, 
\end{align}
and hence the gap between all the bounds is very small.


\section{Energy-Limited Input}
\label{s:fixed-E}

In this section, 
we consider several helper scenarios, 
where the transmitter is subject to a fixed energy constraint \eqref{eq:constraint:energy}.
%
%
%
To that end,
we start by reviewing a known PPM-based scheme and its analysis \cite{BurnashevInfiniteBandwidthExponent1,BurnashevInfiniteBandwidthExponent,TuncelInfinitedBW_SeparationCompanding:Journal} in the absence of a helper.




\begin{scheme}[PPM-based]
\label{scheme:PPM:basic}
\ 

    \textit{Transmitter.}
    \begin{enumerate}
    \item 
    \label{itm:scheme:PPM:basic:Tx:u-->w}
        Quantize $u$ using a uniform quantizer, resulting in a quantized value $\hu_w = Q_M(u)$, 
        and a corresponding index $w \in \left\{ 1, \ldots, n \right\}$.
    \item 
        Send
        \begin{align}
            x_t &= \sqrt{\Energy} \cdot \delta_{t,w} , 
            & t \in \{1, \ldots, n\}, 
        \end{align}
        where $n = M$, and $\delta$ is Kronecker's delta function, viz., it equals 1 if $t = w$, and 0 otherwise. 
    \end{enumerate}

    \textit{Receiver.}
    \begin{enumerate}
    \item
        Receive $\bY$.
    \item 
    \label{itm:scheme:PPM:basic:Rx:recover-w}
        Decode $\hW = \argmax\limits_{t \in \{1, \ldots, n\}} Y_t$.
    \item
        Construct the estimate $\hU = \hu_\hW$ of $u$.
    \end{enumerate}
\end{scheme}

    Since $M = n$ in this scheme, the total budget $L$ of this scheme is $L \triangleq \log M = \log n$.

We first present an upper bound on the error probability of the channel-coding scheme that is encapsulated in \schemeref{scheme:PPM:basic}, viz., with step~\ref{itm:scheme:PPM:basic:Tx:u-->w} of the transmitter replaced with simply receiving a message $W \in \lrc{1, 2, \ldots, M}$, and with the receiver stopping at step~\ref{itm:scheme:PPM:basic:Rx:recover-w}.
This scheme achieves the following performance 
\cite[Ch.~8]{WozencraftJacobsBook}, \cite[Ch.~8]{GallagerBook1968}, \cite[Ch.~2.5]{ViterbiOmuraBook}.

\begin{thm}
\label{thm:PPM:basic:Pe}
    The error probability in decoding $w \in \lrp{1, \ldots, \Exp{L}}$ in \schemeref{scheme:PPM:basic}, with a proper choice of parameters, is bounded from above as\footnote{
    Sub-exponential improvements are available in \cite[Ch.~8]{GallagerBook1968} and \cite{TuncelInfinitedBW_SeparationCompanding:Journal}.}
    \begin{align}
        \PR{\hW \neq w} \leq 
        \begin{cases}
            \Exp{-\lrp{\frac{\ENR}{4} - L}}, & L \leq \frac{\ENR}{8}
         \\ \Exp{-\left( \sqrt{\frac{\ENR}{2}} - \sqrt{L} \right)^2 }, & \frac{\ENR}{8} < L \leq \frac{\ENR}{2} .
        \end{cases}
    \end{align}
\end{thm}

By applying a similar analysis to that of the proof of \thmref{thm:nat-label-scheme} with the error probability bound of \thmref{thm:PPM:basic:Pe}, replacing $nR$ with $L$, setting $n \Rh = 0$ (no helper), and optimizing over $L$, we arrive at the following result that was formerly derived in \cite{BurnashevInfiniteBandwidthExponent1,TuncelInfinitedBW_SeparationCompanding:Journal} (see also the proof of \corref{cor:UB:fixed-E:D:Rh} in the sequel with $\Rh$ set to 0).

\begin{cor}
\label{cor:PPM:basic:D}
    \schemeref{scheme:PPM:basic} achieves an \dist\ that is bounded from above as 
    \begin{align}
        \E{\abs{u - \hU}^\alpha} \leq \Exp{o(\ENR)} \cdot
        \begin{cases}
            \Exp{-\frac{\alpha}{\lrp{1 + \sqrt{\alpha}}^2} \frac{\ENR}{2}}, & \alpha < 1
         \\ \Exp{ -\frac{\alpha}{1 + \alpha} \frac{\ENR}{4}}, & \alpha \geq 1.
        \end{cases}
    \end{align}
\end{cor}

This bound is in fact tight up to sub-exponential terms in $\ENR$ \cite{BurnashevInfiniteBandwidthExponent,EnergyLimitedJSCC:PPM:Lev_Khina:TCOM2022}.

In the remainder of this section, we address the following three scenarios:
\begin{enumerate}
\item 
\label{itm:fixed-E:Rh}
    In \secref{ss:fixed-E:Rh}, we study the setting of help rate $\Rh$ \textit{per time step}. Namely, the helper conveys $n \Rh$ nats to the transmitter.
\item 
    In \secref{ss:crib:Tx-only}, we consider a cribbed helper, namely a helper that knows both the parameter $u$ and the noise sequence $\bZ$ and can convey a total budget of $\Lh$ nats to the transmitter (only), before transmission. 
\item 
    In \secref{ss:crib:two-sided}, we consider a cribbed helper that reveals the same helper message of total budget $\Lh$ to both the transmitter and the receiver.
\end{enumerate}

\subsection{Helper with a Fixed Rate}
\label{ss:fixed-E:Rh}


The following lower bound on $\eps(\alpha)$ is easily obtained by replacing $P$ by $\Energy/n$ in \thmref{thm:Ziv-Zakai-based-LB} and evaluating $nC_0^*$ in the limit of $n \to \infty$.
\begin{cor}
\label{cor:DPT-based-LB:fixed-E}
    The \dist \eqref{eq:def:distortion} is bounded from below as
    \begin{align}
        \eps(\alpha) &\geq \Exp{-\frac{\alpha \ENR}{2}} \cdot \Exp{-\alpha \Rh n + o(n)}.
    \end{align}
\end{cor}

\begin{remark}
    In the energy constrained setting, the DPT~\eqref{eq:DPT:exp} yields 
    \begin{align}
    \label{eq:DPT:dist:energy-constrained}
        \eps(\alpha) &\geq \frac{\alpha^{\alpha-1}}{\lrs{2\Gamma\lrc{\frac{1}{\alpha}}}^\alpha
    \e} \cdot \Exp{-\frac{\alpha \ENR}{2}} \cdot \Exp{-\alpha \Rh n} 
    \end{align}
    after substituting $P = E/n$ and noting that 
    $\lim\limits_{n \to \infty} n C_0 = \ENR/2$.
    Hence, in this setting, the bound of \corref{cor:DPT-based-LB:fixed-E} coincides with that of \eqref{eq:DPT:dist:energy-constrained} for a fixed $\ENR$ up to exponentially negligible terms in~$\Rh$. 
\end{remark}

Somewhat surprisingly, na\"ive replacement of $P$ by $\Energy/n$ in the achievability analysis of \thmref{thm:nat-label-scheme} fails. 
Specifically, the proof of \thmref{thm:FEC:EE:LB} requires more intricate analysis since, in the original proof, the transmission power during the first sub-block in the scheme of \cite[Section IV-A]{Merhav:Tx-assisted-EE:TIT2021} 
needs to be larger than that of the noise. Yet, the result of \thmref{thm:FEC:EE:LB} still holds, as stated in the following theorem; the proof of this result is available in \appref{app:UB:fixed-E:Rh} for both receiver-only and transmitter-only helpers.

\begin{thm}
\label{thm:UB:fixed-E:Pe:Rh}
    The error probability in decoding $w$ in \schemeref{scheme:PPM:basic}, with a proper choice of parameters, under a total energy constraint \eqref{eq:constraint:energy} is bounded from above as
    \begin{align}
    \label{eq:UB:fixed-E:Pe:Rh}
        \PR{\hW \neq w} \leq \Exp{o(n)} \cdot
        \begin{cases}
            \Exp{- n \EE_\infty}, & L < n \Rh
         \\ \Exp{-\lrp{\frac{\ENR}{4} - L + n\Rh}}, & 0 \leq L - n\Rh < \frac{\ENR}{8}
         \\ \Exp{-\left( \sqrt{\frac{\ENR}{2}} - \sqrt{L - n\Rh} \right)^2 }, & \frac{\ENR}{8} \leq L - n\Rh < \frac{\ENR}{2} 
        \end{cases}
    \end{align}
    where $\EE_\infty$ is arbitrarily large for a sufficiently large $n$.

    Moreover, for $L \in [n\Rh, \ENR/2 + n\Rh)$, $L' < n\Rh$ nats can be conveyed 
    with an arbitrarily large error exponent and where $n\Rh - L'$ is arbitrarily small for a sufficiently large $n$, whereas the remaining $nR - L'$ nats---with an error probability of \eqref{eq:UB:fixed-E:Pe:Rh}.
\end{thm}

Using \schemeref{scheme:unequal-error-protection} with $L = \log M$ (in lieu of $nR = \log M$) and following the proof of \thmref{thm:nat-label-scheme} with the error probability bound of \thmref{thm:PPM:basic:Pe}, yields the following upper bound on the \dist\ of this scheme, which is formally proved in \appref{app:UB:fixed-E:Rh}.

\begin{cor}
\label{cor:UB:fixed-E:D:Rh}
    The \dist \eqref{eq:def:distortion} under a total energy constraint \eqref{eq:constraint:energy} is bounded from above as
    \begin{align}
        \epsilon(\alpha) 
        &\leq \Exp{-n \alpha \Rh + o(n)} \cdot 
        \begin{cases}
            \Exp{-\alpha\frac{\ENR}{2(1+\sqrt{\alpha})^2}}, 
            & \alpha\le 1 
         \\ \Exp{-\alpha\frac{\ENR}{4(1+\alpha)}}, 
          & \alpha>1 .
        \end{cases}
    \label{eq:thm:nat-label-scheme:distortion:fixed-E} 
    \end{align}
\end{cor}

We note that the achievability result of \corref{cor:UB:fixed-E:D:Rh} asymptotically matches the exponential decay rate of $\Exp{-\alpha \Rh n}$ of 
\corref{cor:DPT-based-LB:fixed-E}. 
We further note that the proof of \thmref{thm:UB:fixed-E:Pe:Rh} relies on the fact that a help rate of $\Rh$ nats \textit{per time step} is available. It would be interesting to extend the result of \thmref{thm:UB:fixed-E:Pe:Rh} (and consequently also of \corref{cor:UB:fixed-E:D:Rh}) to work for a (possibly large) fixed total budget of $\Lh$ nats in lieu of $n \Rh$.
In fact, for the receiver-assisted scenario this is possible using the channel-coding scheme of Lapidoth and Yan~\cite{Lapidoth-Yan:Entropy:Rx-helper:listsize-capacity:Entropy2021} which allows transmitting $\Lh$ bits error-free at the expense of a small power loss that becomes negligible with~$n$.

\subsection{Transmitter with a Cribbed Helper}
\label{ss:crib:Tx-only}

Consider a helper that has access to both the parameter $u$ and the entire noise sequence $\bZ$, i.e., $\Helper$ is a function of both $u$ and $\bZ$. $\Helper$ of total budget $\Lh = \log \Mh$ is conveyed only to the transmitter prior to the beginning of the transmission. 


We next construct a scheme that quantizes the parameter $u$ and 
uses a variant of \schemeref{scheme:PPM:basic} (the PPM-based scheme) to convey the quantized value.
This variant of the scheme is reminiscent of the scheme of Liu and Viswanath \cite{Liu-Viswanath:Dirty-Paper:Opportunistic-PPM:TIT2006} for writing on dirty paper: For each possible message $w \in \lrc{1, \ldots, M}$, it allocates $\Mh$ time slots, and the helper provides to the transmitter the index of the slot with the maximal noise so that the transmitter can superimpose its entire transmit energy over that noise. This scheme is described as follows.

\begin{scheme}[Transmitter-assisted with cribbing]
\label{scheme:cribbed-helper:Tx-only}
\ 

    \textit{Helper.}
    \begin{enumerate}[i.]
    \item 
    \label{itm:Tx-help:cribbed:helper:generate-W}
        Quantize $u$ using a uniform quantizer, resulting in a quantized value $\hu_w = Q_M(u)$, 
        and a corresponding index $w \in \left\{ 1, \ldots, M \right\}$. 
    \item 
        Determine the most favorable transmit index $\Helper \in \{1, \ldots, \Mh\}$: 
        \begin{align}
        \label{eq:T:decision-rule:cribbed:Tx-only}
            \Helper = \argmax_{i \in \{ 1, \ldots, \Mh \}} Z_{w + (i-1) M}.
        \end{align}

    \item 
        Convey $\Helper$ to the transmitter. 
    \end{enumerate}
    
    \textit{Transmitter.}
    \begin{enumerate}[i.]
    \item
        Receive $\Helper$ from the helper.
    \item 
        Apply step~\ref{itm:Tx-help:cribbed:helper:generate-W} of the helper.
    \item 
        Send
        \begin{align}
            X_t &= \sqrt{\Energy} \cdot \delta_{t, w + (\Helper-1) M} ,
            & t \in \left\{ 1, \ldots, n  \right\}, 
        \end{align}
        where $n = M \cdot \Mh$.
    \end{enumerate}

    \textit{Receiver.}
    \begin{enumerate}[i.]
    \item
        Observe $\bY$.
    \item 
    \label{itm:Tx-help:cribbed:recover-W}
        Decode $\breve{W} \triangleq \argmax\limits_{t \in \{1, \ldots, n \}} Y_t$, and 
        $\hW = \left( \breve{W} - 1 \mod M\right) + 1$.
    \vspace{.3\baselineskip}
    \item
        Construct the estimate $\hU = \hu_\hW$ of $u$.
    \end{enumerate}
\end{scheme}


We first present an upper bound on the error probability of the channel-coding scheme that is encapsulated in \schemeref{scheme:cribbed-helper:Tx-only}, viz., with step~\ref{itm:Tx-help:cribbed:helper:generate-W} of the helper replaced with simply receiving a message $w \in \lrc{1, 2, \ldots, M}$, and with the receiver stopping at step~\ref{itm:Tx-help:cribbed:recover-W}; the proof of this bound is available in \appref{app:cribbed-helper:PPM}.

\begin{thm}
\label{thm:cribbed-helper:PPM:Pe}
    The error probability in decoding $w$ in \schemeref{scheme:cribbed-helper:Tx-only} is bounded from above as
    \begin{align}
        \PR{\hW \neq w} \leq 
            \Exp{- \lrp{\frac{\ENR}{2} + \sqrt{2 \Lh \ENR} - L} + o_{\Lh}(1)} 
    \end{align}
    for $L < \frac{\ENR}{2} + \sqrt{2 \Lh \ENR}$.
\end{thm}


Applying a similar analysis to that in the proof of \thmref{thm:nat-label-scheme} yields the following upper bound on the \dist\ of this scheme; this result is formally proved in \appref{app:cribbed-helper:PPM}.

\begin{cor}
\label{cor:cribbed-helper:PPM:D}
    The \dist \eqref{eq:def:distortion} with a cribbed helper is bounded from above as
    \begin{align}
        \epsilon(\alpha) \leq 
            \Exp{-\frac{\alpha}{1 + \alpha} \lrp{\frac{\ENR}{2} + \sqrt{2 \Lh \ENR}} + o_{\Lh}(1)}. 
    \end{align}
\end{cor}

\subsection{Two-Sided Cribbed Helper}
\label{ss:crib:two-sided}

Consider now a helper that has access to both the parameter $u$ and the entire noise sequence $\bZ$, as in \secref{ss:crib:Tx-only}, i.e., $\Helper$ is a function of both $U$ and $\bZ$. But now, $\Helper$ of total nat budget $\Lh$ is conveyed to both the transmitter and the receiver prior to the beginning of the transmission. 
Consequently, the helper can now use its available nat budget $\Lh$ as a side channel (to simply convey noiselessly $\Lh$ nats of the message), assist directly the transmission over the noisy channel, or both. We next present these alternatives. 
Interestingly, when allocating the helper to assist the transmission over the noisy channel, a \textit{double exponential} decay  of the error probability in $\Lh$ is attained.

\subsubsection{Side channel}

Since the helper knows the message and conveys it to the receiver (and the transmitter), it may 
apply a similar scheme to \schemeref{scheme:unequal-error-protection}, but convey $\wh$---the sub-message that corresponds to the $\Lh \log_2 \e$ MSBs of the quantization of $u$---using $\Helper$ without any noise.
Since these $\Lh \log_2 \e$ MSBs are conveyed to the receiver without error, the support of the effective parameter is limited to $[-\Exp{-\Lh}/2, \Exp{-\Lh}/2)$.
Applying \schemeref{scheme:PPM:basic} to this effective parameter yields the following result, which is an immediate consequence of \corref{cor:PPM:basic:D}.

\begin{thm}
\label{thm:cribbed:two-sided:side-channel}
    The \dist \eqref{eq:def:distortion} is bounded from above as
    \begin{align}
        \epsilon(\alpha) \leq \Exp{-\alpha \Lh + o \lrp{\ENR}} \cdot
        \begin{cases}
            \Exp{-\frac{\alpha}{\lrp{1+\sqrt{\alpha}}^2} \frac{\ENR}{2}},
            & \alpha < 1
         \\ \Exp{-\frac{\alpha}{1+\alpha} \frac{\ENR}{4} },
         & \alpha \geq 1.
        \end{cases}
    \end{align}
\end{thm}
This scheme attains the same performance as in \corref{cor:UB:fixed-E:D:Rh} 
with a simple scheme (thanks to cribbing) for any $\Lh$, and not only for $\Lh = n \Rh$. Note that this scheme does not require knowing $\Helper$ at the transmitter and hence applies also to the setting of a receiver-only helper with cribbing.
    

\subsubsection{Assisted noisy-channel coding}

Since the helper knows both $u$ and $\bZ$ prior to transmission, it knows exactly what the transmitter and the receiver will send and receive. 
Therefore, it can simulate $\Mh$ instances of \schemeref{scheme:PPM:basic} (the PPM-based scheme without assistance) one after the other in time, choose the most favorable one in terms of performance (an instance for which $w$ is decoded correctly, if possible), and convey via $\Helper$ at what time interval to transmit (which of the $\Mh$ schemes to use) to both the transmitter and the receiver. Since the noise sequences in all of these $\Mh$ time intervals are independent, so are their respective error probabilities. This scheme and bounds on its performance are detailed next.

\begin{scheme}[Two-sided cribbed helper]
\label{scheme:cribbed:2sided}
\ 

    \textit{Helper.}
    \begin{enumerate}[i.]
    \item 
    \label{itm:2sided:cribbed:helper}
        Quantize $u$ using a uniform quantizer, resulting in a quantized value $\hu_w = Q_M(u)$, 
        and a corresponding index $w \in \left\{ 1, \ldots, M \right\}$
    \item 
        Determine the most favorable transmit index $\Helper \in \{1, \ldots, \Mh\}$ (an index of the instance of \schemeref{scheme:PPM:basic} for which there will be no detection error of $w$, if possible):
        \begin{align}
        \label{eq:T:decision-rule:cribbed:2sided}
            \Helper = \argmax_{i \in \{ 1, \ldots, \Mh \}} \left( Z_{w + (i-1) M} - \max_{t \in \{1, \ldots, M\}, t \neq w} Z_{t + (i-1)M} \right).
        \end{align}

    \item 
        Convey $\Helper$ to the transmitter and the receiver. 
    \end{enumerate}
    
    \textit{Transmitter.}
    \begin{enumerate}[i.]
    \item
        Receive $\Helper$ from the helper.
    \item 
        Apply step~\ref{itm:2sided:cribbed:helper} of the helper.
    \item 
        Send 
        \begin{align}
            X_t &= \sqrt{\Energy} \cdot \delta_{t, w + (\Helper-1) M} ,
            & t \in \left\{ 1, \ldots, n  \right\}, 
        \end{align}
        where $n = M \cdot \Mh$.
    \end{enumerate}

    \textit{Receiver.}
    \begin{enumerate}[i.]
    \item 
        Observe $\bY$ and receive $\Helper$ from the helper.
    \item 
        Decode $\hW = \argmax\limits_{t \in \{1, \ldots, M\}} Y_{t + (\Helper-1)M}$.
    \label{itm:2sided:Rx:recover-w}
    \item
        Construct the estimate $\hu = \hu_{\hW}$ of $u$.
    \end{enumerate}
\end{scheme}

We first present a simple result that states that the error probability of the information transmission scheme that is encapsulated in \schemeref{scheme:cribbed:2sided} decays \textit{doubly exponentially} with $\Lh$, namely, the error probability of the scheme with step~\ref{itm:2sided:cribbed:helper} of the helper replaced with simply receiving a message $W \in \lrc{1, 2, \ldots, M}$, and with the receiver stopping at step~\ref{itm:2sided:Rx:recover-w}.
This improvement stems from the fact that a decoding error of $w$ happens if and only if all the $\Mh = \Exp{\Lh}$ independent instances of \schemeref{scheme:PPM:basic} (the basic scheme without a helper) fail to correctly decode $w$. Since the instances are independent, the probability of this decoded error is raised to a power of $\Mh = \Exp{\Lh}$ and hence the double exponential decay rate in $\Lh$.

\begin{prop}
\label{prop:2sided:Pe:double-exp}
    Denote by $P_e(L)$ the error probability of \schemeref{scheme:PPM:basic} (PPM without help) for some total nat budget $L$.
    Then, the overall error in decoding $w$, with the same total nat budget $L$ and total help budget $\Lh$, 
    equals 
    \begin{align}
        \PR{\hW \neq w} 
        = \lrs{P_e(L)}^\Exp{\Lh}.
    \end{align}
    In particular, $\PR{\hW \neq w}$ decays doubly exponentially with $\Lh$ for any $L > 0$.
\end{prop}

\begin{IEEEproof}
    Denote the channel error event if the transmitter and receiver (and helper) used instance $i \in \lrc{1, \ldots, \Mh}$ by $\calE_i$. 
    Since the noise sequences corresponding to the different scheme instance $i \in \lrc{1, \ldots, \Mh}$ 
    are i.i.d., 
    \begin{align}
        \PR{\hW \neq w} = \PR{ \bigcap_{i = 1}^{\Mh} \calE_i }
        = \lrs{P_e(L)}^\Exp{\Lh} .
    \end{align}
    Since $P_e < 1$ for \textit{any} $L > 0$, 
    $\PR{\hW \neq w}$ decays doubly exponentially with $\Lh$ for any $L > 0$.\col{\nobreak}{}
\end{IEEEproof}

\begin{remark} 
    The same result holds for any other scheme without help in lieu of \schemeref{scheme:PPM:basic}.
\end{remark}

A straigtforward application of \propref{prop:2sided:Pe:double-exp} with the result of \corref{thm:PPM:basic:Pe} yields 
\begin{align}
\label{eq:2sided:Pe:trivial}
    \PR{\hW \neq w} \leq 
    \begin{cases}
        \Exp{-\lrp{\frac{\ENR}{4} - L} \Exp{\Lh}}, & L < \frac{\ENR}{8}
     \\ \Exp{-\left( \sqrt{\frac{\ENR}{2}} - \sqrt{L} \right)^2 \Exp{\Lh}}, & \frac{\ENR}{8} \leq L < \frac{\ENR}{2} 
     \\ 1, & L \geq \frac{\ENR}{2}.
    \end{cases}
\end{align}

Using this simple bound on the error probability \eqref{eq:2sided:Pe:trivial} and optimizing over $L$ yields the following loose upper bound on the achievable \dist~$\eps(\alpha)$ for a two-sided cribbed helper and transmission under a total power constraint \eqref{eq:constraint:energy}.

\begin{cor}
\label{cor:2sided:PPM-only:D}
    The \dist \eqref{eq:def:distortion} with a cribbed helper that assists both the transmitter and the receiver 
    is bounded from above as
    \begin{align}
        \eps(\alpha) &\leq  
        \begin{cases}
            \Exp{ - \frac{\alpha \, \Exp{\Lh}}{\Exp{\Lh} + \alpha} \frac{\ENR}{4} +  + o_{\ENR}(1)}, & \Lh \leq \log \alpha 
         \\[10pt]
            \Exp{- \frac{\alpha \, \Exp{\Lh}}{\left( \Exp{\Lh/2} + \sqrt{\alpha} \right)^2} \frac{\ENR}{2} +  o_{\ENR}(1)}, & \Lh > \log \alpha .
        \end{cases}
    \label{eq:2sided-cribbed:channel-help}
    \end{align}
\end{cor}

\begin{IEEEproof}
    We bound the \dist\ as follows.
    \begin{subequations}
    \label{eq:2sided-cribbed:channel-help:app}
    \noeqref{eq:2sided-cribbed:channel-help:basic}
    \begin{align}
        \eps(\alpha) &\leq \min_{L :\ \Exp{L} \in \nats} \E{\left| u - \hU \right|^\alpha} 
    \label{eq:2sided-cribbed:channel-help:basic}
     \\ &\leq \min_{L :\ \Exp{L} \in \nats} \lrc{ \CE{\abs{u - \hU}^\alpha}{\hW = w}  + \PR{\hW \neq w} }
    \label{eq:2sided-cribbed:channel-help:bound_by_1}
     \\ &\leq \min_{L :\ \Exp{L} \in \nats} \lrc{ \Exp{-\alpha L} + 
         \begin{Bmatrix}
            \Exp{-\lrp{\frac{\ENR}{4} - L} \Exp{\Lh}}, & L \leq \frac{\ENR}{8}
         \\ \Exp{-\left( \sqrt{\frac{\ENR}{2}} - \sqrt{L} \right)^2 \Exp{\Lh}}, & \frac{\ENR}{8} < L \leq \frac{\ENR}{2} 
        \end{Bmatrix}
        } 
        \quad
    \label{eq:2sided-cribbed:channel-help:single-scheme-bound}
     \\ &\leq \min_{L \geq 0} \lrc{ \Exp{-\alpha L} + 
         \begin{Bmatrix}
            \Exp{-\lrp{\frac{\ENR}{4} - L} \Exp{\Lh} + o_L(1)}, & L \leq \frac{\ENR}{8}
         \\ \Exp{-\left( \sqrt{\frac{\ENR}{2}} - \sqrt{L} \right)^2 \Exp{\Lh} + o_L(1)}, & \frac{\ENR}{8} < L \leq \frac{\ENR}{2} 
        \end{Bmatrix}
        } 
        \quad
    \label{eq:2sided-cribbed:channel-help:int-relax}
    \\ &= 
        \begin{cases}
            \Exp{ - \frac{\alpha \, \Exp{\Lh}}{\Exp{\Lh} + \alpha} \frac{\ENR}{4} + o_{\ENR}(1)}, & \Lh \leq \log \alpha 
         \\[10pt]
            \Exp{- \frac{\alpha \, \Exp{\Lh}}{\left( \Exp{\Lh/2} + \sqrt{\alpha} \right)^2} \frac{\ENR}{2} + o_{\ENR}(1)}, & \Lh > \log \alpha 
        \end{cases}
    \label{eq:2sided-cribbed:channel-help:explicit}
    \end{align}
    \end{subequations}
    where
    \eqref{eq:2sided-cribbed:channel-help:bound_by_1}
    holds since the estimation error and the probability of correct decoding of $w$ are bounded by 1,
    \eqref{eq:2sided-cribbed:channel-help:single-scheme-bound} follows from \eqref{eq:2sided:Pe:trivial}
    and since the estimation error given correct decoding of $w$ is bounded by $1/M = \Exp{-L}$, 
    \eqref{eq:2sided-cribbed:channel-help:int-relax} follows from relaxing the minimization domain,
    and \eqref{eq:2sided-cribbed:channel-help:explicit} follows from solving the optimization with respect to $L$ by equating the exponents. 
\end{IEEEproof}

\begin{remark}
\label{rem:correct-decoding}
    The resulting bound of \eqref{eq:2sided-cribbed:channel-help} is loose when $\Lh$ is large. In particular, in the limit of $\Lh \to \infty$, the bound of \eqref{eq:2sided-cribbed:channel-help} saturates at $2 \Exp{-\alpha \ENR / 2}$, 
    instead of converging to zero. 
    This is also suggested by the result of \thmref{thm:cribbed-helper:PPM:Pe} by noting that the decision rule \eqref{eq:T:decision-rule:cribbed:2sided} majorizes \eqref{eq:T:decision-rule:cribbed:Tx-only}.
    The reason for \eqref{eq:2sided-cribbed:channel-help} being loose for large values of $\Lh$ is that the upper bound on the error probability in decoding $w$ that we have used \eqref{eq:2sided:Pe:trivial} equals 1 at $L = \ENR/2$, 
    whereas the error probability for any $L$ is in fact strictly lower than 1.
    See \secref{s:discussion} for a further discussion.
\end{remark}

\subsubsection{Hybrid}
By allocating a portion $\Lm$ of the total help budget $\Lh$ for side-channel and the remainder $\Lh - \Lm$ for assistance for noisy-channel coding, a more general achievable that subsumes \eqref{eq:2sided-cribbed:channel-help} and the result of \thmref{thm:cribbed:two-sided:side-channel} may be constructed as follows.
Let $\Mm, \Ml, \Mh \in \nats$ where $M = \Mm \cdot \Ml$ and such that $\frac{\Mh}{\Mm} \in \nats$. Equivalently, $\Lm = \log \Mm$, $\Ll = \log \Ml$, $L = \log M$, and $\Lm + \Ll = L$.

\begin{scheme}[two-sided cribbed helper {[hybrid]}]
\label{scheme:cribbed:2sided:hybrid}
\ 

    \textit{Helper.}
    \begin{enumerate}[i.]
    \item 
    \label{itm:2sided:cribbed:helper:hybrid}
        Quantize $u$ using a uniform quantizer \eqref{eq:quantizer}, resulting in a quantized value $\hu_w = Q_M(u)$, 
        and a corresponding index $w \in \left\{ 1, \ldots, M \right\}$
        that is assigned using natural labeling, namely,
        $\hu_1 < \hu_2 < \cdots < \hu_M$. Decompose $w$ as 
        \begin{align}
            w = \Ml \cdot \lrp{\wm - 1} + \wl ,
        \end{align}
        for $\wm \in \lrc{1, \ldots, \Mm}$ and $\wl \in \lrc{1, \ldots, \Ml}$.
    \item 
        Determine the most favorable transmit index $\Helper_\ell \in \{1, \ldots, \Mh / \Mm \}$ (an index of the instance of \schemeref{scheme:PPM:basic} for which there will be no detection error of $\wl$, if possible):
        \begin{align}
        \label{eq:T:decision-rule:cribbed:2sided:hybrid}
            \Helper_\ell = \argmax_{i \in \{ 1, \ldots, M/\Mm \}} \left( Z_{\wl + (i-1) \Ml} - \max_{t \in \{1, \ldots, \Ml\}, t \neq \wl} Z_{t + (i-1)\Ml} \right).
        \end{align}

    \item 
        Convey $\Helper = \Mm \cdot (\Helper_\ell-1) + w_m$ to the transmitter and the receiver. 
    \end{enumerate}
    
    \textit{Transmitter.}
    \begin{enumerate}[i.]
    \item
        Receive $\Helper$ from the helper.
    \item 
        Construct $\Helper_\ell = \floor{\Helper / \Mm} + 1$.
    \item 
        Apply step~\ref{itm:2sided:cribbed:helper:hybrid} of the helper.
    \item 
        Send 
        \begin{align}
            X_t &= \sqrt{\Energy} \cdot \delta_{t, w_\ell + (\Helper_\ell-1) \Ml} ,
            & t \in \left\{ 1, \ldots, n  \right\}, 
        \end{align}
        where $n = \Ml \cdot \Mh / \Mm$.
    \end{enumerate}

    \textit{Receiver.}
    \begin{enumerate}[i.]
    \item 
        Observe $\bY$ and receive $\Helper$ from the helper.
    \item 
        Construct $\Helper_\ell = \floor{\Helper / \Mm} + 1$ and $w_m = (\Helper - 1 \mod \Mm) + 1$.
    \item 
        Decode $\hW_\ell = \argmax\limits_{t \in \{1, \ldots, M_\ell\}} Y_{t + (\Helper_\ell-1)M_\ell}$.
    \label{itm:2sided:hybrid:Rx:recover-w}
    \item 
        Construct $\hW = \Ml \lrp{\wm - 1} + \hW_\ell$
    \item
        Construct the estimate $\hu = \hu_{\hW}$ of $u$.
    \end{enumerate}
\end{scheme}


\schemeref{scheme:cribbed:2sided:hybrid} transmits $\Lm / \log 2$ MSBs essentially error-free (more precisely, with an error probability that decays exponentially fast with $\Lh$ with an arbitrarily large decay rate), resulting in an effective parameter support of $[-\Exp{-\Lm}/2, \Exp{-\Lm}/2)$. Applying the analysis of the result of \corref{cor:2sided:PPM-only:D}
to this effective parameter yields the following upper bound on \dist.

\begin{cor}
\label{cor:2sided:hybrid:PPM:D}
    The \dist \eqref{eq:def:distortion} with a cribbed helper that assists both the transmitter and the receiver 
    is bounded from above as
    \begin{align}
    \label{eq:2sided-cribbed:hybrid}
        \eps(\alpha) 
        &\leq \min_{\Lm \in [0, \Lh]} \Exp{-\alpha \Lm} 
        \begin{cases}
            \Exp{ - \frac{\alpha}{4} \frac{\Exp{\Lh - \Lm}}{\Exp{\Lh - \Lm} + \alpha} \ENR + o_{\ENR}(1)}, & \Lh - \Lm \leq \log \alpha 
         \\[10pt]
            \Exp{- \frac{\alpha}{2} \frac{\Exp{\Lh - \Lm}}{\left( \Exp{\frac{\Lh - \Lm}{2}} + \sqrt{\alpha} \right)^2} \ENR + o_{\ENR}(1)}, & \Lh - \Lm > \log \alpha 
        \end{cases}
    \end{align}
    where the minimum in \eqref{eq:2sided-cribbed:hybrid} may be shown to be attained for one of three possible values: 
    \begin{itemize}
    \item 
        $\Lm = \Lh$: This value corresponds to the side-channel-only upper bound of \thmref{thm:cribbed:two-sided:side-channel}.
    \item 
        $\Lm = 0$: This corresponds to the assisted-noisy-channel-only upper bound of \corref{cor:2sided:hybrid:PPM:D}.
    \item 
        $\Lm = \Lh - \log \alpha$ (when $\log \alpha < \Lh$).
    \end{itemize}
\end{cor}

The proof that the minimum is attained for one of the three values of $\Lm$ that are stated in 
\corref{cor:2sided:hybrid:PPM:D} is technical and is therefore omitted.

Clearly, by improving the bound \eqref{eq:2sided:Pe:trivial} for $L \geq \frac{\ENR}{2}$, 
the bound of \corref{cor:2sided:PPM-only:D}, and consequently also that of \corref{cor:2sided:hybrid:PPM:D}, may be improved as well.

\section{Continuous-Time AWGN Channel with \\ Unconstrained Bandwidth and a Helper}
\label{s:CT}

Up until now, we have concentrated on the setting of transmission over discrete-time AWGN channels. 
We now explain how to transform the results of Sections \ref{s:main} and~\ref{s:fixed-E} 
to apply to continuous-time AWGN channels with an input power constraint and unconstrained bandwidth. In particular, we show that for the transmitter-assisted setting, if the helper is cribbed, namely, it knows both the message and the noise, one can achieve rates that are higher than \eqref{eq:Tx-assisted-capacity:Lapidoth-Marti}, which is achievable, in turn, when the helper knows only the noise but not the message.

We note that the channel-coding rate of the PPM-based schemes of \secref{s:fixed-E} is $R = \log (M) / n < \log(n) / n$; meaning that it decays to zero as $n$ grows.
On the other hand, we have seen in \secref{s:fixed-E} that, to attain an error probability that decays to zero, one must increase the total transmit energy $E$ to infinity. 
Consequently, under a power constraint $P$ \eqref{eq:constraint:power}, increasing the total energy $E = nP$ amounts to increasing the number of utilized channel uses (degrees of freedom) $n$, which in turn means a decaying coding rate $R$ to zero.
In contrast, in continuous time with unconstrained bandwidth, this coupling between the total energy and the rate can be lifted since the number of degrees of freedom for a fixed time period (and total energy) can be made arbitrarily large by taking a large enough bandwidth. Consequently, the results of \secref{s:fixed-E} can be readily applied to the power-limited continuous-time AWGN channel with unconstrained bandwidth as we show in this section.

To avoid ambiguity, we denote by roman (upright) letters ($\Pc$, $\Rc$, $\Cc$, $\EEc$)  continuous-time quantities that are normalized per second to distinguish them from parallel discrete-time quantities that are normalized per sample and are denoted by italicized (slanted) letters ($P$, $R$, $C$, $\EE$). 
We further denote the transmission time in seconds by $\Time$.

We follow the expositions in \cite{Shannon49}, \cite[Chapter~8]{GallagerBook1968} and \cite[Chapter~9]{CoverBook2Edition} (cf. \cite[Chapter~5]{WozencraftJacobsBook}, \cite[Chapters~2 and 3]{ViterbiOmuraBook}) of the problem of transmission over a continuous-time bandlimited AWGN channel.

\textit{Parameter.}
The parameter to be conveyed is $u \in [-1/2, 1/2)$.


\textit{Transmitter.} Maps the parameter $u$ and the helper's description $\Helper \in \lrc{1, \ldots, \Mh}$ to a channel input signal $\lrcm{X(t)}{t \in [0,\Time]}$ that 
is subject to an average power constraint $\Pc$ watts:
\begin{align}
\label{eq:CT:power-constraint}
    \int_{0}^{\Time} \E{X^2(t)} dt \leq \Pc \Time ,
\end{align}
and $\Time$ is the transmission time in seconds.

\textit{Channel.} The signal $X$ is transmitted over a bandlimited continuous-time AWGN channel:
\begin{align}
\label{eq:CT:AWGN-channel}
    Y(t) &= \big(\lrp{ X + Z } * g \big)(t) , & t \in [0, \Time]\,,
\end{align}
where $Y$ and $Z$ are the output and noise signals, respectively;
\begin{align}
    g(t) = 2\BW \cdot \sinc \lrp{2\BW t} \triangleq 
    \begin{cases}
        \frac{\sin(2 \pi \BW t)}{\pi t}, & t \neq 0 
     \\ 2\BW, &t = 0
    \end{cases}
\end{align}
is the impulse response of an ideal low-pass filter 
\begin{align}
    G(f) = \rect \lrp{\frac{f}{2\BW}} \triangleq
    \begin{cases}
        1, & |f| < \BW
     \\ 0, & |f| \geq \BW ;
    \end{cases}
\end{align}
$\sinc$ is the normalized sinc function;
and $Z$ is white Gaussian noise with two-sided spectral density $\Nzero/2$ watts per hertz.

\textit{Helper.}
Knows (non-causally) $\lrcm{(Z * g)(t)}{t \in [0, \Time]}$ and maps it into a finite-rate description $\Helper \in \lrc{1, 2, \ldots, \Mh}$ with $\Mh = \Exp{\Time \Rhc}$ where $\Rhc$ is the help rate in nats per second. $\Helper$ is revealed to the transmitter prior to the beginning of the transmission.

\textit{Receiver.}
Constructs an estimate $\hU$ of the parameter $u$ given the output signal $\lrcm{Y(t)}{t \in [0,\Time]}$.

\textit{Objective.} The same as in \secref{s:model}, namely minimizing the \dist, $\eps(\alpha)$, of \eqref{eq:def:distortion}. The corresponding optimal achievable \dist\ exponent is defined now as 
\begin{align}
    \EEc(\alpha) \triangleq \limsup_{\Time \to \infty} - \frac{1}{\Time} \log \inf \eps(\alpha),
\end{align}
where the infimum is taken over all transmitter--receiver--helper triplets.

As in \secref{ss:background:comm-w-helper}, 
for the parallel channel-coding problem, instead of a parameter, 
a message $W$ that is uniformly distributed over $\lrc{1, 2, \ldots, M}$ is encoded and transmitted by the transmitter, with $M = \Exp{\Time \Rc}$ where $\Rc$ is the rate in nats per second, and decoded by the receiver from the output signal.

\textit{Discrete-time reduction.}
By the Nyquist--Shannon sampling theorem \cite{Shannon49}, 
this problem is essentially reduced to a discrete-time problem with 
\begin{itemize}
\item 
    $n = 2 \BW \Time$ samples with sampling time interval $\frac{1}{2\BW}$, equivalently, $2\BW$ samples per second;
\item 
    $P = \frac{\Pc}{2\BW}$ power per sample;
\item 
    $\sigma^2 = \Nzero/2$ noise variance;
\item 
    Rates $\Rc = 2 \BW R$, $\Cc = 2 \BW C$, $\Cc_0 = 2 \BW C_0$, and $\Rhc = 2 \BW \Rh$ in nats per second.
\end{itemize}
Applying these parameters to \eqref{eq:Tx-assisted-capacity:Lapidoth-Marti}, results in 
the continuous-time analogue \cite{Slepian:bandwidth:ProcIEEE1976} for a transmitter-assisted channel with bandwidth $\BW$ (without cribbing),
\begin{align}
\label{eq:CT:finite-BW:helper-capacity}
    \Cc(\BW) = \Cc_0(\BW) + \Rhc ,
\end{align}
where $\Cc_0(\BW) = \BW \log \lrp{1 + \frac{\Pc}{\Nzero \BW}}$ is the capacity of this channel without assistance. 

For the unconstrained-bandwidth setting, \eqref{eq:CT:finite-BW:helper-capacity}
reduces to 
\begin{align}
    \Cc \triangleq \lim_{\BW \to \infty} \Cc(\BW) = \Cc_0 + \Rhc ,
\end{align}
where $\Cc_0 = \Pc / \Nzero$ is the power-limited capacity with unconstrained bandwidth without assistance
\cite[Chapter~8.2]{CoverBook2Edition}, \cite[Chapter~8.2]{GallagerBook1968}, \cite[Chapter~5.6]{WozencraftJacobsBook}, \cite[Chapters~2.5 and 3.6.1]{ViterbiOmuraBook}, \cite[Chapter~9.3]{CoverBook2Edition}.


Using the discrete-time reduction above and \thmref{thm:FEC:EE:UB}, setting $B \to \infty$ and noting that in this limit the channel becomes very noisy (recall the exposition in \secref{ss:main:compare}), results in the following upper bound on the achievable error exponent $\EEc(\cdot)$:
\begin{align}
    \PR{\hW \neq W} &\leq \Exp{-\Time\, \EEc_a(\Rc)} ,
 \\ \EEc_a(\Rc) &=  
    \begin{cases}
        \infty, & \Rc < \Rhc
     \\ \frac{\Cc_0}{2} - \Rc + \Rhc, & 0 \leq \Rc - \Rhc < \frac{\Cc_0}{4}
     \\ \lrp{\sqrt{\Cc_0} - \sqrt{\Rc - \Rhc}}^2, & \frac{\Cc_0}{4} \leq \Rc - \Rhc < \Cc_0
     \\ 0, & \Rc \geq \Cc_0 + \Rhc .
    \end{cases}
\label{eq:CT:Neri:EE}
\noeqref{eq:CT:Neri:EE}
\end{align}
Unfortunately, \thmref{thm:FEC:EE:LB} does not provide a meaningful lower bound on the error exponent for $\Rc > \Cc$. 

Returning to the problem of parameter transmission, we consider several helper scenarios: a helper that knows the noise process but is oblivious of the message and helps the transmitter is considered in \secref{ss:CT:oblivious}, a (cribbed) helper that knows both the message and the noise and helps only the transmitter is considered in \secref{ss:CT:cribbed:Tx-only}, and, finally, a (cribbed) helper that knows both the message and the noise and helps both the transmitter and the receiver is studied in \secref{ss:CT:cribbed:two-sided}.

\subsection{Message-Oblivious Helper}
\label{ss:CT:oblivious}

We first consider the setting of a helper that knows the noise process non-causally but is oblivious of the message.
We attain the following upper and lower bounds as a corollary of \thmref{thm:nat-label-scheme} (and \corref{cor:UB:fixed-E:D:Rh}) and \thmref{thm:Ziv-Zakai-based-LB}, respectively, using the discrete-time reduction above and by appealing to the results for a very noisy channel of \secref{ss:main:compare}.

\begin{cor}
\label{cor:CT:power-limited-bounds}
    The \dist \eqref{eq:def:distortion} 
    with help rate $\Rhc$ in nats per second, unconstrained bandwidth (arbitrarily large bandwidth), power $\Pc$, two-sided spectral density $\Nzero/2$, and transmission time $\Time$, where the helper assists the transmitter and/or the receiver,
    is bounded from above as
    \begin{align}
        \eps(\alpha) &\leq \Exp{- \Time \alpha \lrp{\Rc_0^* + \Rhc} + o(\Time)} ,
     \\ \Rc_0^* &= 
        \begin{cases}
            \frac{\Cc_0}{ 2 (1 + \alpha)}, & \alpha \ge 1
         \\ \frac{\Cc_0}{\left( 1 + \sqrt{\alpha} \right)^2}, & \alpha < 1
        \end{cases}
    \end{align}
    and from below as 
    \begin{align}
    \epsilon(\alpha)
    \geq \Exp{-\Time \alpha (\Cc_0+\Rhc ) + o(\Time)};
    \end{align}
    where $\Cc_0 = \Pc / \Nzero$.
\end{cor}

In particular, as discussed in \secref{ss:main:compare}, the exponents of the upper and the lower bounds of \corref{cor:CT:power-limited-bounds} coincide in the limit of small $\alpha$.

\subsection{Transmitter with a Cribbed Helper}
\label{ss:CT:cribbed:Tx-only}

We now consider the cribbed-helper scenario, namely, when the helper knows both the message and the noise before transmission.

Using the discrete-time reduction, we obtain the following corollary to \thmref{thm:cribbed-helper:PPM:Pe}. 
\vspace{-.5\baselineskip}
\begin{cor}
\label{cor:CT:capacity-Pe}
    The capacity $\Cc_C$ in nats per second of the transmitter-assisted continuous-time AWGN channel \eqref{eq:CT:AWGN-channel} with help rate $\Rhc$ in nats per second, input power constraint $\Pc$, unconstrained bandwidth (arbitrarily large bandwidth), and two-sided noise spectral density $\Nzero/2$, 
    where the helper knows both the message and the noise, 
    is bounded from below as
    \begin{align}
        \Cc_C &\geq \Cc_0 + 2 \sqrt{\Rhc \Cc_0} ,
    \end{align}
    where $C_0 = P / \Nzero$ is the capacity of this channel without assistance.
    In particular, $\Cc_C > \Cc$ for $\Rhc < 4 \Cc_0$.

    Furthermore, the 
    error probability is bounded from above by 
    \begin{align}
        \PR{\hW \neq W}
        \leq \Exp{- \Time \lrp{\Cc_0 + 2\sqrt{\Rhc \Cc_0} - \Rc} + o(\Time)}
    \end{align}
    for $\Rc < \Cc_0 + 2\sqrt{\Rhc \Cc_0}$\,, where $\Time$ is the total transmission time, namely, an error exponent of 
    \begin{align}
        \EEc_C(\Rc) = \Cc_0 + 2\sqrt{\Rhc \Cc_0} - \Rc
    \end{align}
    is achievable.
\end{cor}

\begin{IEEEproof}
    We follow \cite[Chapter~8.2]{GallagerBook1968} (cf.~\cite[Chapter~5.6]{WozencraftJacobsBook}, \cite[Chapters~2.5 and 3.6.1]{ViterbiOmuraBook}) by applying \schemeref{scheme:cribbed-helper:Tx-only} the discrete-time reduction:
\begin{itemize}
\item 
    $n = 2 \BW \Time$, $P = \frac{\Pc}{2B}$, $\sigma^2 = \Nzero/2$;
\item 
    rates $\Rc = L/\Time = \log (M) / \Time$ and $\Rhc = \Lh/\Time = \log (\Mh) / \Time$ of nats per second.
\end{itemize}
For this choice 
\begin{itemize}
\item 
    $\Energy = \Pc \Time$, where $\Pc$ is the transmit power in Joules per second (cf.~$P$ from \secref{s:model} which was in Joules per channel use);
\item 
    $\ENR = \Energy / \sigma^2 = 2\Pc \Time / \Nzero = 2 \Time \Cc_0$. 
\end{itemize}
We note that for fixed rates $\Rc$ and $\Rhc$, the utilized bandwidth of the scheme with the above choice of parameters,
\begin{align}
    \BW = \frac{\Exp{\lrp{\Rc + \Rhc} \Time}}{2\Time},
\end{align}
grows exponentially with $\Time$.

Substituting these parameters in \thmref{thm:cribbed-helper:PPM:Pe} yields the desired result.
\end{IEEEproof}

\begin{figure}[t]
\centering
    \begin{subfigure}{.49\textwidth}
        \includegraphics[width = \textwidth, trim = {4mm 4mm 24mm 10mm}, clip]{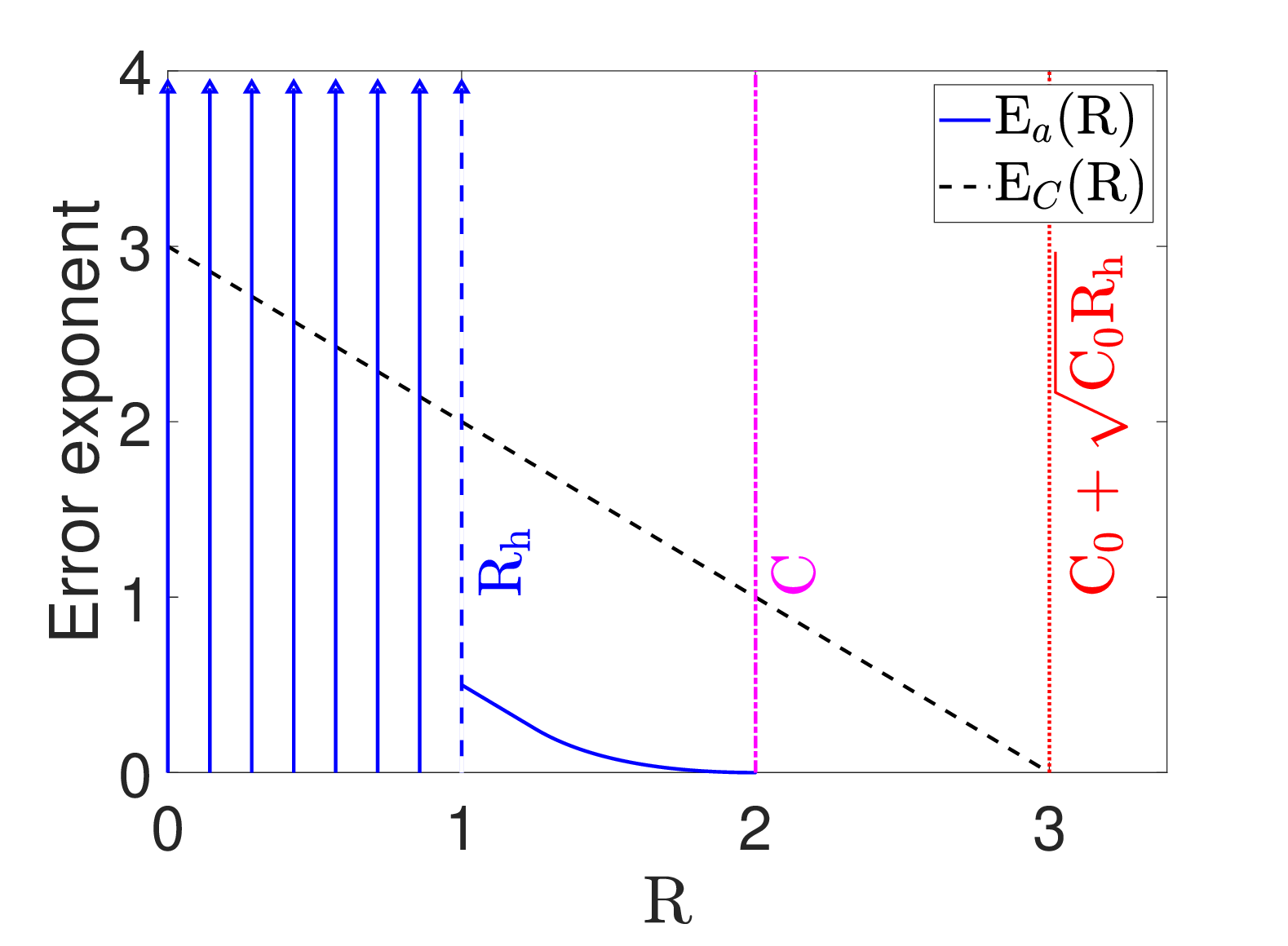}
        \caption{$\Cc_0 = 1, \Rhc = 1$}
        \label{fig:CT:EE:C0=Rh=1}
    \end{subfigure}
    \begin{subfigure}{.49\textwidth}
        \includegraphics[width = \textwidth, trim = {4mm 4mm 24mm 10mm}, clip]{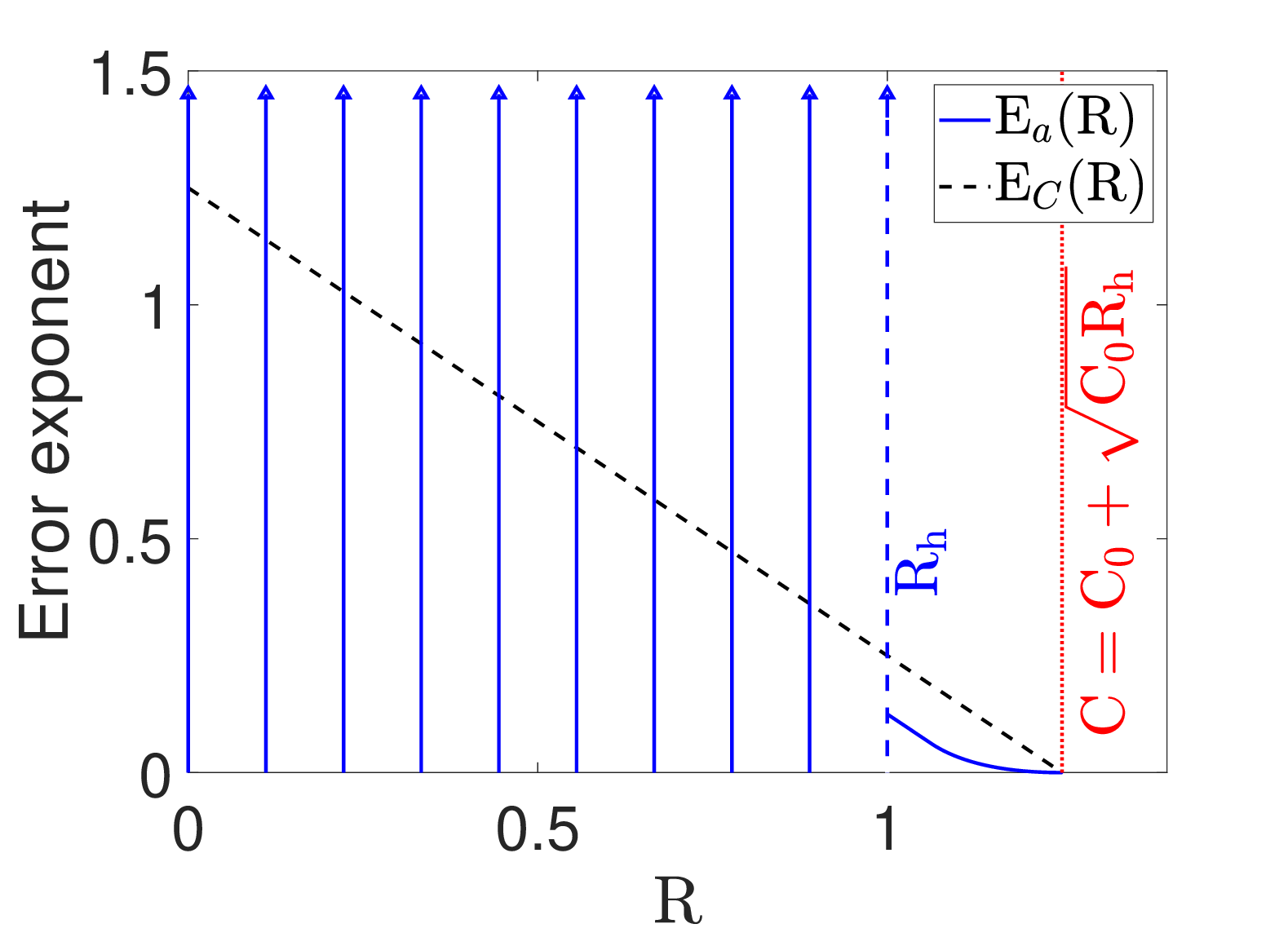}
        \caption{$\Cc_0 = 1/4, \Rhc = 1$}
        \label{fig:CT:EE:C0=1/4_Rh=1}
    \end{subfigure}
    \\ \ \\
    \begin{subfigure}{.49\textwidth}
        \includegraphics[width = \textwidth, trim = {4mm 4mm 24mm 10mm}, clip]{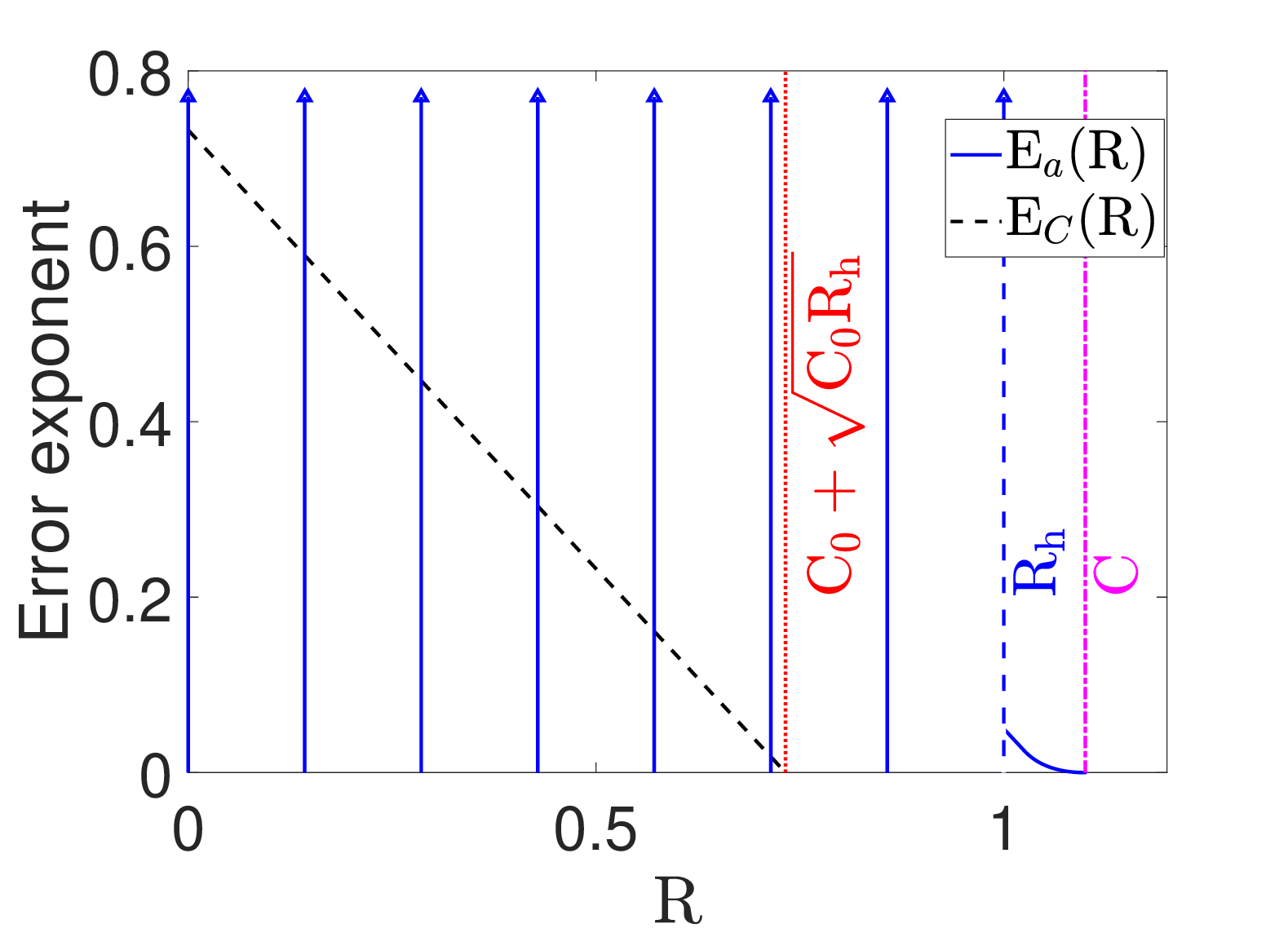}
        \caption{$\Cc_0 = 1/10, \Rhc = 1$}
        \label{fig:CT:EE:C0=1/10_Rh=1}
    \end{subfigure}
    \begin{subfigure}{.49\textwidth}
        \includegraphics[width = \textwidth, trim = {4mm 4mm 24mm 10mm}, clip]{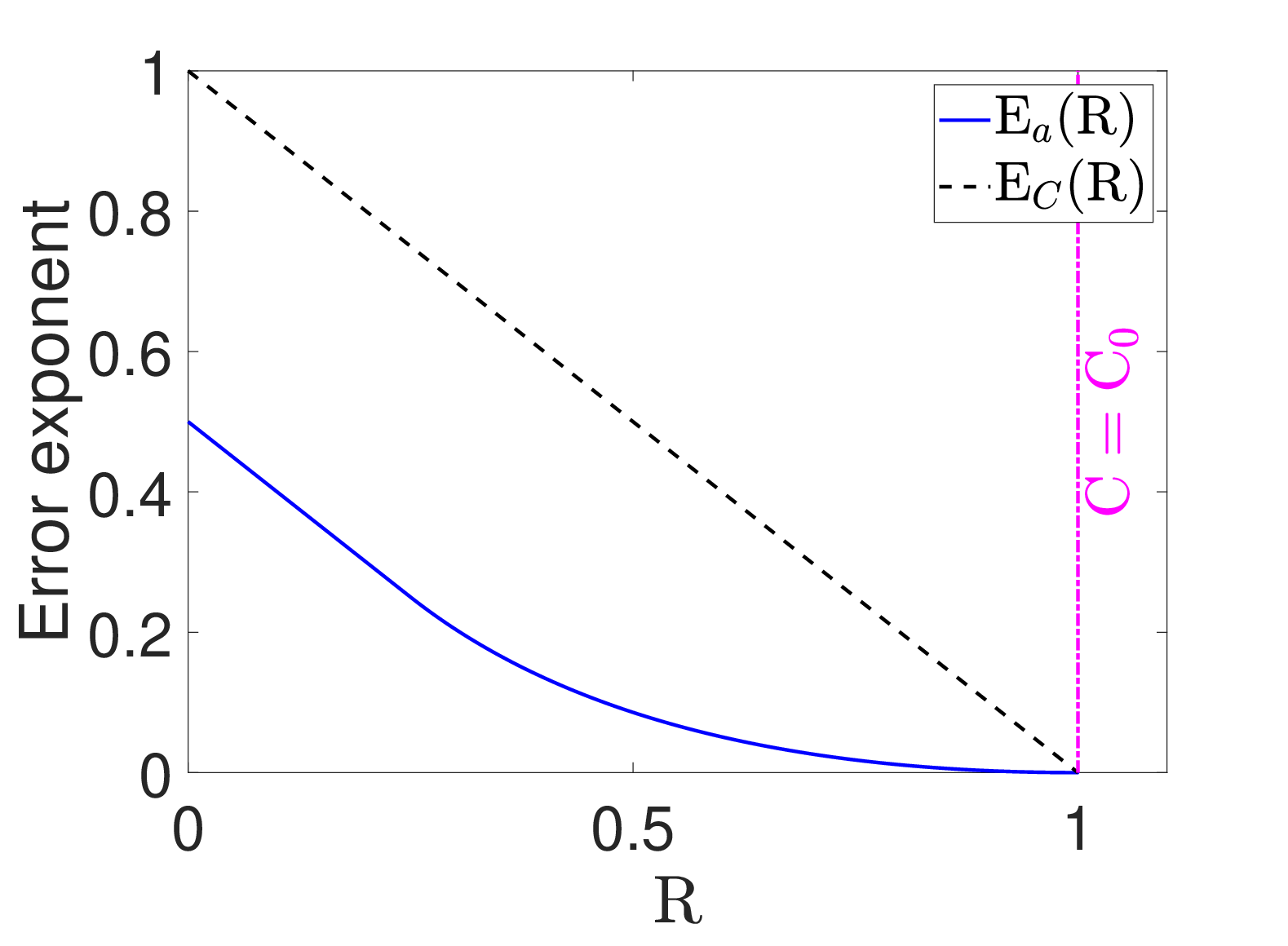}
        \caption{$\Cc_0 = 1, \Rhc = 0$}
        \label{fig:CT:EE:C0=1_Rh=0}
    \end{subfigure}
    \caption{Achievable transmitter-assisted channel-coding error exponents as a function of the rate $\Rc$ for various values of the help rate $\Rhc$ and the capacity without assistance $\Cc_0$: $\EEc_a(\Rc)$---the achievable error exponent of \eqref{eq:CT:Neri:EE} of a helper that knows only the noise, $\EEc_C(\Rc)$---the achievable of \corref{cor:CT:power-limited-bounds} of a helper that knows both the noise and the message.
    The blue vertical arrows denote an arbitrarily large error exponent  $\EEc_a(\Rc)$ for $\Rc < \Rhc$.}
\label{fig:CT:EE}
\end{figure}

We compare the achievable error exponents $\EEc_a(\Rc)$ of \eqref{eq:CT:Neri:EE} and $\EEc_C(\Rc)$ of \corref{cor:CT:capacity-Pe} in \figref{fig:CT:EE}.
First, note that for $\Rc < \Rhc$, $\EEc_a(\Rc)$ is arbitrarily large whereas $\EEc_C(\Rc)$ is bounded. In \figref{fig:CT:EE}, the arbitrarily large error exponent $\EEc_a(\Rc)$ for $\Rc < \Rhc$ is illustrated by blue vertical arrows.
For $\Rc > \Rhc$, on the other hand, both achievable error exponents are bounded. Moreover, whenever $\Rhc \leq 4 \Cc_0$, $\EEc_C(\Rc)$ dominates $\EEc_a(\Rc)$ for $\Rc \in \lrp{\Rhc, \Cc_0 + \Rhc}$. 
This is demonstrated in Figures~\ref{fig:CT:EE:C0=Rh=1} and~\ref{fig:CT:EE:C0=1/4_Rh=1}, wherein the former figure the condition $\Rhc \leq 4 \Cc_0$ is strict and hence achievable rates that exceed $\Cc$
are attainable when the helper is cribbed, whereas in the latter figure
this condition holds with equality and hence the maximal achievable rate of \corref{cor:CT:capacity-Pe} coincides with $\Cc$.

\figref{fig:CT:EE:C0=1/10_Rh=1} depicts the two exponents for the case of $\Rhc \gg 4 \Cc_0$. In this case, the maximal achievable rate of \corref{cor:CT:capacity-Pe} is lower than $\Rh$ and hence $\EEc_a(\Rc)$ dominates $\EEc_C(\Rc)$.

It is interesting to note that there is a discontinuity at $\Rhc = 0$ in the error exponent: 
In the limit of small $\Rhc$, the error exponent of \corref{cor:CT:capacity-Pe} converges to 
\begin{align}
    \lim_{\Rhc \to 0} \EEc_C(\Rc,\Rhc) = \Cc_0 - \Rc, 
\end{align}
which is strictly higher, for all $\Rc < \Cc_0$, than the optimal error exponent without help over the continuous-time AWGN channel with unconstrained bandwidth equals \cite{Wyner:infinite-BW:TIT1967}\footnote{The optimality of this error exponent assumes that \eqref{eq:CT:power-constraint} holds for every message $w$ and not on average. 
While, equal-energy codes (including simplex codes) are suboptimal in general \cite{Steiner:Simplex-Codes:suboptimal:TIT1994} (see also \cite{Dunbridge:Simplex-Codes:Optimality:TIT1967}, \cite[Chapter~4]{Dunbridge:PhD:Simplex-Codes:1966}) they can be shown to achieve the optimal exponential decay with $\Time$ by an expurgation argument with respect to the codeword energies.}

\begin{align}
\label{eq:CT:EE:very-noisy}
    \EEc(\Rc) = 
    \begin{cases}
        \frac{\Cc_0}{2}-\Rc, & \Rc < \frac{\Cc_0}{4}
     \\ (\sqrt{\Cc_0}-\sqrt{\Rc})^2, & \frac{\Cc_0}{4}\le \Rc \le \Cc_0
     \\ 0, & \Rc \ge \Cc_0 
    \end{cases} 
\end{align}
which agrees with $\EEc_a(\Rc)$ for $\Rhc \to 0$.
While surprising, we note that a similar phenomenon happens in the presence of feedback for discrete channels and random transmission duration time, where zero-rate feedback (e.g., stop-feedback) attains a similar achievable straight-line error exponent \cite{Burnashev76,Tchamkerten-Telatar:feedback:BSC:ISIT2002,Tchamkerten-Telatar:feedback:universal:TIT2006,Polyanskiy-Poor-Verdu:feedback:TIT2011}; in fact, since the helper knows both the message and the noise, it can mimic feedback operation.
This is depicted in \figref{fig:CT:EE:C0=1_Rh=0}, where clearly $\EEc_C(\Rc)$ dominates $\EEc(\Rc)$ [which coincides with \eqref{eq:CT:EE:very-noisy}].

\begin{figure}[t]
\centering
    \psfrag{Neri-UB-no-cribbing-----------}[bl]{\footnotesize Cribless: converse}
    \psfrag{Neri-LB-no-cribbing}[bl]{\footnotesize Cribless: achiev.}
    \psfrag{PPM-UB-cribbed}[bl]{\footnotesize Cribbed: achiev.}
    \psfrag{Hybrid-2sided}[bl]{\footnotesize Cribbed two-sided: achiev.}
    \begin{subfigure}{.49\textwidth}
        \includegraphics[width = \textwidth, trim = {4mm 0 22mm 10mm}, clip]{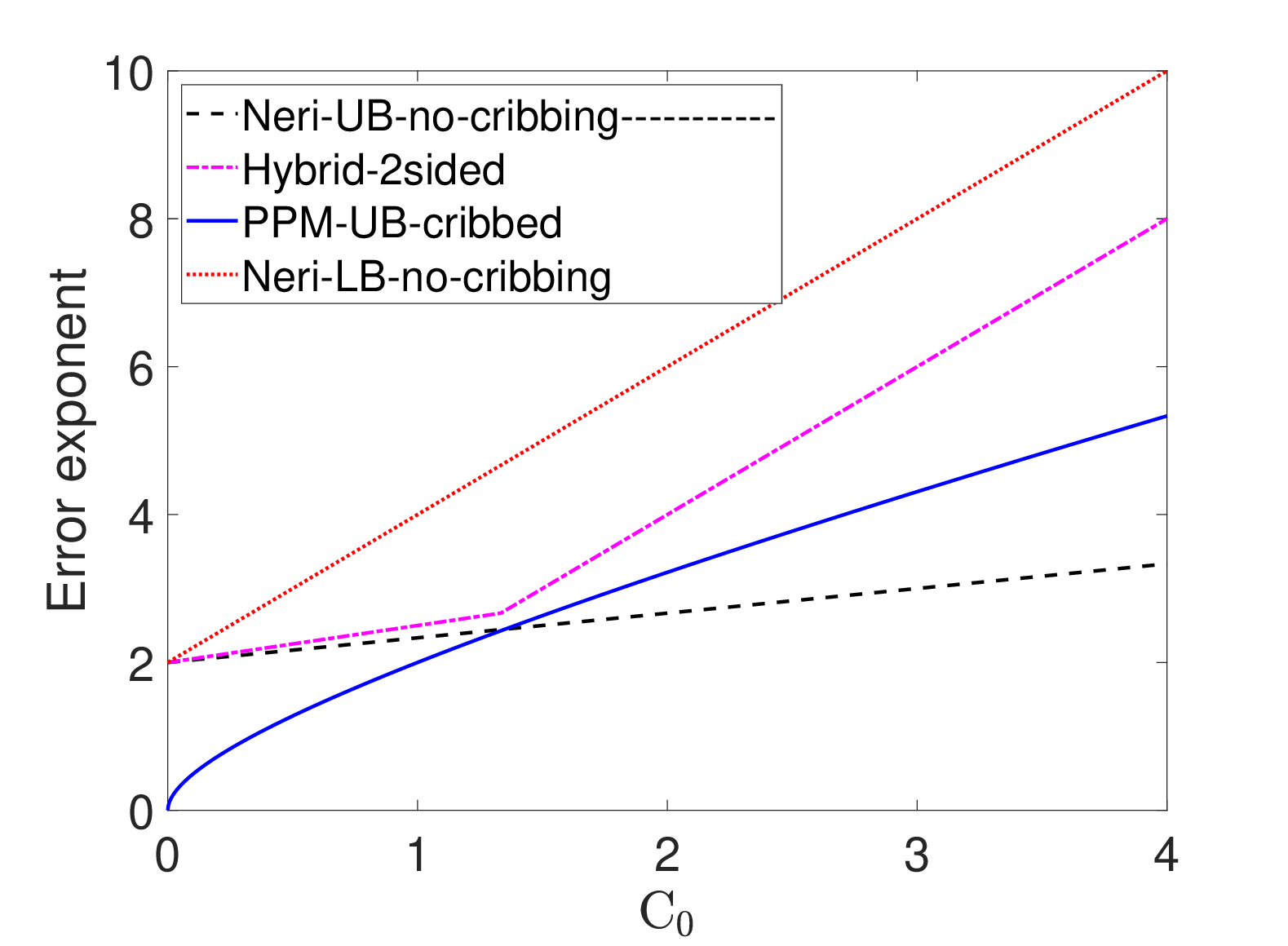}
        \caption{$\alpha = 2, \Rhc = 1$}
        \label{fig:CT:dist:varying_C0:Rh=1_alpha=2}
    \end{subfigure}
    \begin{subfigure}{.49\textwidth}
        \includegraphics[width = \textwidth, trim = {4mm 0 22mm 10mm}, clip]{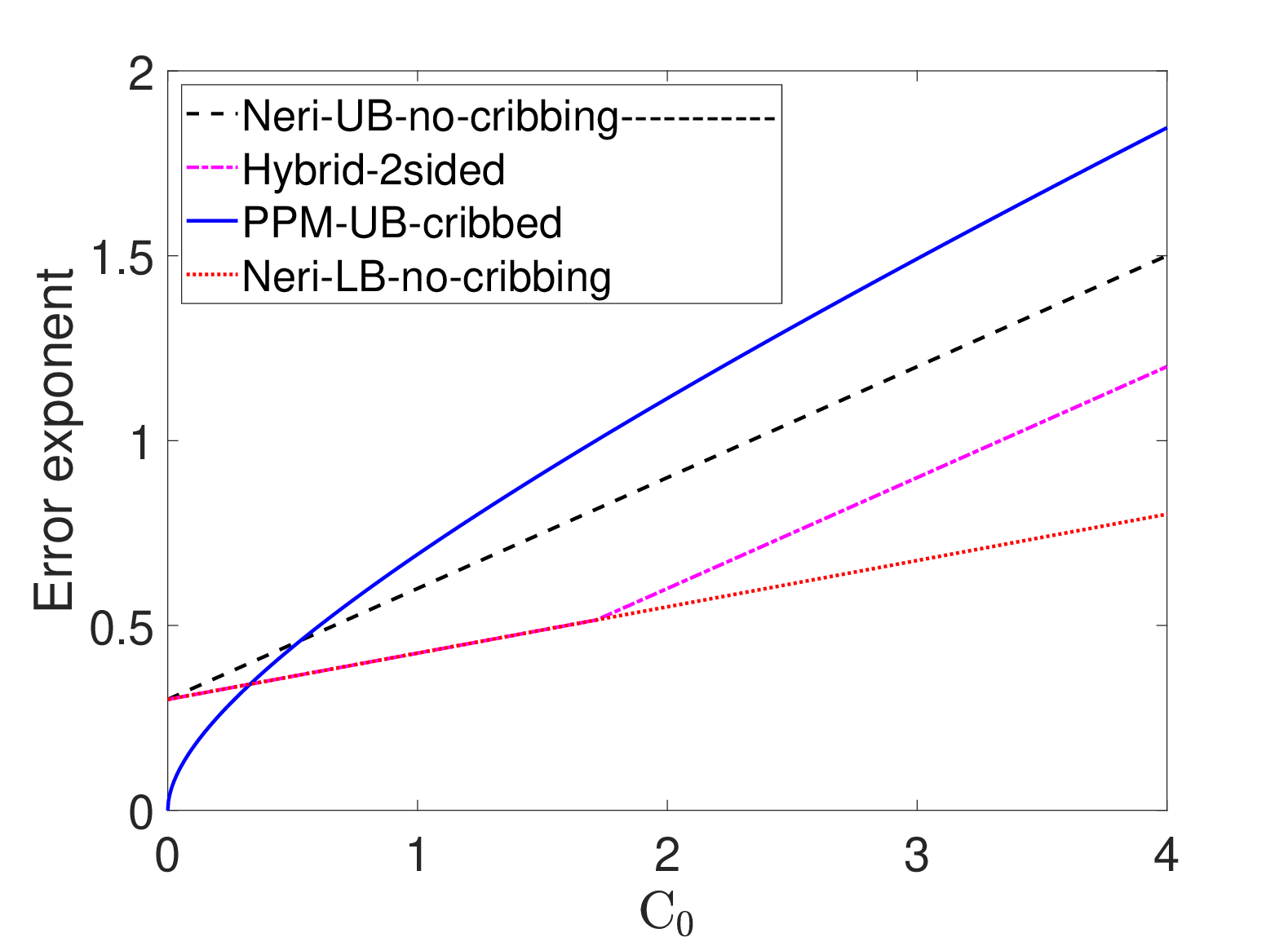}
        \caption{$\alpha = 0.3, \Rhc = 1$}
        \label{fig:CT:dist:varying_C0:Rh=1_alpha=0.3}
    \end{subfigure}
    \caption{Achievable \dist\ exponents as a function of the capacity without assistance $\Cc_0$ for help rate $\Rhc = 1$ and $\alpha = 2, 0.3$: 
    ``Cribless: converse'' and ``Cribless: achievable'' denote the upper and lower bounds on the \dist\ exponent of \corref{cor:CT:power-limited-bounds} for the scenario of a helper that knows only the noise and assists only the transmitter,
    ``Cribbed: achiev.'' denotes the achievable error exponent of \corref{cor:CT:cribbed-helper:PPM:D} for the scenario of a cribbed helper that assists only the transmitter, 
    and ``Cribbed two-sided: achiev.'' denotes the achievable error exponent of \corref{cor:CT:2sided:PPM:D} for the scenario of a cribbed helper that assists both the transmitter and the receiver.}
\label{fig:CT:dist:varying_C0}
\end{figure}

Returning to the parameter transmission problem with a cribbed helper, 
we attain the following bound on the achievable \dist\ upon applying the discrete-time reduction to \corref{cor:cribbed-helper:PPM:D}.

\begin{cor}
\label{cor:CT:cribbed-helper:PPM:D}
    The \dist \eqref{eq:def:distortion} 
    with help rate $\Rhc$ in nats per second, unconstrained bandwidth (arbitrarily large bandwidth), power $\Pc$, two-sided spectral density $\Nzero/2$, transmission time $\Time$, 
    and a helper that knows both the message and the noise and assists the transmitter only,
    is bounded from above as
    \begin{align}
        \epsilon(\alpha) \leq 
            \Exp{-\frac{\alpha}{1 + \alpha} \lrp{\Cc_0 + 2 \sqrt{\Rhc \Cc_0}} \Time + o(\Time)},
    \end{align}
    where $\Cc_0 = \Pc / \Nzero$.
\end{cor}

Comparing the upper and lower bounds on the \dist\ exponent of \corref{cor:CT:power-limited-bounds} (cribless helper) to the achievable \dist\ exponent of \corref{cor:CT:cribbed-helper:PPM:D} (cribbed helper), 
we see that the achievable cribbed-helper \dist\ exponent outperforms the achievable error exponent without cribbing of \corref{cor:CT:power-limited-bounds} for a sufficiently large $\Cc_0$ for a fixed value of $\Rhc$. This behavior is illustrated in \figref{fig:CT:dist:varying_C0} and agrees with the behavior observed for channel coding in \figref{fig:CT:EE}.

\begin{figure}[t]
\centering
    \psfrag{Neri-UB-no-cribbing-----------}[bl]{\footnotesize Cribless: converse}
    \psfrag{Neri-LB-no-cribbing}[bl]{\footnotesize Cribless: achiev.}
    \psfrag{PPM-UB-cribbed}[bl]{\footnotesize Cribbed: achiev.}
    \psfrag{Hybrid-2sided}[bl]{\footnotesize Cribbed two-sided: achiev.}
    \begin{subfigure}{.49\textwidth}
        \includegraphics[width = \textwidth, trim = {4mm 0 22mm 10mm}, clip]{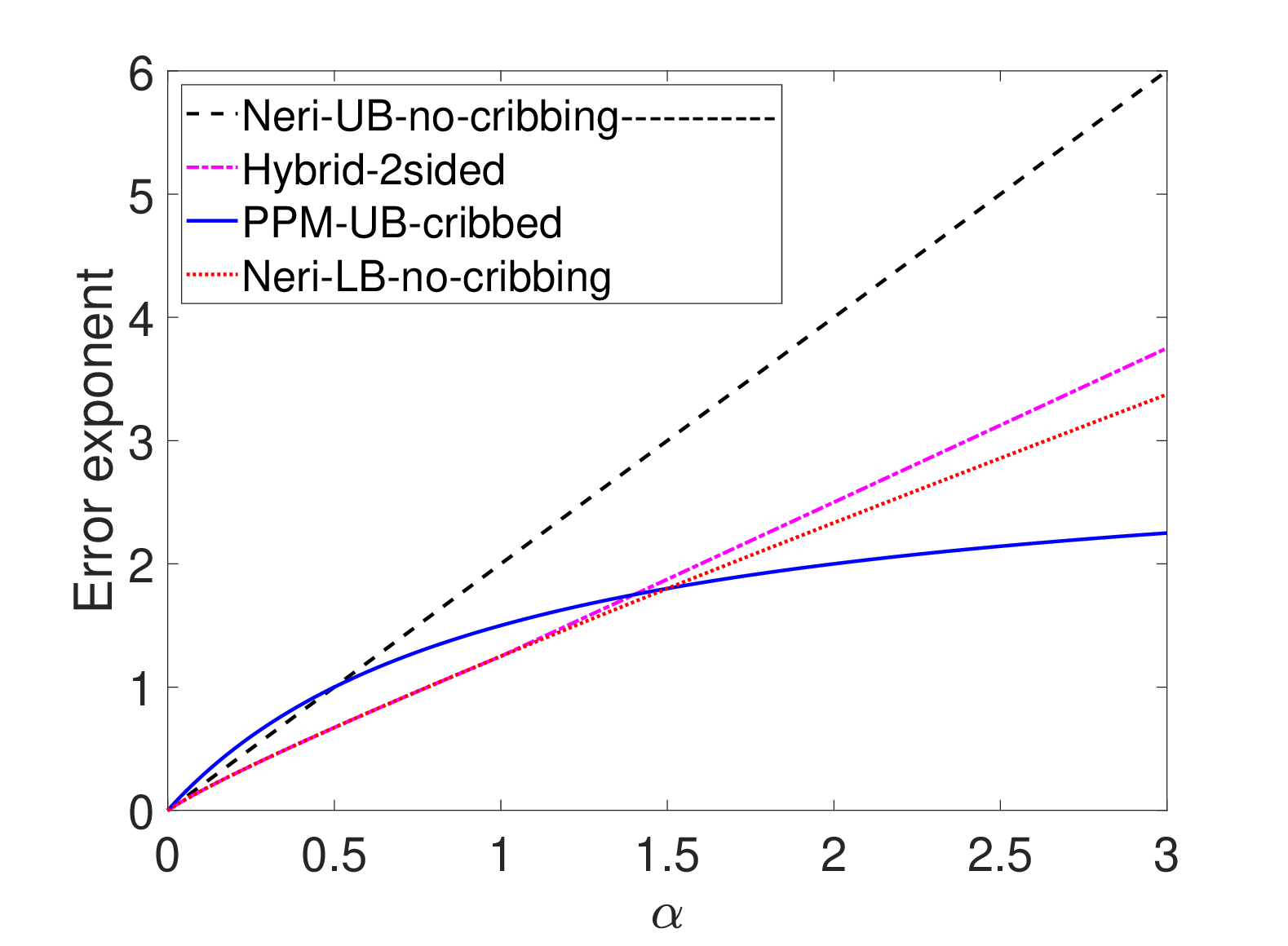}
        \caption{$\Cc_0 = 1, \Rhc = 1$}
        \label{fig:CT:dist:varying_alpha:Rh=1_C0=1}
    \end{subfigure}
    \begin{subfigure}{.49\textwidth}
        \includegraphics[width = \textwidth, trim = {4mm 0 22mm 10mm}, clip]{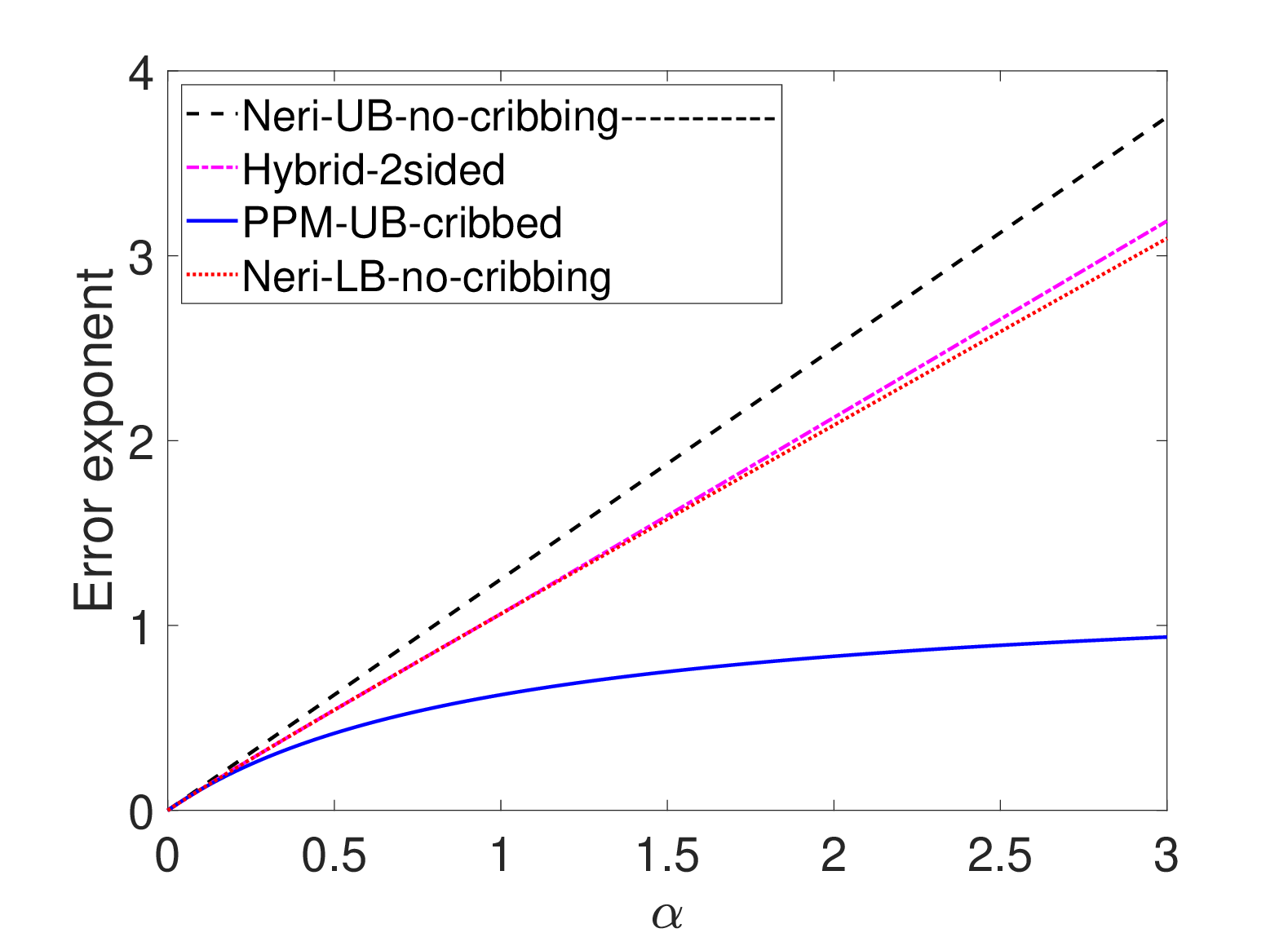}
        \caption{$\Cc_0 = 1/4, \Rhc = 1$}
        \label{fig:CT:dist:varying_alpha:Rh=1_C0=1/4}
    \end{subfigure}
    \caption{Achievable \dist\ exponents as a function of $\alpha$ for help rate $\Rhc = 2, 0.3$ and two values of the capacity without assistance $\Cc_0 = 1, 1/4$: 
    ``Cribless: converse'' and ``Cribless: achievable'' denote the upper and lower bounds on the \dist\ exponent of \corref{cor:CT:power-limited-bounds} for the scenario of a helper that knows only the noise and assists only the transmitter,
    ``Cribbed: achiev.'' denotes the achievable error exponent of \corref{cor:CT:cribbed-helper:PPM:D} for the scenario of a cribbed helper that assists only the transmitter, 
    and ``Cribbed two-sided: achiev.'' denotes the achievable error exponent of \corref{cor:CT:2sided:PPM:D} for the scenario of a cribbed helper that assists both the transmitter and the receiver.}
\label{fig:CT:dist:varying_alpha}
\end{figure}

Interestingly, although the upper (converse) bound on the achievable \dist\ exponent of \corref{cor:CT:power-limited-bounds} is somewhat weak since it relies on the DPT, the achievable cribbed-helper \dist\ exponent of \corref{cor:CT:cribbed-helper:PPM:D} exceeds it for low values of $\alpha$ and sufficiently large help-rate values, as
is evident from Figures~\ref{fig:CT:dist:varying_C0:Rh=1_alpha=0.3} and~\ref{fig:CT:dist:varying_alpha:Rh=1_C0=1}.

For $\Cc_0 = 1/4$ and $\Rhc = 1$, we have seen that the channel-coding error exponent of a cribbed-helper dominates the optimal achievable error exponent without cribbing in \figref{fig:CT:EE:C0=1/4_Rh=1} for $\Rc > \Rhc$ (which is the relevant rate region for the derivation of the \dist\ performance). 
Thus, one may expect the cribbed-helper \dist\ exponent to dominate the cribless-helper \dist\ exponent. Yet, \figref{fig:CT:dist:varying_alpha:Rh=1_C0=1/4} clearly illustrates the opposite. The reason for that seeming contradiction is the unequal protection that is offered by \thmref{thm:FEC:EE:LB}---where a rate of $\Rhc$ nats per second can be conveyed essentially error-free---and is leveraged by \schemeref{scheme:unequal-error-protection}, whereas the channel-coding error exponent of the message $w$ in \schemeref{scheme:cribbed-helper:Tx-only} is finite for all $\Rc$ values.

\subsection{Two-Sided Cribbed Helper}
\label{ss:CT:cribbed:two-sided}

When the cribbed helper is available both at the transmitter and the receiver, \propref{prop:2sided:Pe:double-exp} suggests an achievable doubly-exponential error probability decay rate with time $\Time$. 
This is formulated in the next corollary, whose proof is very similar to that of \corref{cor:CT:capacity-Pe}, 
and is therefore omitted.
\begin{cor}
\label{cor:CT:double-exp:Pe:double-exp}
    Let $P_e(\Rc,\Time)$ by the error probability of 
    a channel-coding scheme with rate $\Rc$ nats per second and transmission time $\Time$ seconds over the continuous-time AWGN channel \eqref{eq:CT:AWGN-channel} (without a helper) with input power constraint $\Pc$, unconstrained bandwidth (arbitrarily large bandwidth), and two-sided noise spectral density $\Nzero/2$.
    Then, there exists a channel-coding scheme with the same rate $\Rc$ and transmission time $\Time$, with a helper with help rate $\Rhc$ that knows both the message and the noise, and assists both the transmitter and the receiver,
    that attains an error probability
    of 
    \begin{align}
        \PR{\hW \neq w} 
        = \left[ P_e(\Rc,\Time) \right]^\Exp{\Lh} .
    \end{align}
%
    In particular, the error probability decays doubly exponentially with $\Rhc$ for any $\Rc$, and it decays doubly exponentially with $\Time$ for $\Rc < \Cc_0$.
\end{cor}

Applying the discrete-time reduction to \corref{cor:2sided:hybrid:PPM:D} yields the following result.

\begin{cor}
\label{cor:CT:2sided:PPM:D}
    The \dist \eqref{eq:def:distortion} 
    with help rate $\Rhc$ in nats per second, unconstrained bandwidth (arbitrarily large bandwidth), power $\Pc$, two-sided spectral density $\Nzero/2$, transmission time $\Time$, 
    and a helper that knows both the message and the noise and assists both the transmitter and the receiver 
    is bounded from above as
    \begin{align}
        \epsilon(\alpha) \leq 
            \Exp{-\Time \alpha \max \lrc{\Cc_0, \Rhc + \frac{\Cc_0}{\min\lrp{4, \lrc{1 + \sqrt{\alpha}}^2}}} + o_T(1)},
    \end{align}
    where $\Cc_0 = \Pc / \Nzero$.
\end{cor}

\begin{IEEEproof}
    Applying the discrete-time reduction of \secref{s:CT}
    with $\Lh = \Time \Rhc$ and $\ENR = 2 \Time \Cc_0$ to \eqref{eq:2sided-cribbed:hybrid} of \corref{cor:2sided:hybrid:PPM:D} results in the following upper bound on the \dist.
    \begin{align}
    \label{eq:CT:2sided-cribbed:hybrid}
        \eps(\alpha)
        &\leq \min_{\Rmc \in [0, \Rhc]} 
        \begin{cases}
            \Exp{ - \Time \alpha \lrs{ \Rmc + \frac{\Exp{\Time \lrp{\Rhc - \Rmc}}}{\Exp{\Time \lrp{\Rhc - \Rmc}} + \alpha} \frac{\Cc_0}{2}} + o_T(1)}, & \Rhc - \Rmc \leq \frac{\log \alpha}{\Time}
         \\[10pt]
            \Exp{- \Time \alpha \lrs{ \Rmc + \frac{\Exp{\Time \lrp{\Rhc - \Rmc}}}{\left( \Exp{\frac{\Time \lrp{\Rhc - \Rmc}}{2}} + \sqrt{\alpha} \right)^2} \Cc_0 } + o_T(1)}, & \Rhc - \Rmc > \frac{\log \alpha}{\Time} 
        \end{cases}
    \end{align}
    where the minimum is attained at $\Rmc = \Rhc$, $\Rmc = 0$, or $\Rmc = \Rhc - \log (\alpha) / \Time$. Next, we evaluate the bound of \eqref{eq:CT:2sided-cribbed:hybrid} for these three values.
    \begin{itemize}
    \item 
        \underline{$\Rmc = \Rhc$:} 
        The conditions of the first and second cases on the right-hand side of \eqref{eq:CT:2sided-cribbed:hybrid} reduce to $\alpha \geq 1$ and $\alpha < 1$, respectively.
        Therefore, the resulting achievable exponent is 
        \begin{align}
            \begin{cases}
                \alpha \lrs{ \Rhc + \frac{\Cc_0}{2(1 + \alpha)}}, & \alpha \geq 1
             \\ \alpha \lrs{\Rhc + \frac{\Cc_0}{\lrp{1 + \sqrt{\alpha}}^2}}, & \alpha < 1.
            \end{cases}
        \end{align}
    \item 
        \underline{$\Rmc = 0$:}
        Now, $\Rhc - \Rmc = \Rhc > \frac{\log \alpha}{\Time}$ for a sufficiently large $\Time$. 
        Therefore, the resulting achievable exponent is 
        $\alpha \Cc_0$, 
        since 
        \begin{align}
            \lim_{\Time \to \infty} \frac{\Exp{\Time \lrp{\Rhc - \Rmc}}}{\left( \Exp{\frac{\Time \lrp{\Rhc - \Rmc}}{2}} + \sqrt{\alpha} \right)^2} = 1
        \end{align}
        for any fixed $\alpha > 0$.
    \item 
        \underline{$\Rmc = \Rhc - \log (\alpha) / \Time$:}
        In this case both of the cases on the right-hand side of \eqref{eq:CT:2sided-cribbed:hybrid} yield the same achievable error exponent 
        $\alpha \lrp{\Rhc + \Cc_0/4}$.
    \end{itemize}
    Noting that $2(1 + \alpha), \lrp{1 + \sqrt{\alpha}}^2 \geq 4$ for $\alpha \geq 1$ and that $2(1 + \alpha), \lrp{1 + \sqrt{\alpha}}^2 < 4$ for $\alpha < 1$ concludes the proof.
    \end{IEEEproof}

The lower bound on the \dist\ exponent of \corref{cor:CT:2sided:PPM:D} for the two-sided cribbed helper scenario coincides with the lower bound of \corref{cor:CT:power-limited-bounds} for a cribless helper that is available only at the transmitter for $\alpha \leq 1$ and $\Rhc \geq 0.75 \Cc_0$, 
and is strictly better otherwise. This is also illustrated in Figures~\ref{fig:CT:dist:varying_C0} and~\ref{fig:CT:dist:varying_alpha}.

However, when comparing the lower bound on the \dist\ exponent of \corref{cor:CT:2sided:PPM:D} for the two-sided cribbed helper scenario with that of \corref{cor:CT:cribbed-helper:PPM:D} for the scenario of a cribbed helper that assists only the transmitter, 
we see in Figures~\ref{fig:CT:dist:varying_C0} and~\ref{fig:CT:dist:varying_alpha} that the latter exceeds the former when $\alpha$ is low and $\Cc_0$ is large enough compared to $\Rhc$.
Since \schemeref{scheme:cribbed:2sided} (all the more so \schemeref{scheme:cribbed:2sided:hybrid}) majorizes \schemeref{scheme:cribbed-helper:Tx-only}, this means that bound of \corref{cor:CT:2sided:PPM:D} is loose in these regimes as the bound of \corref{cor:CT:cribbed-helper:PPM:D} serves also as a lower bound on the \dist\ exponent for the two-sided cirbbed helper scenario.


\section{Summary and Discussion}
\label{s:discussion}

In this work, we derived exponential upper and lower bounds on the optimal achievable $\alpha$-th moment absolute error in transmitting a parameter over an AWGN channel with a helper that knows the noise before transmission begins. 
The lower (impossibility) bound utilized the Ziv--Zakai bounding technique, 
whereas the upper (achievability) bound used judiciously the unequal error protection property of the transmitter-assisted channel coding technique of \cite{Merhav:Tx-assisted-EE:TIT2021}. 
These bounds coincide for low values of $\alpha$, but a gap remains for other values. 

We further considered the setting of energy-limited channel coding and parameter transmission over this channel, for which the previously proposed schemes do not work as is, and refined the achievability technique \cite{Lapidoth-Marti:Tx-assisted:TIT2020,Merhav:Tx-assisted-EE:TIT2021} and its analysis. 
For the energy-constrained setting where the helper knows both the message (``cribbing") and the noise before transmission begins and reveals it to the transmitter, we proposed a PPM-based scheme that outperforms these results under a condition between the help rate and the available energy. When such a helper assists both the transmitter and the receiver, we showed that a doubly-exponential error probability decay rate with the help rate is achievable.

Finally, we translated the results for both the power- and energy-limited settings to results for the setting of transmission over a continuous-time AWGN noise with an input power constraint and unconstrained bandwidth. 
In particular, for transmitter-assisted channel coding with a helper that knows both the message and the noise, we derived achievable rates that are higher than the achievable rate $\Cc_0 +\Rhc$ of a helper that knows only the noise but not the message \cite{Lapidoth-Marti:Tx-assisted:TIT2020}.
Moreover, an achievable error exponent in the limit of zero help rate, $\Rh \to 0$, was shown to equal $C_0 - R$, which is reminiscent of Burnashev's error exponent \cite{Burnashev76}.

The results of \propref{prop:2sided:Pe:double-exp}
and \corref{cor:CT:double-exp:Pe:double-exp} rely on error probability results for channel coding without a helper, with most upper and lower bounds becoming trivial for $R \geq C_0$.
To attain meaningful results for $R \geq C_0$, more delicate bounds for this rate region need to be employed, such as the bounds on correct decoding above capacity \cite{Oohama:AWGN:strong-converse:ISIT2017,Oohama:input-cost:strong-converse:IEICE2018,Tridenski--Somekh-Baruch:AWGN-method-of-types:Arxiv2023}. This is left for future research.



\appendices 


\section{Proofs for \secref{ss:fixed-E:Rh}}
\label{app:UB:fixed-E:Rh}

\begin{IEEEproof}[Proof of \thmref{thm:UB:fixed-E:Pe:Rh}]
    We refine the analysis in the proof of \thmref{thm:FEC:EE:LB} in \cite[Section~IV-A]{Merhav:Tx-assisted-EE:TIT2021}
    with a refined analysis and choice of design parameters that depend on $n$.
    Consider first the case of a receiver-only helper, in which the quantized noise is subtracted by the receiver.
    In particular, instead of the fixed in $n$ value $\tau$, 
    let $\tau_n\in[0,1]$ be a sequence with the following two properties:
    (i) $\lim_{n\to\infty}n\tau_n=\infty$; (ii) 
    $\lim_{n\to\infty}\tau_n\log n=0$. In other words, $\tau_n$ decays faster than
    $1/\log n$ but slower than $1/n$, e.g., $\tau_n=1/\log^2n$ or
    $\tau_n=1/\sqrt{n}$.
    Divide the block of length $n$ into two sub-blocks: a sub-block of length $n\tau_n$, for which the 
    receiver receives a quantized noise description from the helper at rate $n\Rh$, whereas over the remaining part of the block, of
    length $n(1-\tau_n)$, no help is provided at all.
    A uniform scalar quantizer is used
    for all $t \in \lrc{1, \ldots, \tau_n n}$.
    In the limit $n \to \infty$, $\tau_n\to 0$, $\Rh /\tau_n\to\infty$, meaning that the quantizer operates in the high-resolution regime.
    More precisely, consider the $\tau_n n$-dimensional ball of radius
    $\sqrt{\tau_n n \sigma^2 \lrp{1+\frac{B}{\tau_n}}}$, centered at the origin, in the space of noise sequences, $\{\bZ_{1:\tau_n n}\}$, where $B > 0$
    is a design parameter, to be chosen later. The helper's lossy compression
    scheme is based on partitioning this ball into hypercubes of size
    $\Delta_n > 0$ and quantizing $\bZ_{1:\tau_n n}$ into the center of the hypercube
    to which it belongs. If $\bZ_{1:\tau_n n}$ falls outside the ball, then the
    compression fails and an overload error occurs. Accordingly,
    the step size $\Delta_n$ of the uniform scalar quantizer is chosen such that
    the logarithm of the number of hypercubes whose union fully covers the
    ball would not exceed $n\Rh$.
    In other words, $\Delta_n$ should be chosen such that $\Exp{n\Rh}$
    hypercubes must fully cover the ball.
    This amounts to the following relationship:
    \begin{subequations}
    \label{eq:hypercube:covering}
    \noeqref{eq:hypercube:covering:approx}
    \begin{align}
        n \Rh &= \left\lceil \log \frac{\mbox{Vol}\lrc{ \calB_{\tau_n n} \lrp{
        \sqrt{\tau_n n \sigma^2 \lrp{1 + \frac{B}{\tau_n}}}+\sqrt{t}\cdot \Delta_n} }} {\Delta_n^{\tau_n n}} \right\rceil
    \label{eq:hypercube:covering:vol}
     \\ &= \frac{n\tau_n}{2}\log \frac{2\pi
    \e \lrp{ \sigma\sqrt{1 + \frac{B}{\tau_n}}+\Delta_n }^2}{\Delta_n^2} + o(n),
    \label{eq:hypercube:covering:approx}
    \end{align}
    \end{subequations}
    where the right-hand side of \eqref{eq:hypercube:covering:vol} is an upper bound on the number of covering hypercubes, with
    $\sqrt{\tau_n n \sigma^2(1+B/\tau_n)}+\sqrt{\tau_n n}\cdot \Delta_n$ being an upper bound on the
    radius of a larger ball
    that contains (bounds from above) all hypercubes whose
    union completely covers\footnote{The second term, $\sqrt{\tau_n n}\cdot \Delta_n$, is
    the length of the main diagonal of a $\tau_n n$-dimensional hypercube of size
    $\Delta_n$, which is the largest excess radius that each covering
    hypercube can add to the original ball.}
    the original ball of radius $\sqrt{\tau_n n \sigma^2(1+B/\tau_n)}$.
    By demanding that $\Delta_n$ would be large enough such that the first equality
    holds, we guarantee that the rate budget of the
    helper is enough to support the implementability of this scheme.
    Equivalently, from the second equality, we find $\Delta_n$ to be given by
    \begin{align}
            \label{Delta}
            \Delta_n=\frac{\sqrt{2\pi e\sigma^2 \lrp{1+\frac{B}{\tau_n}}}}{\Exp{\frac{\Rh - o_n(1)}{\tau_n}}-\sqrt{2\pi \e}},
    \end{align}
    where for sufficiently small $\tau_n > 0$, the denominator is positive.
    Let $\bZ'_{1:\tau_n n} = q\lrp{\bZ_{1:\tau_n n}}$ be the quantized version of $\bZ_{1:\tau_n n}$ using this quantizer.
    
    During the first sub-block of length $\tau_n n$, the transmitter
    sends $\bX_{1:\tau_n n}$ that depends only
    on the message. The corresponding segment of the received signal, after subtracting $\bZ'_{1: \tau_n n}$, is then
    \begin{align}
        \bY_{1: \tau_n n} - \bZ'_{1: \tau_n n} &= \bX_{1: \tau_n n} + \bZ_{1: \tau_n n} - \bZ'_{1: \tau_n n}
     \\ &= \bX_{1: \tau_n n} + \tb''Z_{1: \tau_n n},
    \end{align}
    where $\tbZ_{1: \tau_n n} \triangleq \bZ_{1: \tau_n n} - \bZ'_{1: \tau_n n}$ is the residual quantization noise.
    As long as the norm of $\bZ_{1: \tau_n n}$ is less than $\sqrt{\tau_n n \sigma^2(1+B/\tau_n)}$, the quantization error sequence $\tbZ_{1: \tau_n n}$,
     lies within the hypercube 
    $\lrs{-\Delta_n / 2, \Delta_n/ 2}^{\tau_n n}$.
    Therefore, if the transmitter uses a quantizer with step size $\Delta_n$ for each coordinate, the transmission in this segment will be error-free (beyond the overload error event), as the residual noise sequence cannot cause
    a passage to the hypercube of any other codeword provided it falls within the ball.
    Such a lattice code can therefore support essentially error-free transmission of $nR'$
    information nats, where
    \begin{subequations}
    \label{eq:rprime}
    \noeqref{eq:rprime:3}
    \begin{align}
        nR' &\ge \frac{n\tau_n}{2}\log\left(\frac{2\pi \e
    \Energy}{n\tau_n \Delta_n^2}\right)-o(n)-n\tau\epsilon(\Delta_n)
    \label{eq:rprime:1}
     \\ &= \frac{n\tau_n}{2}\log \frac{2\pi \e \Energy \cdot (\exp\{(\Rh - o_n(1))/\tau_n\}-\sqrt{2\pi e})^2}{2\pi n\tau_n \e\sigma^2(1+B/\tau_n)} - o(n) -n\tau_n\epsilon(\Delta_n)
    \label{eq:rprime:2}
     \\ &= n[\Rh - o_n(1)]+\frac{n\tau_n}{2}\log\frac{\Energy}{n\tau_n\sigma^2(1+B/\tau_n)}
     \\ &\quad + n\tau_n\log\left(1-\sqrt{2\pi \e}\cdot\exp\{-(\Rh - o_n(1))/\tau_n\}\right)-o(n)-n\tau_n\epsilon(\Delta_n),
    \label{eq:rprime:3}
    \end{align}
    \end{subequations}
    where $\epsilon(\Delta_n)>0$ is a function with the property $\lim_{\Delta_n\to
    0}\epsilon(\Delta_n)/\Delta_n < \infty$,
    \eqref{eq:rprime:1} is proved in the appendix of
    \cite{Merhav:Tx-assisted-EE:TIT2021}, and \eqref{eq:rprime:2} follows from 
    substituting \eqref{Delta} in the main term of the
    resulting expression.
    Clearly, since $\tau_n$ tends to zero such that $\tau_n\log n\to 0$,
    then $R'$ approaches a limit that is not less than $\Rh $. In
    other words, we can transmit
    at any rate, arbitrarily close to $\Rh $ nats per channel use,
    error-free, provided that $\bZ_{1:\tau_n n} \in \calB_{\tau_n n} \lrp{ \tau_n n\sigma^2(1+B/\tau_n) }$.
    An error will occur, in this segment, only if
    $\bZ_{1:\tau_n n}$ falls outside the aforementioned union of hypercubes (in the space of
    noise sequences),
    which in turn implies that $\bZ_{1:\tau_n n} \notin \calB_{\tau_n n} \lrp{ \tau_n n\sigma^2(1+B/\tau_n) }$ as those hypercubes cover the sphere of radius
    $\sqrt{\tau_n n \sigma^2(1+B/\tau_n)}$;
    the probability of this event, namely, $\{\sum_{i=1}^{\tau_n n} Z_i^2 > \tau_n n \sigma^2(1+B/\tau_n)\}$, 
    is easily upper-bounded by the Chernoff bound:
    \begin{eqnarray}
    \mbox{Pr}\left\{\sum_{i=1}^{\tau_n n}Z_i^2 > \tau_n n \sigma^2\left(1+\frac{B}{\tau_n}\right)\right\}
    \le
    \exp\left\{-\frac{n}{2}\left[B-\tau_n\log\left(1+\frac{B}{\tau_n}\right)\right]\right\},
    \end{eqnarray}
    which behaves like $\Exp{-nB/2}$ as $\tau_n\to 0$. Hence, we can make the exponential
    decay of this probability as fast as desired by choosing $B$ sufficiently
    large.
    
    The number of information nats that 
    we can encode in the first sub-block (of length $n\tau_n$) is therefore 
    \begin{subequations}
    \label{eq:app:R'}
    \noeqref{eq:app:R':1}
    \begin{align}
            nR'&\ge n \Rh + \frac{n\tau_n}{2}\log\frac{\Energy}{n\tau_n \sigma^2 (1+B/\tau_n)}+
            n\tau_n\log\left(1-\sqrt{2\pi \e}\cdot \Exp{-\frac{\Rh - o_n(1)}{\tau_n}}\right) 
    \\ & \quad - o(n)-n\tau_n\epsilon(\Delta_n)
    \label{eq:app:R':1}
     \\ &= n \Rh + \frac{n\tau_n}{2}\log\frac{\Energy}{n \sigma^2(B+\tau_n)}+
             n\tau_n\log\left(1-\sqrt{2\pi e}\cdot \Exp{-\frac{\Rh - o_n(1)}{\tau_n}}\right)
    \\ & \quad - o(n)-n\tau_n\epsilon(\Delta_n),
    \label{eq:app:R':2}
    \end{align}
    \end{subequations}
    which is arbitrarily close $n\Rh $ for a sufficiently small
    $\tau_n$, as the second and the third terms in \eqref{eq:app:R':2}
     tend to zero as $\tau_n\log n\to 0$.
    Thus, we can transmit at a rate arbitrarily close to $\Rh $ nats
    with an error
    exponent that is as large as desired. 
    
    Finally, we use this channel code in cascade with uniform quantization of $U$,
    as before. The overall $\alpha$-th moment of the estimation error is of the order of
    $\Exp{-n\alpha\Rh} + \Exp{-nB/2}$, where the second term can be made 
    exponentially negligible as $n$ grows by selecting $B> 2\alpha\Rh$.
    
    Finally, if only the transmitter receives help, we can still approach the DPT converse
    bound by letting $\tau_n$ tend to zero such that $n\tau_n$ is a constant,
    as long as it is a reasonably large constant. In this case, the quantized
    noise sequence, $q \lrp{\bZ_{1:\tau_n n}}$, can be subtracted from the codeword, provided that the allowed energy, $\Energy$, is at least as large as
    $\sigma^2 \tau_n n$. All the above calculations continue to hold with a constant $\tau_n n$
    in mind, as the Stirling approximation applies reasonably well to
    $\Gamma(t+1/2)$ (which plays a role as a factor in the formula of the volume of a
    $\tau_n n$-dimensional ball).
\end{IEEEproof}

\begin{IEEEproof}[Proof of \corref{cor:UB:fixed-E:D:Rh}]
    The resulting \dist\ of 
    \schemeref{scheme:unequal-error-protection} with $L = \log M$
    is bounded for any $u \in [-1/2, 1/2)$ as follows.
    \begin{subequations}
    \label{eq:proof:nat-label-scheme:fixed-E}
    \begin{align}
        &\E{\abs{\hat{U}-u}^\alpha}
        \leq \frac{1}{M^\alpha} + \PR{\hWh \neq \wh} + \frac{1}{\lrp{M'}^\alpha} \PR{\hW_\ell \neq w_\ell}
    \label{eq:proof:nat-label-scheme:fixed-E:bound_by_1}
     \\ &\leq \Exp{-\alpha L} + \Exp{-n \Eh}
     \\ &\quad + \Exp{-n \alpha \lrp{\Rh - \eps}+ o(n)} \cdot
        \begin{cases}
            \Exp{- n \EE_\infty}, & L < n \Rh
         \\ \Exp{-\lrp{\frac{\ENR}{4} - L + n\Rh}}, & 0 \leq L - n\Rh < \frac{\ENR}{8}
         \\ \Exp{-\left( \sqrt{\frac{\ENR}{2}} - \sqrt{L - n\Rh} \right)^2 }, & \frac{\ENR}{8} \leq L - n\Rh < \frac{\ENR}{2} 
        \end{cases}
    \label{eq:proof:nat-label-scheme:fixed-E:EE}
     \\ &\leq \Exp{n \eps + o(n)} \cdot
        \begin{cases}
            \Exp{- \alpha L}, & L < n \Rh
         \\ \Exp{- \min \lrc{\alpha L, \frac{\ENR}{4} - L + \lrp{1 + \alpha} n\Rh}}, & 0 \leq L - n\Rh < \frac{\ENR}{8}
         \\ \Exp{-\min \lrc{ \alpha L, \frac{\ENR}{2} - \sqrt{2 \ENR \lrp{L - n \Rh}} + L } }, & \frac{\ENR}{8} \leq L - n\Rh < \frac{\ENR}{2} 
        \end{cases}
        \quad
    \label{eq:proof:nat-label-scheme:fixed-E:final}
    \end{align} 
    \end{subequations}
    where 
    \eqref{eq:proof:nat-label-scheme:fixed-E:bound_by_1} follows from \eqref{eq:proof:nat-label-scheme:bound_by_1};
    and \eqref{eq:proof:nat-label-scheme:fixed-E:EE}
    follows from \eqref{eq:UB:fixed-E:Pe:Rh}.
    
    Optimization of \eqref{eq:proof:nat-label-scheme:fixed-E:final} over $L$, which is essentially the same as that in \eqref{eq:nat-label-scheme:R*} and \eqref{eq:very-noisy:D}, concludes the proof.
\end{IEEEproof}

\section{Proofs for \secref{ss:crib:Tx-only}}
\label{app:cribbed-helper:PPM}

\begin{IEEEproof}[Proof of \thmref{thm:cribbed-helper:PPM:Pe}]
    Without loss of generality, assume $\sigma^2 = 1$; consequently, $\ENR = \Energy$.
    Denote 
    \begin{align}
        \calI_w \triangleq \left\{w, w + M, w + 2M, \ldots, w + \left( \Mh - 1 \right) M \right\} 
    \end{align}
    and recall that $n = M \cdot \Mh$.
    Denote further $G_w = \max_{\tau \in \calI_w} Z_\tau$. Clearly, the cumulative distribution function (c.d.f.), $F_G$, and probability distribution function (p.d.f.), $f_G$, of $G_w$ equal
    \begin{subequations}
    \label{eq:G}
    \noeqref{eq:G:CDF}
    \begin{align}
        F_G(x) &= F_Z^{\Mh}(x) = \lrs{1 - Q(x)}^{\Mh}, 
    \label{eq:G:CDF}
     \\ f_G(x) &= \Mh F_Z^{\Mh-1}(x) f_Z(x) = \Mh \lrs{1 - Q(x)}^{\Mh - 1} f_Z(x),
    \label{eq:G:PDF}
    \end{align}
    \end{subequations}
    respectively.
    Hence, the error probability may be bounded as follows.
    \begin{subequations}
    \label{eq:cribbed:Tx-only:Pe:bound}
    \noeqref{eq:cribbed:Tx-only:Pe:bound:def}
    \noeqref{eq:cribbed:Tx-only:Pe:bound:pedestrian}
    \noeqref{eq:cribbed:Tx-only:Pe:bound:Q-def}
    \begin{align}
        \Pr& \left( \hW \neq w \right) = \Pr \left( \max_{t \notin \calI_w } Z_t \geq \sqrt{\Energy} + \max_{\tau \in \calI_w} Z_\tau \right)
    \label{eq:cribbed:Tx-only:Pe:bound:def}
     \\ &= \Pr \left( \max_{t \notin \calI_w } Z_t \geq \sqrt{\Energy} + G_w \right)
    \label{eq:cribbed:Tx-only:Pe:bound:G-def}
     \\ &= \int_{-\infty}^\infty \PR{\max_{t \notin \calI_w } Z_t \geq \sqrt{\Energy} + x} f_G(x) dx
    \label{eq:cribbed:Tx-only:Pe:bound:iid}
     \\ &= \int_{-\infty}^\infty \PR{\bigcup_{t \notin \calI_w } Z_t \geq \sqrt{\Energy} + x} 
        \Mh \lrs{1 - Q(x)}^{\Mh - 1} f_Z(x) dx
    \label{eq:cribbed:Tx-only:Pe:bound:fG}
     \\ &\leq \int_{-\infty}^\infty \sum_{t \notin \calI_w} \PR{Z_t \geq \sqrt{\Energy} + x}
        \Mh \lrs{1 - Q(x)}^{\Mh - 1} f_Z(x) dx
    \label{eq:cribbed:Tx-only:Pe:bound:UB}
     \\ &\leq \Mh^2 M \int_{-\infty}^\infty \Q{\sqrt{\Energy} + x}
        \Exp{- \lrp{\Mh - 1} Q(x)} f_Z(x) dx
    \label{eq:cribbed:Tx-only:Pe:bound:e>1+x}
     \\ &\leq \lrs{1 + o_{\Lh}(1)} \Mh^2 M \int_{\sqrt{2 (1-\eps) \Lh}}^\infty \Q{\sqrt{\Energy} + x} \frac{1}{\sqrt{2\pi\e}} \Exp{-\frac{x^2}{2}} dx
    \label{eq:cribbed:Tx-only:Pe:bound:Q>=exp}
     \\ &\leq \lrs{1 + o_{\Lh}(1)} \Mh^2 M \int_{\sqrt{2 (1-\eps) \Lh}}^\infty \Exp{-\frac{\lrp{\sqrt{\Energy} + x}^2}{2}} \frac{1}{\sqrt{2\pi}} \Exp{-\frac{x^2}{2}} dx
    \label{eq:cribbed:Tx-only:Pe:bound:chernoff}
     \\ &= \lrs{1 + o_{\Lh}(1)} \frac{\Mh^2 M}{\sqrt{2}} \Exp{-\frac{\Energy}{4}} \int_{\sqrt{2 (1-\eps) \Lh}}^\infty \frac{1}{\sqrt{\pi}} \Exp{-\lrp{x + \frac{\sqrt{\Energy}}{2}}^2} dx
    \label{eq:cribbed:Tx-only:Pe:bound:pedestrian}
     \\ &= \lrs{1 + o_{\Lh}(1)} \frac{\Mh^2 M}{\sqrt{2}} \Exp{-\frac{\Energy}{4}} \Q{\sqrt{2}\lrs{\sqrt{2 (1-\eps) \Lh} + \frac{\sqrt{\Energy}}{2}}}
    \label{eq:cribbed:Tx-only:Pe:bound:Q-def}
     \\ &\leq \lrs{1 + o_{\Lh}(1)} \Exp{2\Lh + L} 
     \Exp{-\frac{\Energy}{4}} \Exp{-\lrp{\sqrt{2(1 - \eps) \Lh} + \frac{\sqrt{\Energy}}{2}}^2}
    \label{eq:cribbed:Tx-only:Pe:bound:chernoff2}
     \\ &= \Exp{-\lrp{\frac{\Energy}{2} + \sqrt{2 (1 - \eps) \Energy \Lh} - L - 2 \eps \Lh} + o_{\Lh}(1)}
    \label{eq:cribbed:Tx-only:Pe:bound:final}
    \end{align}
    \end{subequations}
    where 
    \eqref{eq:cribbed:Tx-only:Pe:bound:G-def} holds by the definition of $G_w$;
    \eqref{eq:cribbed:Tx-only:Pe:bound:iid} holds since the entries of $\bZ$ are i.i.d.;
    \eqref{eq:cribbed:Tx-only:Pe:bound:fG} follows from \eqref{eq:G:PDF}
    \eqref{eq:cribbed:Tx-only:Pe:bound:UB} follows from the union bound;
    \eqref{eq:cribbed:Tx-only:Pe:bound:e>1+x} follows from the inequality $1 - x \leq \Exp{-x}$ and since the entries of $\bZ$ are i.i.d.\ standard Gaussian;
    \eqref{eq:cribbed:Tx-only:Pe:bound:Q>=exp} 
    holds for any $\eps > 0$, however small, 
    for a sufficiently large $n$, 
    and follows from 
    \begin{subequations}
    \label{eq:(Mh-1)Q>=}
    \noeqref{eq:(Mh-1)Q>=:final}
    \begin{align}
        (\Mh - 1) \Q{x}
        &\geq \lrp{\Exp{\Lh}-1} \Q{\sqrt{2 (1-\eps) \Lh}}
    \label{eq:(Mh-1)Q>=:monotone}
     \\ &\geq \lrp{\Exp{\Lh}-1} \frac{\sqrt{2 (1-\eps) \Lh}}{\sqrt{2\pi} \lrp{1 + \sqrt{2 (1-\eps) \Lh}}^2} \Exp{-\lrp{\Lh - \eps}}
    \label{eq:(Mh-1)Q>=:Q-LB}
     \\ &\geq \Exp{\eps \Lh/2}
    \label{eq:(Mh-1)Q>=:final}
    \end{align}
    \end{subequations}
    for all $x \leq \sqrt{2 (1-\eps) \Lh}$ for any $\eps > 0$ for a sufficiently large $n$, 
    where 
    \eqref{eq:(Mh-1)Q>=:monotone} follows from the monotonicity of the Q function, 
    and \eqref{eq:(Mh-1)Q>=:Q-LB} follows from the lower bound on the Q function \cite{Q-function:approx:com1979}
        $\Q{x} \geq \frac{x}{\sqrt{2\pi} (1+x)^2} \Exp{-x^2/2}$ for $x > 0$,
    meaning that 
    \begin{align}
        \Mh^2 M &\int_{-\infty}^{\sqrt{2 (1-\eps) \Lh}} \Q{\sqrt{\Energy} + x}
        \Exp{- \lrp{\Mh - 1} Q(x)} f_Z(x) dx
     \\ &\leq \Exp{2\Lh + L} \Exp{-\Exp{\eps \Lh/2}}
        \int_{-\infty}^{\sqrt{2 (1-\eps) \Lh}} \Q{\sqrt{\Energy} + x} f_Z(x) dx
     \\ &\leq \Exp{2\Lh + L} \Exp{-\Exp{\eps \Lh/2}}
    \end{align}
    decays \textit{doubly exponentially} with $\Lh$ uniformly for all $E \geq 0$;
    \eqref{eq:cribbed:Tx-only:Pe:bound:chernoff} and \eqref{eq:cribbed:Tx-only:Pe:bound:chernoff2} follow from the Chernoff bound \cite{Q-function:approx:com1979}
    $\Q{x} \leq \Exp{-x^2/2}$;
    and \eqref{eq:cribbed:Tx-only:Pe:bound:final} holds since $1 + o_x(1) = \Exp{o_x(1)}$.
\end{IEEEproof}

\begin{IEEEproof}[Proof of \corref{cor:cribbed-helper:PPM:D}]
    The resulting \dist\ of 
    \schemeref{scheme:cribbed-helper:Tx-only}
    is bounded for any $u \in [-1/2, 1/2)$ as follows.
    \begin{subequations}
    \label{eq:proof:cribbed-helper:D}
    \begin{align}
        \E{\abs{\hat{U}-u}^\alpha} &\leq \PR{\hW \neq w} + \CE{\abs{\hU - u}^\alpha}{\hW = w}
    \label{eq:proof:cribbed-helper:D:bound_by_1}
     \\ &\leq \Exp{- \lrp{\frac{\ENR}{2} + \sqrt{2 \Lh \ENR} - L} + o_{\Lh}(1)} 
        + \Exp{-\alpha L}
    \label{eq:proof:cribbed-helper:D:Pe-bound}
     \\ &= \Exp{- \min \lrc{\frac{\ENR}{2} + \sqrt{2 \Lh \ENR} - L, \alpha L} + o_{\Lh}(1)}
    \label{eq:proof:cribbed-helper:D:min}
    \end{align} 
    \end{subequations}
    where \eqref{eq:proof:cribbed-helper:D:bound_by_1} 
    holds since the estimation error and the probability of correct decoding of $w$ are bounded by 1,
    and \eqref{eq:proof:cribbed-helper:D:Pe-bound} follows from \thmref{thm:cribbed-helper:PPM:Pe}.
    
    Optimization of \eqref{eq:proof:nat-label-scheme:fixed-E:final} is achieved for the value $L$ for which the two operands in the minimum in \eqref{eq:proof:cribbed-helper:D:min} are equal; its substitution concludes the proof.
\end{IEEEproof}



\end{document}